  \documentclass[prd,aps,floats,letterpaper,floatfix,
groupedaddress,superscriptaddress
,nofootinbib
]{revtex4}

  \usepackage{amsthm}
  \usepackage{graphicx}
  \usepackage{amsmath}
\usepackage{amsfonts}

\newcommand{\cl}[1]{{\centerline{#1}}}

\begin{document}

 \title{Was Einstein Right? A Centenary Assessment}

  \author{Clifford M. Will\\
  Department of Physics \\
        University of Florida \\
        Gainesville FL 32611 U.S.A.}


\begin{abstract}
This article is an overview of 100 years of testing general relativity, to be published in the book {\em General Relativity and Gravitation: A Centennial Perspective}, to commemorate the 100th anniversary of general relativity.  It is effectively an abridged version of the recent update of the author's {\em Living Review in Relativity}.
\end{abstract}

\maketitle

\section{Introduction}
\label{S1}

When general relativity was born 100 years ago, experimental
confirmation was almost a side issue.  Admittedly, Einstein did calculate observable
effects of general relativity, such as the perihelion advance of Mercury,
which he knew to be an unsolved problem, and the deflection of light,
which was subsequently verified.  But compared to the inner consistency and
elegance of the theory, he regarded such empirical questions as almost
secondary.   He famously stated that if the measurements of light deflection disagreed with the theory he would ``feel sorry for the dear Lord, for the theory {\em is} correct!''.  

By contrast, today at the centenary of Einstein's towering theoretical achievement, experimental gravitation is a major component of
the field, characterized by continuing efforts to test the theory's
predictions, both in the solar system and in the astronomical world, to detect gravitational waves from astronomical sources, and to search for possible gravitational imprints of phenomena originating in the quantum, high-energy or cosmological realms.

The modern history of experimental relativity can be divided roughly into
four periods: Genesis, Hibernation, a Golden Era, and the Quest for Strong
Gravity. The Genesis (1887\,--\,1919) comprises the period of the two great
experiments which were the foundation of relativistic physics -- the
Michelson--Morley experiment and the E\"otv\"os experiment -- and the two
immediate confirmations of general relativity -- the deflection of light and the
perihelion advance of Mercury.  Following this was a period of Hibernation
(1920\,--\,1960) during which theoretical work temporarily outstripped
technology and experimental possibilities, and, as a consequence, the
field stagnated and was relegated to the backwaters of physics and
astronomy.

But beginning around 1960, astronomical discoveries (quasars, pulsars,
cosmic background radiation) and new experiments pushed general relativity to the
forefront. Experimental gravitation experienced a Golden Era (1960\,--\,1980)
during which a systematic, world-wide effort took place to understand the
observable predictions of general relativity, to compare and contrast them with the
predictions of alternative theories of gravity, and to perform new
experiments to test them.  New technologies -- atomic clocks, radar and laser ranging, space probes, cryogenic capabilities, to mention only a few -- played a central role in this golden era.  The period began with an experiment to confirm
the gravitational frequency shift of light (1960) and ended with the
reported decrease in the orbital period of the Hulse-Taylor binary pulsar
at a rate consistent with the general relativistic prediction of gravitational-wave energy loss (1979). The results all supported general relativity, and most
alternative theories of gravity fell by the wayside (for a popular review,
see~\cite{WER}).

Since that time, the field has entered what might be termed a Quest for Strong
Gravity. 
Much like modern art, the term ``strong'' means different things to different people.  To one steeped in general relativity, the principal figure of merit
that distinguishes strong from weak gravity is the quantity $\epsilon \sim
GM/Rc^2$, where $G$ is the Newtonian gravitational constant, $M$ is the
characteristic mass scale of the phenomenon, $R$ is the characteristic
distance scale, and $c$ is the speed of light.  Near the event horizon of
a non-rotating black hole, or for the expanding observable universe,
$\epsilon \sim 1$; for neutron stars, $\epsilon \sim 0.2$. These are
the regimes of strong gravity. For the solar system, $\epsilon < 10^{-5}$;
this is the regime of weak gravity.   

An alternative view of ``strong'' gravity comes from the world of particle physics.  Here the figure of merit is $GM/R^3c^2 \sim \ell^{-2}$, where the Riemann curvature of spacetime associated with the phenomenon, represented by the left-hand-side, is comparable to the inverse square of a favorite length scale $\ell$.  If $\ell$ is the Planck length, this would correspond to the regime where one expects conventional quantum gravity effects to come into play.   Another possible scale for $\ell$ is the TeV scale associated with many models for unification of the forces, or models with extra spacetime dimensions.   From this viewpoint, strong gravity is where the curvature is comparable to the inverse length squared.  Weak gravity is where the curvature is much smaller than this.  The universe at the Planck time is strong gravity.  Just outside the event horizon of an astrophysical black hole is weak gravity. 

Considerations of the possibilities for new physics from either point of view have led to a wide range of questions that will motivate new tests of general relativity as we move into its second century:

\begin{itemize}

\item
Are the black holes that are in evidence throughout the universe truly the black holes of general relativity?

\item
Do gravitational waves propagate with the speed of light and do they contain more than the two basic polarization states of general relativity?

\item
Does general relativity hold on cosmological distance scales?

\item
Is Lorentz invariance strictly valid, or could it be violated at some detectable level?

\item
Does the principle of equivalence break down at some level?

\item
Are there testable effects arising from the quantization of gravity?

\end{itemize}  

In this centenary assessment of the experimental basis of general relativity, we will summarize the current status of
experiments, and attempt to chart the future of the subject.  We will not
provide complete references to early work done in this field but instead
will refer the reader to selected recent papers and to the appropriate review articles and monographs.  For derivations of many of the effects discussed in this article we will refer to \textit{Theory and Experiment in Gravitational
Physics}~\cite{tegp}, hereafter referred to as TEGP;  references to
TEGP will be by chapter or section, e.g., ``TEGP~8.9''.  For a more comprehensive review, we refer readers to the author's ``Living Review'' \cite{WillLivrev} (or its recent update~\cite{2014arXiv1403.7377W})\footnote{This article was prepared roughly in parallel with the author's Living Review update, and may be considered a streamlined version of the latter.  This will account for essentially identical prose between the two papers in numerous places}.   The ``Resource Letter'' by the author \cite{2010AmJPh..78.1240W}, contains 100 key references for experimental gravity.

\section{The foundations of gravitation theory}
\label{S2}

\subsection{The Einstein equivalence principle}
\label{eep}

The principle of equivalence has historically played an important role in
the development of gravitation theory.  Newton regarded this principle as
such a cornerstone of mechanics that he devoted the opening paragraph of
the \textit{Principia} to it. In 1907, Einstein used the principle as a
basic element in his development of
general relativity.  We now regard the principle of
equivalence as the foundation, not of Newtonian gravity or of general relativity, but of
the broader idea that spacetime is curved.
Much of
this viewpoint can be traced back to Robert Dicke, who contributed crucial
ideas
about the foundations of gravitation theory between 1960 and 1965. These
ideas were summarized in his influential Les Houches lectures of
1964~\cite{dicke64},  and resulted in what has come to be called the
Einstein equivalence principle (EEP).

One elementary equivalence principle is the kind Newton had in mind when
he stated that the property of a body called ``mass'' is proportional to
the ``weight'', and is known as the weak equivalence principle (WEP).
An alternative statement of WEP is that the trajectory of a freely
falling ``test'' 
body (one not acted upon by such forces as electromagnetism,
too small to be affected by tidal gravitational forces, and spinless) is independent of
its internal structure and composition. In the simplest case of dropping
two different bodies in a gravitational field, WEP states that the bodies
fall with the same acceleration.  This is often termed the Universality of
Free Fall, or UFF.

The Einstein equivalence principle (EEP) is a more powerful and
far-reaching concept; it has three components:
\begin{enumerate}
\item 
{\em Weak Equivalence Principle}.  The trajectory of a freely
falling ``test'' 
body is independent of
its internal structure and composition.
\item 
{\em Local Lorentz Invariance}.  The outcome of any local non-gravitational 
experiment is
  independent of the velocity of the freely-falling reference frame in
  which it is performed.
\item 
{\em Local Position Invariance}.  The outcome of any local non-gravitational 
experiment is
  independent of where and when in the universe it is performed.
\end{enumerate}
The Einstein equivalence principle is the heart and soul of gravitational
theory, for it is possible to argue convincingly that if EEP is valid,
then gravitation must be a ``curved spacetime'' phenomenon, in other
words,  gravity must be governed by a ``metric theory of gravity'', whose postulates are:
\begin{enumerate}
\item Spacetime is endowed with a symmetric metric.
\item The trajectories of freely falling test bodies are geodesics of
  that metric.
\item In local freely falling reference frames, the non-gravitational
  laws of physics are those written in the language of special
  relativity.
\end{enumerate}

General relativity is a metric theory of gravity, but then so are
many others, including the Brans--Dicke theory and its generalizations.  
It is not uncommon for modern variants of general relativity, especially those motivated by quantum gravity, unification, or extra dimensions, to introduce weak, {\em effective} violations of the metric postulates.
Accordingly, it is important to test the various aspects of
the Einstein equivalence principle thoroughly.
We first survey the experimental tests, and describe some of the theoretical
formalisms that have been developed to interpret them.

\subsubsection{Tests of the weak equivalence principle}
\label{wep}

A direct test of WEP is the comparison of the acceleration of two
laborat\-ory-sized bodies of different composition in an external
gravitational field.
If the principle were violated, then the
accelerations of different bodies would differ. 
A measurement or limit
on the fractional difference in acceleration between two bodies
then yields a quantity called the ``E\"otv\"os ratio'' given by
$ \eta \equiv 2 |a_1 - a_2|/|a_1 + a_2|$. 

Many high-precision E\"otv\"os-type experiments have been performed, from
the pendulum experiments of Newton, Bessel, and Potter to the classic
torsion-balance measurements of E\"otv\"os,
Dicke, Braginsky, and their collaborators.
In modern torsion-balance experiments, two objects of different
composition are connected by a rod or placed on a tray and suspended in a
horizontal orientation by a fine wire. If the gravitational acceleration
of the bodies differs, and this difference has a component
perpendicular to
the suspension wire, there will be a torque induced on the 
wire, related to the angle between the wire and the direction of the
gravitational acceleration. 
Beginning in
the late 1980s, numerous experiments were carried out primarily to search
for a ``fifth force'', but their null
results also constituted tests of WEP.  The ``E\"ot-Wash''
experiments carried out at the University of Washington used a
sophisticated torsion balance tray to compare the accelerations of various
materials toward local topography on Earth, movable laboratory masses, the
Sun and the galaxy, and have reached levels of
$2 \times 10^{-13}$~\cite{2008PhRvL.100d1101S}. 

The recent development of atom interferometry has yielded tests of WEP, albeit to modest accuracy, comparable to that of the original E\"otv\"os experiment.  In these experiments, one measures the local acceleration of the two separated wavefunctions of an atom such as Cesium by studying the the interference pattern when the wavefunctions are combined, and compares that with the acceleration of a nearby macroscopic object of different composition~\cite{2010Metro..47L...9M,2010Natur.463..926M}.
A claim
\cite{2010Natur.463..926M} that these experiments test the gravitational redshift was subsequently shown to be incorrect~\cite{2011CQGra..28n5017W}.  Further improvements are anticipated~\cite{2009aosp.conf..411H}.

\begin{figure}[t]
  \includegraphics[scale=0.45]{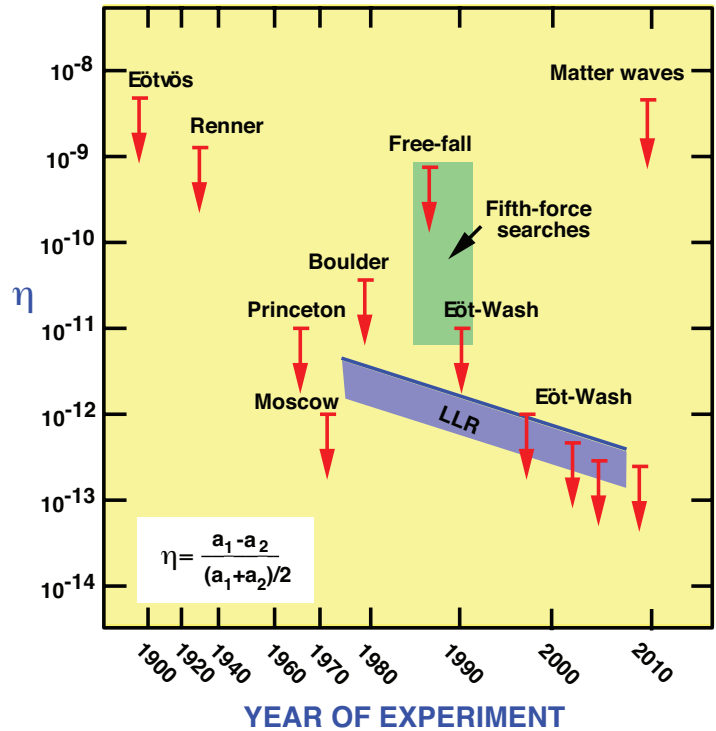}
  \caption{Selected tests of the weak equivalence principle,
    showing bounds on $\eta$.
    The free-fall and
    E\"ot-Wash experiments were originally performed to search for a
    fifth force (green region, representing many experiments). 
    The blue band shows evolving bounds on $\eta$ for
    gravitating bodies from lunar laser ranging (LLR).}
  \label{wepfig}
\end{figure}

The resulting upper limits on $\eta$ are summarized
in Figure~\ref{wepfig} (for an extensive bibliography of
experiments up to 1991, see~\cite{fischbach92}).

A number of projects are in the development or planning stage to push
the bounds on $\eta$ even lower. The project MICROSCOPE
is
designed to test WEP to $10^{-15}$. It is being developed
by the French space agency CNES for  launch in late 2015,
for a one-year mission.
The drag-compensated 
satellite will be in a Sun-synchronous polar orbit at 700~km
altitude, with a payload consisting of two differential
accelerometers, one with elements made of the same material
(platinum), and another with elements made of different materials
(platinum and titanium).  Other concepts for future improvements include advanced space experiments (Galileo-Galilei, STEP), lunar laser ranging (see Sec.\ \ref{Nordtvedteffect}), binary pulsar observations, and experiments with anti-hydrogen.  For a review of past and future tests of WEP, see Vol.\ 29, Issue 18 of {\em Classical and Quantum Gravity} \cite{0264-9381-29-18-180301} and the review by Adelberger et al.~\cite{2009PrPNP..62..102A}.


\subsubsection{Tests of local Lorentz invariance}
\label{lli}

Although special relativity itself never benefited from the kind of
``crucial'' experiments, such as the perihelion advance of Mercury and
the deflection of light, that contributed so much to the initial acceptance
of
general relativity and to the fame of Einstein, the steady
accumulation of experimental support, together with the successful
merger of special relativity
with quantum mechanics, led to its being accepted
by mainstream physicists by the late 1920s, ultimately to become part of
the standard toolkit of every working physicist.
This accumulation included
\begin{itemize}
\item the classic Michelson--Morley experiment and its
  descendents;
\item the Ives--Stillwell, Rossi--Hall, and other tests of
  time-dilation;
\item tests of the independence of the speed of light of the velocity
  of the source, using both binary X-ray stellar sources and
  high-energy pions;
\item tests of the isotropy of the speed of light.
\end{itemize}

In addition to these direct experiments, there was the Dirac equation
of quantum mechanics and its prediction of anti-particles and spin;
later would come the stunningly successful relativistic theory of
quantum electrodynamics.  For a pedagogical review on the occasion of the 2005 centenary of special relativity, see~\cite{2006eins.book...33W}.

In 2015, on the 110th anniversary of special relativity,
one might ask ``what is there left to test?'' Special relativity has
been so thoroughly
integrated into the fabric of modern physics that its validity is
rarely
challenged, except by cranks and crackpots.
It is ironic then, that during the past 15 years, a vigorous
theoretical and experimental effort has been launched to find violations of special relativity.
The motivation for this effort is not a desire
to repudiate Einstein, but to look for
evidence of new physics ``beyond'' Einstein, such as apparent, or ``effective''
violations
of Lorentz invariance that might result from certain models of quantum
gravity.
Quantum gravity asserts that there is a fundamental length scale
given by the Planck length, $\ell_\mathrm{Pl} = (\hbar G/c^3)^{1/2} = 1.6 \times
10^{-33} \mathrm{\ cm}$, but since length is not an invariant quantity
(Lorentz--FitzGerald contraction), then there could be a violation of
Lorentz
invariance at some level in quantum gravity. In brane world
scenarios, while
physics may be locally Lorentz invariant in the higher dimensional
world,
the confinement of the interactions of normal physics to our
four-dimensional ``brane'' could induce apparent Lorentz violating
effects.
And in models such as string theory, the presence of additional
vector and tensor long-range fields that couple to matter of the
standard
model could induce effective violations of Lorentz symmetry.
These and other ideas have motivated
a serious
reconsideration of how to test Lorentz invariance with better
precision and
in new ways.

A simple and useful way of interpreting some of these modern
experiments, called the $c^2$-formalism,  is to suppose that the
electromagnetic interactions suffer a slight violation of Lorentz
invariance, through a change in the speed of electromagnetic radiation $c$
relative to the limiting speed of material test particles ($c_0$, made 
to take the value unity via a choice of units), in other words, $c \ne 1$. Such a violation necessarily selects a preferred
universal rest frame, presumably that of the cosmic background radiation,
through which we are moving at about $370 \mathrm{\ km\ s}^{-1}$. Such a
Lorentz-non-invariant electromagnetic interaction would cause shifts in
the energy levels of atoms and nuclei that depend on the orientation of
the quantization axis of the state relative to our universal velocity
vector, and on the quantum numbers of the state. The presence or absence
of such energy shifts can be examined by measuring the energy of one such
state relative to another state that is either unaffected or is affected
differently by the supposed violation.
The magnitude of these ``clock
anisotropies'' turn out to be proportional to
$\delta \equiv | c^{-2}-1|$ (see TEGP 2.6(a) for details).

\begin{figure}[t]
  \includegraphics[scale=0.45]{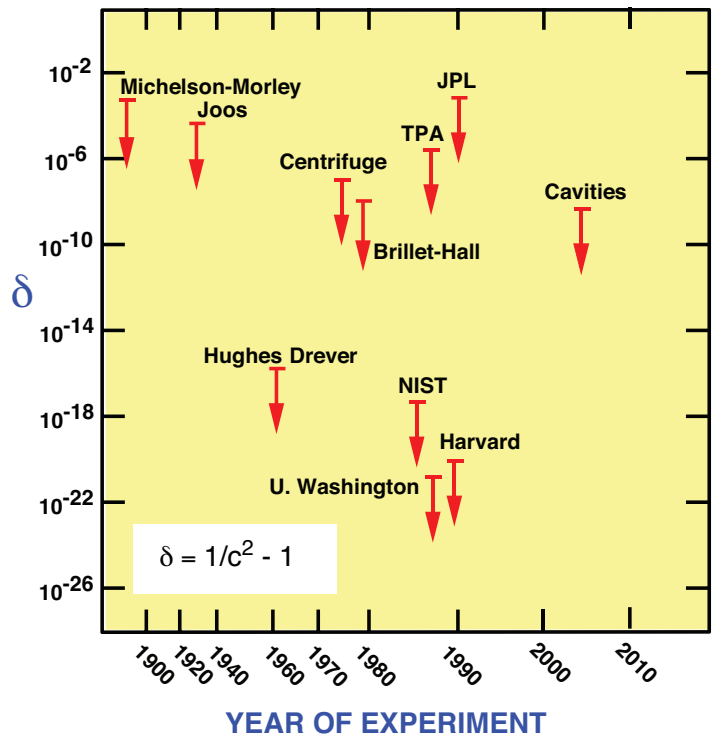}
  \caption{Selected tests of local Lorentz invariance showing the
    bounds on the parameter $\delta$, which measures the degree of
    violation of Lorentz invariance in electromagnetism. The
    Michelson--Morley, Joos, Brillet--Hall and cavity experiments test the
    isotropy of the round-trip speed of light. 
    The centrifuge, two-photon absorption (TPA) and
    JPL experiments test the isotropy of light speed using one-way
    propagation. The most precise experiments test isotropy of
    atomic energy levels.}
  \label{llifig}
\end{figure}

The earliest clock anisotropy experiments were the
Hughes--Drever experiments, performed in the period
1959\,--\,60, yielding limits
on the parameter $\delta$ shown in Fig.~\ref{llifig}. 
Dramatic improvements were made in the 1980s using
laser-cooled trapped atoms and ions, which made it possible
to reduce the broading of resonance lines caused by collisions.
Also
included for comparison in Fig.~\ref{llifig} is the corresponding limit obtained from
Michelson--Morley type experiments  and their modern variants such as the Brillet-Hall experiment, and comparisons between cavity oscillators and atomic clocks.

The $c^2$ framework focusses exclusively on classical electrodynamics.
It has been extended to the entire standard model of particle physics by
Kosteleck\'y and collaborators.  Called the Standard Model Extension (SME), it takes the standard
$\mathrm{SU(3)} \times \mathrm{SU(2)} \times \mathrm{U(1)}$ field theory of particle physics, and
modifies the terms in the action by inserting a variety of tensorial
quantities in the quark, lepton, Higgs, and gauge boson sectors 
that could explicitly violate LLI.   The modified terms split naturally 
into those that are
odd under CPT (i.e.\ that violate CPT) and terms that are even under CPT.
The result is a rich and complex framework, with many parameters to be
analyzed and tested by experiment.  Experimentalists have risen to the challenge, and have placed interesting bounds on many of the SME parameters.  
A variety of clock anisotropy experiments have been carried out to
bound the electromagnetic parameters of the SME 
framework.  Other testable effects of Lorentz-invariance violation include arrival-time variations in TeV gamma-ray bursts from blazars,  threshold
effects in particle reactions, birefringence in photon propagation,
gravitational Cerenkov radiation, and neutrino oscillations.
The details of the SME and other approaches to testing LLI are  beyond the scope of this article; the reader is referred to the reviews by
Mattingly~\cite{mattingly}, Liberati~\cite{Liberati2013} and Kosteleck\'y and Russell~\cite{RevModPhys.83.11}.  The last article gives ``data tables'' showing experimental bounds on all the various parameters of the model.
The SME has also been extended to a parametrization of local Lorentz violations in gravity (see, for example, \cite{PhysRevD.83.016013}).

\subsubsection{Tests of local position invariance}
\label{lpi}

The principle of local position invariance, the third part of
EEP, can be tested by the gravitational redshift
experiment,
 the first experimental test of gravitation proposed by Einstein, eight years before the full theory of general relativity. Despite
the fact that Einstein regarded this as a crucial test of general relativity, we now
realize that it does not distinguish between general relativity and any other metric
theory of gravity, but is only a test of EEP. A typical gravitational
redshift experiment measures the frequency or wavelength shift $Z \equiv
\Delta \nu / \nu = - \Delta \lambda / \lambda = {\Delta U}/{c^2}$ between two identical
frequency standards (clocks) placed at rest at different heights in a
static gravitational potential $U$. 
If LPI is not
valid, then it turns out that the shift can be written
\begin{equation}
  Z = (1 + \alpha) \frac{\Delta U}{c^2},
  \label{E4}
\end{equation}
where the parameter $\alpha$ may depend upon the nature of the
clock whose shift is being measured (see TEGP~2.4~(c) for
details).

\begin{figure}[t]

 \includegraphics[scale=0.45]{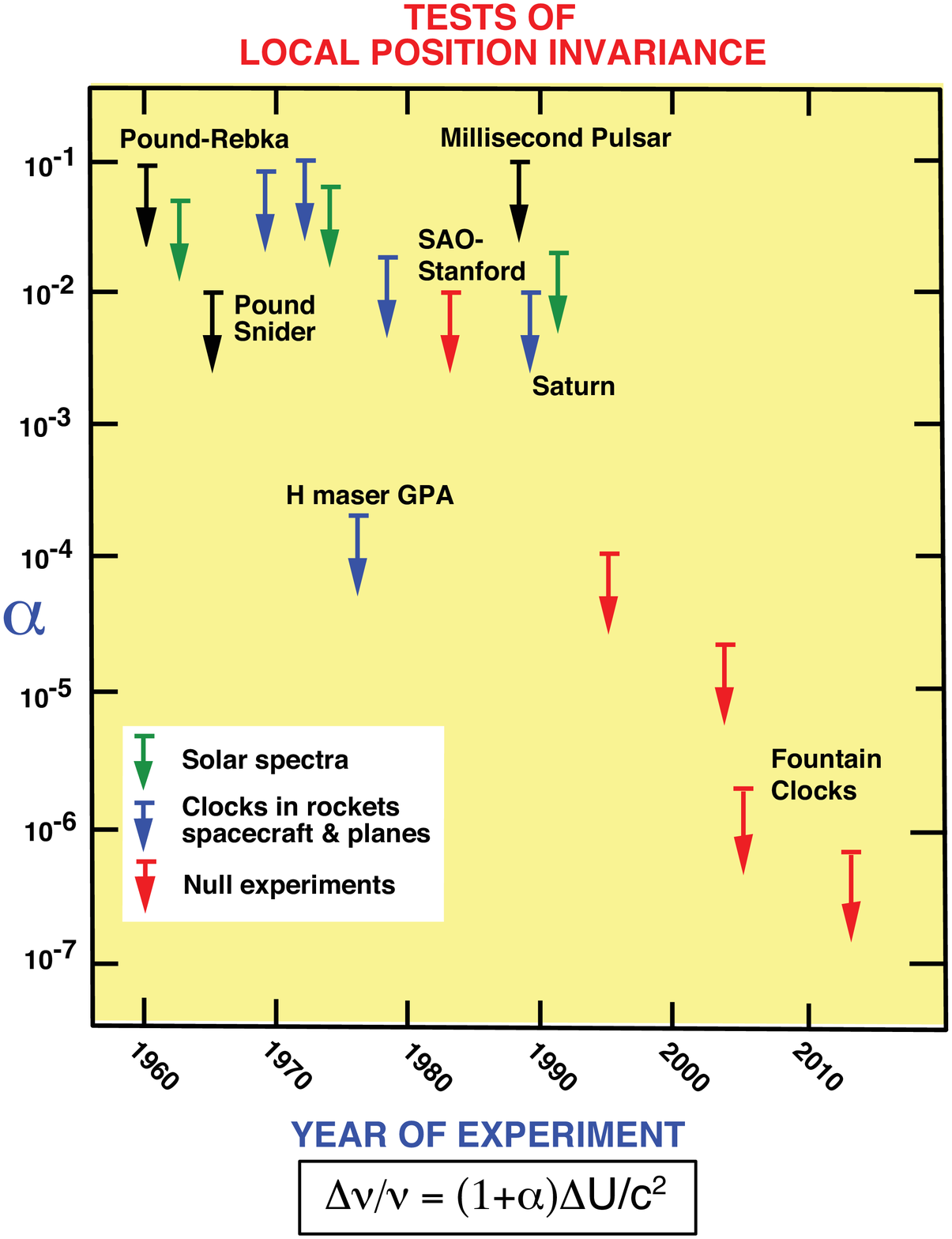}
  \caption{Selected tests of local position invariance via
    gravitational redshift experiments, showing bounds on $\alpha$,
    which measures degree of deviation of redshift from the formula
    $\Delta \nu / \nu = \Delta U/c^2$. In null redshift experiments, the
    bound is on the difference in $\alpha$ between different kinds of
    clocks.}
  \label{lpifig}
\end{figure}

The first successful, high-precision redshift measurement was the
series of Pound--Rebka--Snider experiments of 1960\,--\,1965 that
measured the frequency shift of gamma-ray photons from ${}^{57} \mathrm{Fe}$
as they ascended or descended the Jefferson Physical Laboratory
tower at Harvard University. The high accuracy
achieved -- one percent -- was obtained by making
use of the M\"ossbauer effect to produce a narrow resonance line
whose shift could be accurately determined. Other experiments
since 1960 measured the shift of spectral lines in the Sun's
gravitational field and the change in rate of atomic clocks
transported aloft on aircraft, rockets and satellites. Figure~\ref{lpifig}
summarizes the important test of LPI that have been
performed since 1960.

After almost 50 years of inconclusive or contradictory measurements, the
gravitational redshift of solar spectral lines was finally  measured
reliably. During the early years of general relativity, the failure to
measure this effect in solar lines
was siezed upon by some as reason to doubt the theory.
Unfortunately, the measurement is not simple.
Solar spectral lines are subject to the ``limb effect'', a variation
of line wavelengths between the center of the solar disk and
its edge or ``limb''; this effect is actually a Doppler shift caused by complex
convective and turbulent motions in the photosphere and lower
chromosphere, and is expected to be minimized by observing at the
solar limb, where the motions are predominantly transverse. The
secret is to use strong, symmetrical lines, leading to unambiguous
wavelength measurements. Successful measurements were finally made
in 1962 and 1972. In 1991, LoPresto et
al.
measured the solar shift in agreement with LPI to about 2~percent by 
observing the oxygen triplet lines both in
absorption in the limb and in emission just off the limb~\cite{lopresto91}.

The most precise standard redshift
test to date was the Vessot--Levine rocket
experiment known as Gravity Probe-A (GPA) that took place in June 1976.  A
hydrogen-maser clock was flown on a rocket to an altitude of
about 10,000~km and its frequency compared to a similar clock on
the ground. 
Analysis of the data yielded a limit
$| \alpha | < 2 \times 10^{-4}$.

A ``null'' redshift experiment performed in 1978 tested whether the {\it
relative} rates of two different clocks  --- hydrogen masers and cavity stabilized oscillators --- depended upon the diurnal and annual variations of the solar gravitational potential at the location of the laboratory. 
This bound was subsequently improved using more stable frequency
standards, such as atomic fountain clocks.  The best
current bounds, from comparing various pairings of Rubidium, Cesium, Hydrogren and isotopes of Dysprosium~\cite{2012PhRvL.109h0801G,2013PhRvA..87a0102P,2013PhRvL.111f0801L} hover around the one part per million mark. 

The Atomic Clock Ensemble in Space (ACES) project will place both a cold trapped atom clock based on Cesium called PHARAO (Projet d'Horloge Atomique par Refroidissement d'Atomes en Orbit), and an advanced hydrogen maser clock on the International Space Station to measure the gravitational redshift to parts in $10^6$, as well as to carry out a number of fundamental physics experiments and to enable improvements in global timekeeping.  Launch is currently scheduled for May 2016.  

Modern navigation using Earth-orbiting atomic clocks and
accurate time-transfer must routinely take gravitational redshift and
time-dilation effects into account. For example, the Global Positioning
System (GPS) provides absolute positional
accuracies of around 15~m (even better in
its military mode), and 50 nanoseconds
in time transfer accuracy, anywhere on Earth.
Yet the difference in rate between
satellite and ground clocks as a result of 
relativistic effects is a whopping 39~\emph{microseconds} per day
($46 \mathrm{\ \mu s}$ from the gravitational redshift, and $-7
\mathrm{\ \mu s}$
from time dilation). If these effects were not accurately accounted for,
GPS would fail to function at its stated accuracy. This represents a
welcome practical application of general relativity!  For the role of general relativity in GPS,
see~\cite{ashby2}; for a popular essay, see~\cite{physicscentral}.

Local position invariance also refers to position in time. If LPI is
satisfied, the fundamental constants of non-gravitational physics
should be constants in time. Table~\ref{varyconstants} shows current bounds on
cosmological variations in selected dimensionless constants. For
discussion and references to early work, see TEGP~2.4~(c)
or~\cite{dyson72}. For a comprehensive recent review 
both of experiments and of 
theoretical ideas that underly proposals for
varying constants, 
see~\cite{2011LRR....14....2U}.  

\begin{table}[t]
  \caption[Bounds on cosmological variation of fundamental
    constants of non-gravitational physics.]{Bounds on cosmological
    variation of fundamental constants of non-gravitational
    physics. For an in-depth review, see~\cite{2011LRR....14....2U}.}
  \label{varyconstants}
  \renewcommand{\arraystretch}{1.2}
  \centering
  \begin{tabular}{p{3.0 cm}|rlp{4.0 cm}}
    \hline \hline
    Constant $k$ &
    \multicolumn{1}{c}{Limit on $\rule{0 em}{1.2 em}\dot{k} / k$} &
    \multicolumn{1}{c}{Redshift} &
    \multicolumn{1}{c}{Method} \\ 
    &
    \multicolumn{1}{c}{($ \mathrm{yr}^{-1} $)} \\
    \hline \hline
    Fine structure  &
    $  1.3 \times 10^{-16} $ &
    $ 0 $ &
    Clock comparisons  \\
    constant& $  0.5 \times 10^{-16} $ &
    $ 0.15 $ &
    Oklo Natural Reactor 
    \\
   ($ \alpha_\mathrm{EM} = e^2 / \hbar c $)  & $  3.4 \times 10^{-16} $ &
    $ 0.45 $ &
    $ {}^{187}\mathrm{Re} $
    decay in meteorites  \\
    & $  1.2 \times 10^{-16} $ &
    $ 0.4 \mbox{\,--\,} 2.3 $ &
    Spectra in distant quasars  \\
    \hline
    Weak interaction  &$  1 \times 10^{-11} $ &
    $ 0.15 $ &
    Oklo Natural Reactor \\
    constant& $  5 \times 10^{-12} $ & $ 10^9 $ &
    Big Bang nucleosynthesis  \\
    ($ \alpha_\mathrm{W} = G_\mathrm{f} m_\mathrm{p}^2 c / \hbar^3 $)&\\
    \hline
    e-p mass ratio &
    $  3.3 \times 10^{-15}$ & 0 &
    Clock comparisons \\
    &$  3 \times 10^{-15} $ &
    $ 2.6 \mbox{\,--\,} 3.0 $ &
    Spectra in distant quasars\\
    \hline \hline
  \end{tabular}
  \renewcommand{\arraystretch}{1.0}
\end{table}

Experimental bounds on varying constants come in two types: bounds on
the present rate of variation, and bounds on the difference between
today's value and a value in the distant past. The main example of
the former type is the clock comparison test, in which highly stable
atomic clocks of different fundamental type are intercompared over
periods ranging from months to years (variants of the null redshift
experiment). 
The second type of bound involves measuring the relics of or signal from
a process that occurred in the distant
past and comparing the inferred value of the constant with the value
measured in the laboratory today. 
One sub-type uses astronomical
measurements of spectral lines at large redshift, while the other uses
fossils of nuclear processes on Earth to infer values of constants
early in geological history, notably from the Oklo natural nuclear reactor in Gabon. 
Despite continuing reports by one group~\cite{2012MNRAS.422.3370K} of a variation of the fine structure constant over cosmic time as seen in spectral lines, most groups find no significant variation (eg.~\cite{2012ApJ...746L..16K,2013MNRAS.430.2454L}).

\subsubsection{EEP, particle physics, and the search for new interactions}
\label{newinteractions}

There is mounting 
theoretical evidence to suggest that EEP is 
likely to be violated at some level, whether by quantum gravity
effects, by effects arising from string theory, or by hitherto
undetected interactions.  As a result, EEP and related tests are now viewed as ways 
to discover or place
constraints on new physical interactions, or as a branch of
``non-accelerator particle physics'', searching for the possible imprints
of high-energy particle effects in the low-energy realm of gravity.
One example of this is the ``fifth-force'' episode of the middle 1980s, in which E\"otv\"os-type experiments and tests of the gravitational inverse-square law were used to limit the existence of intermediate-range forces that could arise from particle interactions beyond the standard model.  Tests of the inverse square law at sub-mm distance scales continue in an effort to search for evidence of extra dimensions or light scalar particles~\cite{2009PrPNP..62..102A}.    Anomalies in the orbit of the Pioneer 10 and 11 spacecraft at 20 to 70 astronomical units from the Sun were touted for a time as evidence for new physics, but were ultimately shown to be the result of anisotropic emission of heat from the spacecraft~\cite{2010LRR....13....4T,2012PhRvL.108x1101T}.

\subsection{Metric theories of gravity and the strong equivalence principle}
\label{metrictheories}

The  Einstein
equivalence principle is strongly supported by empirical evidence, as we have seen.
This tells us that the only theories of
gravity that have a hope of being viable are metric
theories, or possibly theories that at worst admit very weak
or short-range non-metric couplings.  Therefore for
the remainder of this article, we will turn our attention
exclusively to metric theories of gravity.

The property that all non-gravitational fields should couple in
the same manner to a single gravitational field is sometimes
called ``universal coupling''.  Because of it, one can discuss the
metric as a property of spacetime itself rather than as a field
over spacetime. This is because its properties may be measured
and studied using a variety of different experimental devices,
composed of different non-gravitational fields and particles,
and, because of universal coupling, the results will be
independent of the device. Thus, for instance, the proper time
between two events is a characteristic of spacetime and of the
location of the events, not of the clocks used to measure it.

Mathematically, if EEP is valid, the non-gravitational laws of
physics may be formulated by taking their special relativistic
forms in terms of the Min\-kowski metric {\boldmath $\eta$} and simply
``going over'' to new forms in terms of the curved spacetime
metric {\boldmath $g$}, using the mathematics of differential geometry.
The details of this ``going over'' can be found in standard
textbooks (see~\cite{2009fcgr.book.....S}, TEGP~3.2.).



In any metric theory of gravity, matter and non-gravitational
fields respond only to the spacetime metric {\boldmath $g$}. In
principle, however, there could exist other gravitational fields
besides the metric, such as scalar fields, vector fields, and so
on.  The role of such fields is to mediate
the manner in which matter and non-gravitational fields generate
gravitational fields and produce the metric; once determined,
however, the metric alone acts back on the matter in the manner
prescribed by EEP.

What distinguishes one metric theory from another, therefore, is
the number and kind of gravitational fields it contains in
addition to the metric, and the equations that determine the
structure and evolution of these fields.   The fields could be dynamical, in that they obey their own set of dynamical equations, or they could be ``fixed'' or non-dynamical --- an example of the latter is a background Minkowski  metric.

General relativity is a purely dynamical theory since it contains
only one gravitational field, the metric itself, and its
structure and evolution are governed by partial differential
equations (Einstein's equations). Brans--Dicke theory and its
generalizations are purely
dynamical theories; the field equation for the metric involves the
scalar field (as well as the matter as source), and that for the
scalar field involves the metric. 

Consider now a local gravitating system, as described by a chosen metric theory.
 The system could be a star, a black hole,
the solar system, or a Cavendish experiment. 
 Because the metric is coupled directly or
indirectly to the other fields of the theory, the gravitational
environment in which the local gravitating system resides can
influence the metric generated by the local system via the
boundary values of the auxiliary fields.  Those boundary values
typically depend on the distribution of matter external to the system.
Consequently, the
results of local gravitational experiments may depend on the
location and velocity of the frame relative to the external
environment. Of course, local \emph{non}-gravitational
experiments are unaffected since the gravitational fields they
generate are assumed to be negligible, and since those
experiments couple only to the metric, whose form can always be
made locally Minkowskian at a given spacetime event.

 A theory which contains only the metric {\boldmath $g$} (such as general relativity)
 yields
  local gravitational physics which is independent of the location and
  velocity of the local system. This follows from the fact that the
  only field coupling the local system to the environment is
  {\boldmath $g$}, and it is always possible to find a coordinate
  system in which {\boldmath $g$} takes the Minkowski form at the
  boundary between the local system and the external environment
  (neglecting inhomogeneities in the external gravitational
  field).  Thus, apart from standard tidal effects, the external environment cannot act back on the local gravitating system.

 A theory which contains the metric {\boldmath $g$} and dynamical
  scalar fields $\varphi_A$ yields local gravitational physics which
  may depend on the location of the frame but which is independent of
  the velocity of the frame. This follows from the asymptotic Lorentz
  invariance of the Minkowski metric and of the scalar fields, but now
  the asymptotic values of the scalar fields may depend on the
  location of the frame.   For example in scalar-tensor theories, the effective 
  gravitational coupling strength depends on the scalar field, which may vary in space 
  and time, depending on the external environment.

A theory which contains the metric {\boldmath $g$} and
  additional dynamical vector or tensor fields or prior-geometric
  fields yields local gravitational physics which may have both
  location and velocity-dependent effects.  
These ideas can be summarized in the strong equivalence principle
(SEP), which states that:
\begin{enumerate}
\item WEP is valid for self-gravitating bodies as well as for test
  bodies.
\item The outcome of any local test experiment is independent of the
  velocity of the (freely falling) apparatus.
\item The outcome of any local test experiment is independent of where
  and when in the universe it is performed.
\end{enumerate}
The distinction between SEP and EEP is the inclusion of bodies
with self-gravitational interactions (planets, stars) and of
experiments involving gravitational forces (Cavendish
experiments, gravimeter measurements). Note that SEP contains
EEP as the special case in which local gravitational forces are
ignored.

General relativity seems to be the only viable
metric theory that embodies SEP completely. In
Section~\ref{septests}, we will discuss experimental evidence for the
validity of SEP.


\section{The parametrized post-Newtonian formalism}
\label{ppn}

Despite the possible existence of long-range gravitational fields
in addition to the metric,  matter and
non-gravitational fields are completely oblivious to them in metric theories of gravity. The
only gravitational field that enters the equations of motion is
the metric {\boldmath $g$}.
Thus the metric and the equations of motion for matter become the
primary entities for calculating observable effects,
and all that distinguishes one
metric theory from another is the particular way in which matter
and possibly other gravitational fields generate the metric.

The comparison of metric theories of gravity with each other and
with experiment becomes particularly simple when one takes the
slow-motion, weak-field limit. This approximation, known as the
post-Newtonian limit, is sufficiently accurate to encompass most
solar-system tests that can be performed in the foreseeable
future. In this limit, the spacetime metric
{\boldmath $g$} predicted by a broad class of metric theories of gravity has the
same structure. It can be written as an expansion about the
Minkowski metric ($ \eta_{\mu\nu} = {\rm diag} (-1,1,1,1)$) in terms
of dimensionless gravitational potentials of varying degrees of
smallness.
These potentials are constructed from the matter
variables, such as mass density $\rho$, energy density $\rho \Pi$, pressure $p$, four-velocity $u^\alpha$, ordinary velocity $v^j \equiv u^j/u^0$, etc., in imitation of the Newtonian gravitational
potential
\begin{equation}
  U ({\bf x}, t) \equiv G \int \rho ({\bf x}', t)
  |{\bf x} - {\bf x}'|^{-1} \, d^3 x'.
  \label{E20}
\end{equation}
The ``order of smallness'' is determined according to the rules
$U/c^2 \sim~(v/c)^2 \sim \Pi/c^2 \sim p/ \rho c^2 \sim \epsilon$,
$v^i/c \sim | d/d(ct) | / | d/dx | \sim \epsilon^{1/2}$, and so on.

A consistent post-Newtonian limit requires determination of $g_{00}$
correct through ${\cal O} (\epsilon^2)$,
$g_{0i}$ through ${\cal O} (\epsilon^{3/2})$, and $g_{ij}$
through ${\cal O} (\epsilon)$ (for details see TEGP~4.1 or \cite{PW2014}). The only way that
one metric theory differs from another is in the numerical values
of the coefficients that appear in front of the metric
potentials. The parametrized post-Newtonian (PPN) formalism
inserts parameters in place of these coefficients, parameters
whose values depend on the theory under study. In the current
version of the PPN  formalism, ten
parameters are used, chosen in such a manner that they measure or
indicate general properties of metric theories of gravity
(see Table~\ref{ppnmeaning}). Under reasonable assumptions about the
kinds of potentials that can be present at post-Newtonian order
(basically only Poisson-like potentials), one finds that ten PPN
parameters exhaust the possibilities.

\begin{table}[t]
  \caption{The PPN Parameters and their significance (note that
    $\alpha_3$ has been shown twice to indicate that it is a measure
    of two effects).}
  \label{ppnmeaning}
  \renewcommand{\arraystretch}{1.2}
  \centering
  \begin{tabular}{p{0.5 cm}|p{4.0 cm}ccc}
    \hline \hline
    \multicolumn{2}{p{4.5 cm}}{Parameter and what it measures relative to GR} &
    \multicolumn{1}{p{1.2 cm}}{\centering Value in GR} &
    \multicolumn{1}{p{2.2 cm}}{\centering Value in semiconservative
      theories} &
    \multicolumn{1}{p{2.2 cm}}{\centering Value in fully conservative
      theories} \\
    \hline \hline
    $\gamma$ &
    How much space-curva\-ture produced by unit rest mass? &
    $1$ & $\gamma$ & $\gamma$ \\
    \hline
    $\beta$ &
    How much ``nonlinearity'' in the superposition law for gravity? &
    $1$ & $\beta$ & $\beta$ \\
    \hline
    $\xi$ &
    Preferred-location effects? &
    $0$ & $\xi$ & $\xi$ \\
    \hline
    $\alpha_1$ &
    Preferred-frame effects? &
    $0$ & $\alpha_1$ & $0$ \\
    $\alpha_2$ & &
    $0$ & $\alpha_2$ & $0$ \\
    $\alpha_3$ & &
    $0$ & $0$ & $0$ \\
    \hline
    $\alpha_3$ &
    Violation of conservation &
    $0$ & $0$ & $0$ \\
    $\zeta_1$ &
    of total momentum? &
    $0$ & $0$ & $0$ \\
    $\zeta_2$ & &
    $0$ & $0$ & $0$ \\
    $\zeta_3$ & &
    $0$ & $0$ & $0$ \\
    $\zeta_4$ & &
    $0$ & $0$ & $0$ \\
    \hline \hline
  \end{tabular}
  \renewcommand{\arraystretch}{1.0}
\end{table}

The parameters $\gamma$ and $\beta$ are the usual
Eddington--Robertson--Schiff parameters used to describe the
``classical'' tests of general relativity, and are in some sense the most important; they
are the only non-zero parameters in general relativity and scalar-tensor gravity.
The parameter $\xi$ is non-zero in
a class of theories of gravity that predict preferred-location effects
such as a galaxy-induced anisotropy in the local gravitational
constant $G_\mathrm{L}$ (also called ``Whitehead'' effects);
$\alpha_1$, $\alpha_2$, $\alpha_3$ measure whether or
not the theory predicts post-Newtonian preferred-frame
effects; $\alpha_3$, $\zeta_1$, $\zeta_2$,
$\zeta_3$, $\zeta_4$ measure whether or not the theory
predicts violations of global conservation laws for total
momentum. 
In Table~\ref{ppnmeaning} we show the values these
parameters take in general relativity;
in any theory of gravity that possesses conservation laws for
  total momentum, called ``semi-conservative'' (any theory that is
  based on an invariant action principle is semi-conservative); and
in any theory that in addition possesses six global conservation
  laws for angular momentum, called ``fully conservative'' (such
  theories automatically predict no post-Newtonian preferred-frame
  effects).
Semi-conservative theories have five free PPN  parameters ($\gamma$,
$\beta$, $\xi$, $\alpha_1$, $\alpha_2$) while fully conservative
theories have three ($\gamma$, $\beta$, $\xi$).

The PPN  formalism was pioneered by
Kenneth Nordtvedt~\cite{nordtvedt2}, who studied the post-Newtonian
metric of a system of gravitating point masses, extending earlier
work by Eddington, Robertson and Schiff (TEGP~4.2). 
Will~\cite{Will71a} generalized the framework to perfect fluids. A
general and unified version of the PPN  formalism was developed by
Will and Nordtvedt~\cite{willnordtvedt72}. The canonical version, with
conventions altered to be more in accord with standard textbooks
such as~\cite{MTW}, is discussed in detail in TEGP, Chapter 4. 


\section{Competing theories of gravity}
\label{theories}

One of the important applications of the PPN formalism is the
comparison and classification of alternative metric theories of
gravity. The population of viable theories has fluctuated over
the years as new effects and tests have been discovered, largely
through the use of the PPN  framework, which eliminated many
theories thought previously to be viable. The theory population
has also fluctuated as new, potentially viable theories have been
invented.

In this review, we will focus on general relativity, the
general class of scalar-tensor modifications of it, of which the
Jordan--Fierz--Brans--Dicke theory (Brans--Dicke, for short)
is the classic example, and vector-tensor or scalar-vector-tensor theories. 
The reasons are several-fold:
\begin{itemize}
\item A full compendium of alternative theories circa 1981 is given in
  TEGP, Chapter 5.
\item Many alternative metric theories developed during the 1970s and
  1980s could be viewed as ``straw-man'' theories, invented to prove
  that such theories exist or to illustrate particular properties. Few
  of these could be regarded as well-motivated theories from the point
  of view, say, of field theory or particle physics.
\item A number of theories fall into the class of ``prior-geometric''
  theories, with absolute elements such as a flat background metric in
  addition to the physical metric. Most of these theories predict
  ``preferred-frame'' effects, that have been tightly constrained by
  observations (see Section~\ref{preferred}). An example is Rosen's
  bimetric theory.
\item A large number of alternative theories of gravity predict
  gravitational wave emission substantially different from that of
  general relativity, in strong disagreement with observations of the
  binary pulsar (see Section~\ref{S5}).
\item Scalar-tensor modifications of general relativity have become very popular in
  unification schemes such as string theory, and in cosmological model
  building. Because the scalar fields could be massive, the potentials
  in the post-Newtonian limit could be modified by Yukawa-like terms.
\item Theories that also incorporate vector fields have attracted recent attention, in the
  spirit of the SME (see Section~\ref{lli}), as models for violations
  of Lorentz invariance in the gravitational sector, and as potential candidates to account for phenomena such as galaxy rotation curves without resorting to dark matter.
\end{itemize}


\subsection{General relativity}
\label{generalrelativity}

The metric {\boldmath $g$} is the sole
dynamical field, and the theory contains no arbitrary functions or
parameters, apart from the value of the Newtonian coupling constant
$G$, which is measurable in laboratory experiments. Throughout this
article, we ignore the cosmological constant $\Lambda_\mathrm{C}$. 
We do this despite recent evidence, from supernova data, of an accelerating
universe, which would indicate either a non-zero
cosmological constant or a dynamical
``dark energy'' contributing about 70~percent of the critical density.
Although
$\Lambda_\mathrm{C}$ has significance for quantum field theory, quantum
gravity, and cosmology, on the scale of the solar-system or of stellar
systems its effects are negligible, for the values of $\Lambda_\mathrm{C}$
inferred from supernova observations.  On the other hand, the conundrum of accelerated expansion has motivated the development of alternative theories of gravity, notably of the so-called $f(R)$ type.

The field equations of general relativity are derivable from an invariant action
principle $\delta I=0$, where
\begin{equation}
  I = \frac{c^3}{16 \pi G} \int R (-g)^{1/2} \, d^4 x +
  I_\mathrm{m} (\psi_\mathrm{m}, g_{\mu\nu}),
  \label{E21}
\end{equation}
where $R$ is the Ricci scalar, and $I_\mathrm{m}$ is the matter action, which
depends on matter fields $\psi_\mathrm{m}$ universally coupled to the metric
{\boldmath $g$}. By varying the action with respect to $g_{\mu\nu}$, we
obtain the field equations
\begin{equation}
  G_{\mu\nu} \equiv R_{\mu\nu} - \frac{1}{2} g_{\mu\nu} R = \frac{8 \pi G}{c^4} T_{\mu\nu},
  \label{E22}
\end{equation}
where $T_{\mu\nu}$ is the matter energy-momentum tensor. General
covariance of the matter action implies the equations of motion
${T^{\mu\nu}}_{;\nu}=0$; varying $I_\mathrm{m}$ with respect to
$\psi_\mathrm{m}$
yields the matter field equations of the Standard Model. 
By virtue of the \emph{absence} of
prior-geometric elements, the equations of motion are also a
consequence of the field equations via the Bianchi identities
${G^{\mu\nu}}_{;\nu}=0$.

The general procedure for deriving the post-Newtonian limit of metric
theories is spelled out in TEGP~5.1, and is described in
detail for general relativity in TEGP~5.2, or Chapter 8 of~\cite{PW2014}. 
The PPN parameter values are
listed in Table~\ref{ppnvalues}.

\begin{table}[t]
  \caption[Metric theories and their PPN parameter values.]{Metric
    theories and their PPN parameter values ($\alpha_3 = \zeta_i=0$
    for all cases). The parameters $\gamma^\prime$, $\beta^\prime$,
    $\alpha_1^\prime$, and $\alpha_2^\prime$ denote complicated
    functions of the arbitrary constants and matching parameters.}
  \label{ppnvalues}
  \centering
  \begin{tabular}{lccccccc}
    \hline \hline
    Theory &Arbitrary&Cosmic&\multicolumn{5}{p{5 cm}}{\cl{PPN parameters}}\\[-1 em]
    &functions&matching&\multicolumn{5}{p{5 cm}}{\cl{\hrulefill}}\\[-1 em]
    &or constants&parameters&$\gamma$ & $\beta$ & $\xi$ & 
           $\alpha_1$ & $\alpha_2$ \\
    \hline \hline
    General relativity
    & none &
    none &
    $1$ &
    $1$ &
    $0$ &
    $0$ &
    $0$ \\
    \hline
    Scalar-tensor \\
    \quad Brans--Dicke &
    $\omega_\mathrm{BD}$ &
    $\phi_0$ &
    $ \displaystyle \frac{1+\omega_\mathrm{BD}}{2+\omega_\mathrm{BD}}$ &
    $1$ &
    $0$ &
    $0$ &
    $0$ \\
    \quad General, $f(R)$ &
    $A(\varphi)$,
    $V(\varphi)$ &
    $\varphi_0$ &
    $ \rule{0 cm}{0.6cm} \displaystyle \frac{1+\omega}{2+\omega}$ &
    $1+ \displaystyle\frac{\lambda}{4+2\omega}$ &
    $0$ &
    $0$ &
    $0$ \\ [0.6 em]
    \hline
    Vector-tensor \\
    \quad Unconstrained &
    $\omega, c_1, c_2, c_3, c_4$ &
    $u$ &
    $\gamma^\prime$ &
    $\beta^\prime$ &
    $0$ &
    $\alpha_1^\prime$ &
    $\alpha_2^\prime$ \\
    \quad Einstein-{\AE}ther &
    $c_1, c_2, c_3, c_4$ &
    none &
    $1$ &
    $1$ &
    $0$ &
    $\alpha_1^\prime$ &
    $\alpha_2^\prime$ \\ 
    \quad Khronometric&
    $\alpha_K, \beta_K, \lambda_K$ & none&
    $1$ & $1$ & $0$ & 
    $\alpha_1^\prime$ &
    $\alpha_2^\prime$
    \\ [0.6 em]
    \hline
    Tensor-Vector-Scalar
    &$k,c_1, c_2, c_3, c_4$&
    $\phi_0$&
    $1$ &
    $1$ &
    $0$ &
    $\alpha_1^\prime$ &
    $\alpha_2^\prime$ \\
       \hline \hline
  \end{tabular}
  \renewcommand{\arraystretch}{1.2}
\end{table}


\subsection{Scalar-tensor theories}
\label{scalartensor}

These theories contain the metric {\boldmath $g$}, a
scalar field $\varphi$, a potential  function $V(\varphi)$, and a
coupling function $A(\varphi)$ (generalizations to more than one scalar
field have also been carried out~\cite{DamourEspo92}).
For some purposes, the action is conveniently written in a non-metric
representation, sometimes denoted the ``Einstein frame'', in which the
gravitational action looks exactly like that of general relativity:
\begin{eqnarray}
  \tilde I &=& \frac{c^3}{16 \pi G} \int \left[ \tilde{R} -
  2 \tilde{g}^{\mu\nu} \partial_\mu \varphi \, \partial_\nu \varphi -
  V (\varphi) \right] (- \tilde{g})^{1/2} \, d^4 x \nonumber \\
  && \quad + I_\mathrm{m}
  \left( \psi_\mathrm{m}, A^2 (\varphi) \tilde{g}_{\mu\nu} \right),
  \label{E23}
\end{eqnarray}
where $\tilde R \equiv \tilde g^{\mu\nu} \tilde R_{\mu\nu}$ is the
Ricci scalar of the
``Einstein'' metric $\tilde g_{\mu\nu}$. 
 This representation is a
``non-metric'' one because the matter fields $\psi_\mathrm{m}$ couple to a
combination of $\varphi$ and $\tilde g_{\mu\nu}$.
Despite appearances, however,
it is a metric theory, because it can be put
into a metric representation by identifying the ``physical metric''
\begin{equation}
  g_{\mu\nu} \equiv A^2 (\varphi) \tilde g_{\mu\nu}.
  \label{E24}
\end{equation}
The action can then be rewritten in the metric form
\begin{eqnarray}
  I &=& \frac{c^3}{16 \pi G} \int \left[ \phi R - \phi^{-1} \omega(\phi)
  g^{\mu\nu} \partial_\mu \phi \partial_\nu \phi - \phi^2 V \right]
  (-g)^{1/2} \, d^4 x 
  \nonumber \\
  && \quad + I_\mathrm{m} (\psi_\mathrm{m}, g_{\mu\nu}),
  \label{E25}
\end{eqnarray}
where
\begin{equation}
  \begin{array}{rcl}
    \phi & \equiv & A (\varphi)^{-2},
    \\ [0.5 em]
    3 + 2 \omega (\phi) & \equiv & \alpha (\varphi)^{-2},
    \\ [0.5 em]
    \alpha (\varphi) & \equiv & \displaystyle
    \frac{d(\ln A(\varphi))}{d\varphi}.
  \end{array}
  \label{E26}
\end{equation}
The Einstein frame is useful for discussing general characteristics of
such theories, and for some cosmological applications, while the metric
representation is most useful for calculating observable effects.
The field equations, post-Newtonian limit and PPN  parameters are
discussed in TEGP~5.3, and the values of the PPN  parameters are
listed in Table~\ref{ppnvalues}.

The parameters that enter the post-Newtonian limit are
\begin{equation}
  \omega \equiv \omega(\phi_0),
  \qquad
  \lambda \equiv \left[ \frac{\phi \, {d\omega}/{d\phi}}{
  (3 + 2 \omega) (4 + 2 \omega)} \right]_{\phi_0}\!\!\!,
  \label{E27}
\end{equation}
where $\phi_0$ is the value of $\phi$ today far from the
system being studied, as determined by appropriate cosmological boundary
conditions.
In Brans--Dicke theory ($\omega(\phi) \equiv \omega_\mathrm{BD}= \mathrm{const.}$),
the larger the value of
$\omega_\mathrm{BD}$, the smaller the effects of the scalar field, and in the
limit $\omega_\mathrm{BD} \to \infty$ ($\alpha_0 \to 0$),
the theory becomes indistinguishable from
general relativity in all its predictions. In more general
theories, the function $\omega ( \phi )$ could have the
property that, at the present epoch, and in weak-field situations,
the value of the scalar field $\phi_0$ is such that
$\omega$ is very large and $\lambda$ is very small (theory almost
identical to general relativity today), but that for past or
future values of $\phi$, or in strong-field regions such as the
interiors of neutron stars, $\omega$ and $\lambda$ could take on values
that would lead to significant differences from general relativity. 

Damour and Esposito-Far\`ese~\cite{DamourEspo92} have adopted an alternative
paramet\-rization of scalar-tensor theories, in which  one
expands $\ln A(\varphi)$
about a cosmological background field value $\varphi_0$:
\begin{equation}
  \ln A (\varphi) = \alpha_0 (\varphi -\varphi_0) +
  \frac{1}{2} \beta_0 (\varphi - \varphi_0)^2 + \dots
  \label{alphaexpand}
\end{equation}
A precisely linear coupling function produces Brans--Dicke theory, with
$\alpha_0^2 = 1/(2 \omega_\mathrm{BD} +3)$, or  
$1/(2+\omega_\mathrm{BD})=2\alpha_0^2/(1+\alpha_0^2)$.
The function $\ln A(\varphi)$ acts
as a potential for the scalar field $\varphi$ within matter, 
and, if $\beta_0
>0$, then during cosmological evolution,
the scalar field naturally evolves toward the minimum of the
potential, i.e.\ toward $\alpha_0 \approx 0$, 
$\omega \to \infty$, or toward
a theory close to, though not precisely general relativity~\cite{DamourNord93a, DamourNord93b}.
Estimates of the
expected relic deviations from general relativity today in such theories depend on the
cosmological model, but range from $10^{-5}$ to a few times $10^{-7}$
for $|\gamma-1|$.

Negative
values of $\beta_0$ correspond to a ``locally unstable'' 
scalar potential (the overall theory is still stable in the sense of having
no tachyons or ghosts).
In this case, objects such as neutron stars can experience a
``spontaneous scalarization'', whereby the interior values of $\varphi$
can take on values very different from the exterior values, through
non-linear interactions between strong gravity and the scalar field,
dramatically affecting the stars' internal structure and leading to
strong violations of SEP. On the other hand, in the case 
$\beta_0 <0$, one must confront that fact that, 
with an unstable $\varphi$ potential,
cosmological evolution would presumably drive the system away from the
peak where $\alpha_0 \approx 0$,
toward parameter values that could be excluded
by solar system experiments. 

Scalar fields coupled to gravity or matter are also
ubiquitous in particle-physics-inspired models of unification, such as
string theory.
In some models, the coupling to matter may lead to
violations of EEP, which could be  tested or bounded
by the experiments described in Section~\ref{eep}. In
many models the scalar field could be massive; if the Compton wavelength is
of macroscopic scale, its effects are those of a ``fifth force''.
Only if the theory can be cast as a metric theory with a
scalar field of infinite range or of range long compared to the scale
of the system in question (solar system) can the PPN  framework be
strictly
applied. If the mass of the scalar field is sufficiently large that its
range is microscopic, then, on solar-system scales, the scalar field is
suppressed, and the theory is essentially equivalent to general
relativity.   For a detailed review of scalar-tensor theories see~\cite{2007sttg.book.....F}.

\subsection{$f(R)$ theories}
\label{f(R)}

These are theories whose action has the form
\begin{equation}
  I = \frac{c^3}{16 \pi G} \int f(R) (-g)^{1/2} \, d^4 x +
  I_\mathrm{m} (\psi_\mathrm{m}, g_{\mu\nu}),
  \label{fRaction}
\end{equation}
where $f$ is a function chosen so that at cosmological scales, the universe will experience accelerated expansion without  resorting to either a cosmological constant or dark energy.  However, it turns out that such theories are equivalent to scalar-tensor theories:  replace $f(R)$ by $f(\chi) - f_{,\chi} (\chi) (R-\chi)$, where $\chi$ is a dynamical field.  Varying the action with respect to $\chi$ yields $f_{,\chi\chi}(R-\chi) = 0$, which implies that $\chi = R$ as long as $f_{,\chi\chi} \ne 0$.  Then defining a scalar field $\phi \equiv f_{,\chi} (\chi) $ one puts the action into the form of a scalar-tensor theory given by Eq.\ (\ref{E25}), with $\omega(\phi) =0$ and $\phi^2 V = \phi \chi(\phi) - f(\chi(\phi))$.  As we will see, this value of $\omega$ would ordinarily strongly violate solar-system experiments, but it turns out that in many models, the potential $V(\phi)$ has the effect of giving the scalar field a large effective mass in the presence of matter (the so-called ``chameleon mechanism''), so that the scalar field is suppressed at distances that extend outside bodies like the Sun and Earth.  In this way, with only modest fine tuning, $f(R)$ theories can claim to obey standard tests, while providing interesting, non general-relativistic behavior on cosmic scales.  For detailed reviews of this class of theories, see \cite{2010RvMP...82..451S,lrr-2010-3}.


\subsection{Vector-tensor theories}
\label{vectortensor}

These theories contain the metric {\boldmath $g$} and a dynamical, 
typically timelike, four-vector
field $u^\mu$. In some models, the four-vector is unconstrained, while
in others, called Einstein-{\AE}ther theories it is constrained to be timelike
with unit norm. The most general action for such theories
that is quadratic in derivatives of the vector is given by
\begin{eqnarray}
  I &=& \frac{c^3}{16 \pi G} \int \left[ (1 + \omega u_\mu u^\mu) R -
  K^{\mu\nu}_{\alpha\beta} \nabla_\mu u^\alpha \nabla_\nu u^\beta +
  \lambda (u_\mu u^\mu + 1) \right] 
  \nonumber \\
  && \quad \quad \times (-g)^{1/2} \, d^4x +
  I_\mathrm{m} (\psi_\mathrm{m}, g_{\mu\nu}),
  \label{aetheraction}
\end{eqnarray}
where
\begin{equation}
  K^{\mu\nu}_{\alpha\beta} = c_1 g^{\mu\nu} g_{\alpha\beta} +
  c_2 \delta^\mu_\alpha \delta^\nu_\beta +
  c_3 \delta^\mu_\beta\delta^\nu_\alpha -
  c_4 u^\mu u^\nu g_{\alpha\beta}.
  \label{ktensor}
\end{equation}
The coefficients $c_i$ are arbitrary. In the unconstrained theories,
$\lambda \equiv 0$ and  $\omega$ is arbitrary. In the constrained theories,
$\lambda$ is a Lagrange multiplier, and by virtue of the constraint
$u_\mu u^\mu = -1$, the factor $\omega u_\mu u^\mu$
in front of the Ricci scalar can be
absorbed into a rescaling of $G$; equivalently, in the constrained theories,
we can set $\omega=0$. Note that the possible term $u^\mu u^\nu
R_{\mu\nu}$ can be shown under integration by parts to be equivalent to a
linear combination of the terms involving $c_2$ and $c_3$. 

Unconstrained theories were studied during the 1970s as ``straw-man''
alternatives to general relativity. In addition to having up to four
arbitrary parameters, they also left the magnitude of the vector field 
arbitrary, since it satisfies a linear homogenous vacuum field equation
of the form ${\cal L} u^\mu =0$ ($c_4=0$ in all such cases studied). 
Indeed, this latter fact was one of most serious defects of these theories.

\begin{description}
\item[General vector-tensor theory; \boldmath $\omega$, $\tau$,
  $\epsilon$, $\eta$ {(TEGP~5.4)}]~\\
  The gravitational Lagrangian for this class of theories had the form
  $R + \omega u_\mu u^\mu R + \eta u^\mu u^\nu R_{\mu\nu}-\epsilon
  F_{\mu\nu}F^{\mu\nu} + \tau \nabla_\mu u_\nu \nabla^\mu u^\nu$,
  where $F_{\mu\nu} = \nabla_\mu u_\nu -\nabla_\nu u_\mu$,
  corresponding to the values $c_1 = 2\epsilon - \tau$, $c_2 = -\eta$,
  $c_1+c_2+c_3= -\tau$, $c_4=0$ and $\lambda=0$. 
  In these theories $\gamma$, $\beta$,
  $\alpha_1$, and $\alpha_2$ are complicated functions of the
  parameters and of $u^2 = -u^\mu u_\mu$, while the rest vanish.
\end{description}

\begin{description}
\item[Will--Nordtvedt theory~{\cite{willnordtvedt72}}]~\\
  This is the special case $c_1 = -1$, $c_2 = c_3 = c_4 =\lambda=0$. In this
  theory, the PPN parameters are given by $\gamma = \beta = 1$,
  $\alpha_2 = u^2/(1+u^2/2)$, and zero for the rest.
\end{description}

\begin{description}
\item[Hellings--Nordtvedt theory; \boldmath $\omega$
  {\cite{hellings73}}]~\\
  This is the special case $c_1 =2$, $c_2 =2\omega$,
  $c_1 + c_2 +c_3 = 0 = c_4$, $\lambda=0$. Here $\gamma$, $\beta$, $\alpha_1$ and
  $\alpha_2$ are complicated functions of the parameters and of $u^2$,
  while the rest vanish.
\end{description}

\begin{description}
\item[Einstein-{\AE}ther theories;  \boldmath $c_1$, $c_2$, $c_3$, $c_4$ 
{\cite{jacobson01, mattingly02, jacobson04, eling04, foster05}}]~\\
These theories were motivated in part by a desire to explore
possibilities for violations of Lorentz invariance in gravity, in parallel
with similar studies in matter interactions, such as the SME.  Here $\gamma = \beta =1$, $\alpha_1$ and $\alpha_2$ are complicated functions of the $c_k$ parameters, and the rest vanish.
By requiring that gravitational wave modes have real (as
opposed to imaginary) frequencies, one can impose the bounds
$c_1/(c_1+c_4) \ge 0$ and $(c_1+c_2+c_3)/(c_1+c_4) \ge 0$. 
Considerations of positivity of energy impose the constraints $c_1 > 0$,
$c_1+c_4>0$ and $c_1+c_2+c_3 >0$. 

\item[Khronometric theory; \boldmath $\alpha_K, \, \beta_K, \, \lambda_K$
\cite{2009PhRvD..79h4008H,2010PhRvL.104r1302B,2011JHEP...04..018B,2014PhRvD..89h1501J}]~\\
This is the low-energy limit of ``Ho\v{r}ava gravity'', a proposal for a gravity theory that is power-counting renormalizable.
The vector field is required to be hypersurface orthogonal, so that higher spatial derivative terms could be introduced to effectuate renormalizability.   A ``healthy'' version of the theory can be shown to correspond to the values $c_1 = -\epsilon$
 $c_2 = \lambda_K$, $ c_3 = \beta_K + \epsilon$ and $c_4 = \alpha_K + \epsilon$, where the limit $\epsilon \to \infty$ is to be taken.
\end{description}

\subsection{Tensor-vector-scalar (TeVeS) theories}
\label{sec:TeVeS}

This class of theories was invented to provide a fully relativistic theory of gravity that could mimic the phenomenological behavior of so-called Modified Newtonian Dynamics (MOND), whereby in a weak-field regime, Newton's laws hold, namely $a = Gm/r^2$ where $m$  is the mass of a central object, as long as $a$ is small compared to some fundamental scale $a_0$, but in a regime where $a > a_0$, the equations of motion would take the form $a^2/a_0 = Gm/r^2$.  With such a behavior, the rotational velocity of a particle far from a central mass would have the form $v \sim \sqrt{ar} \sim (Gma_0)^{1/4}$, thus reproducing the flat rotation curves observed for spiral galaxies, without invoking a distribution of dark matter.

Devising such a theory turned out to be no simple matter, and the final result, TeVeS was rather complicated~\cite{PhysRevD.70.083509}.  Furthermore, it was shown to have unexpected singular behavior that was most simply cured by incorporating features of the Einstein-{\AE}ther theory~\cite{PhysRevD.77.123502}.   The extended theory is based on an ``Einstein'' metric $\tilde{g}_{\mu\nu}$, related to the physical metric ${g}_{\mu\nu}$ by
\begin{equation}
{g}_{\mu\nu} \equiv e^{-2\phi} \tilde{g}_{\mu\nu} - 2 u_\mu u_\nu \sinh (2\phi) \,,
\end{equation}
where $u^\mu$ is a vector field, and $\phi$ is a scalar field.   The action for gravity is the standard general relativity action of Eq.\ (\ref{E21}), but defined using the Einstein metric $\tilde{g}_{\mu\nu}$, while the matter action is that of a standard metric theory, using ${g}_{\mu\nu}$.  These are supplemented by the vector action, given by that of Einstein-{\AE}ther theory, Eq.\ (\ref{aetheraction}), and a scalar action, given by 
\begin{eqnarray}
I_S &=& -\frac{c^3}{2k^2 \ell^2 G} \int {\cal F} (k\ell^2 h^{\mu\nu} \phi_{,\mu} \phi_{,\nu} ) 
 (-g)^{1/2} \, d^4x \,,
\end{eqnarray}
where $k$ is a constant, $\ell$ is a distance, and $h^{\mu\nu} \equiv \tilde{g}^{\mu\nu} - u^\mu u^\nu$, indices being raised and lowered using the Einstein metric. The function ${\cal F}(y)$ is chosen so that $\mu(y) \equiv d{\cal F}/dy$ is unity in the high-acceleration, or normal Newtonian regime, and nearly zero in the MOND regime.  

The PPN parameters of the theory~\cite{PhysRevD.80.044032}
 have the values $\gamma = \beta =1$ and $\xi = \alpha_3 = \zeta_i = 0$, while the parameters $\alpha_1$ and $\alpha_2$ are complicated functions of $k$, $c_k$ and the asymptotic value $\phi_0$ of the scalar field.  

For reviews of TeVeS, MOND and their confrontation with the dark-matter paradigm, see~\cite{2009CQGra..26n3001S,2012LRR....15...10F}.

\subsection{Other theories}
\label{sec:Other}

Numerous alternative theories of gravity have been developed and studied extensively in recent years, most motivated from the direction of particle physics, quantum gravity or unification.  

Massive gravity theories attempt to give the putative ``graviton'' a mass.  The simplest attempts fall afoul of the so-called  van Dam-Veltman-Zakharov discontinuity, which leads to a violation of experiment.  Attempts to avoid this problem involve treating non-linear aspects of the theory at the fundamental level; many models incorporate a second tensor field in addition to the metric. For a recent review, see~\cite{2012RvMP...84..671H}.  Quadratic gravity is a recent incarnation of an old idea of adding to the action of general relativity terms quadratic in the Riemann and Ricci tensors or the Ricci scalar, as an effective model for quantum gravity corrections.  Chern-Simons gravity adds a parity-violating term proportional to $^*R^{\alpha\beta\gamma\delta}R_{\alpha\beta\gamma\delta}$ to the action of general relativity, where $^*R^{\alpha\beta\gamma\delta}$ is the dual Riemann tensor.

\section{Tests of post-Newtonian gravity}
\label{sec:PNtests}

\subsection[Tests of the parameter $\gamma$]{Tests of the parameter
  \boldmath $\gamma$}
\label{gamma}

With the PPN formalism in hand, we are now ready to confront
gravitation theories described in Sec.\ \ref{theories} with the results of solar-system
experiments. In this section we focus on tests of the parameter
$\gamma$, consisting of the deflection of light and the time delay
of light.


\subsubsection{The deflection of light}
\label{deflection}

A light ray (or photon) which passes the Sun at a distance $d$ is
deflected by an angle
\begin{equation}
  \delta \theta = \frac{1}{2} (1 + \gamma) \frac{4G M_\odot}{dc^2}
  \frac{1 + \cos \Phi}{2}
  \label{E28}
\end{equation}
(TEGP~7.1), where $M_\odot$ is the mass of the Sun
and $\Phi$ is the angle between the Earth-Sun line and the
incoming direction of the photon.
For a grazing ray,
$d \approx d_\odot$, $\Phi \approx 0$, and
\begin{equation}
  \delta \theta \approx \frac{1}{2} (1 + \gamma) 1.''7505,
  \label{E29}
\end{equation}
independent of the frequency of light.  In practice, one measures how the relative angular separation
between an observed source of light and a nearby reference source evolves
as the Sun moves across the sky, as seen from Earth.

It is interesting to note that the classic derivations of the
deflection that use only 
the corpuscular theory of light, by Cavendish in 1784 and von Soldner
in 1803~\cite{willcavendish}, or that use only
the principle of equivalence, by Einstein in 1911,
yield only the ``1/2'' part of the coefficient in front of the
expression in Eq.~(\ref{E28}). But the result of these calculations
is the deflection of light relative to local straight lines, as
established for example by rigid rods; however, because of space
curvature around the Sun, determined by the PPN  parameter
$\gamma$, local straight lines are bent relative to asymptotic
straight lines far from the Sun by just enough to yield the
remaining factor ``$\gamma /2$''. The first factor ``1/2''
holds in any metric theory, the second ``$\gamma /2$'' varies
from theory to theory. Thus, calculations that purport to derive
the full deflection using the equivalence principle alone are
incorrect.

\begin{figure}[t]
 \includegraphics[scale=0.45]{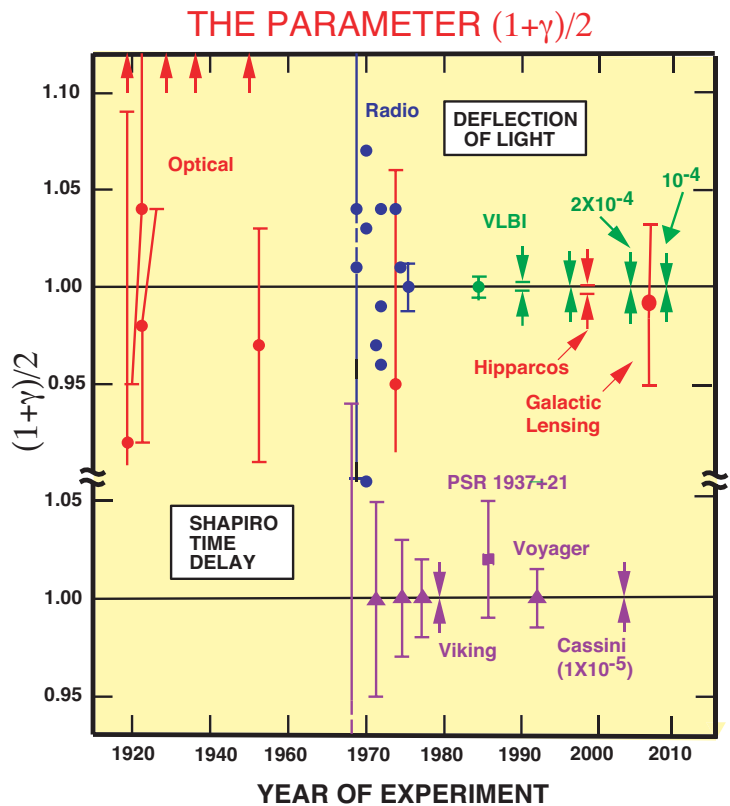}
  \caption{Measurements of the coefficient $(1 + \gamma )/2$ from
    light deflection and time delay measurements. Its general relativity
    value is unity. The arrows at the top denote anomalously large
    values from early eclipse expeditions. The Shapiro time-delay
    measurements using the Cassini spacecraft yielded an agreement with general relativity
    to $10^{-3}$~percent, and VLBI light deflection measurements have
    reached 0.02~percent. Hipparcos denotes the optical astrometry
    satellite, which reached 0.1~percent.}
  \label{gammavalues}
\end{figure}

The prediction of the full bending of light by the Sun was one of
the great successes of Einstein's general relativity.
Eddington's confirmation of the bending of optical starlight
observed during a solar eclipse in the first days following World
War I helped make Einstein famous. However, the experiments of
Eddington and his co-workers had only 30~percent accuracy, and
succeeding experiments were not much better:  The results were
scattered between one half and twice the Einstein value
(see Figure~\ref{gammavalues}), and the accuracies were low.

However, the development of radio interferometery, and later
of very-long-baseline radio interfer\-om\-etry (VLBI), produced
greatly improved determinations of
the deflection of light. These techniques now have the capability
of measuring angular separations and changes in angles
to accuracies better than 
100 microarcseconds. Early measurements took advantage
of the fact that certain groups of strong quasistellar radio sources annually
pass
very close to the Sun (as seen from the Earth), and by 1975 were reaching accuracies at the level of a percent.

In recent years, transcontinental and intercontinental VLBI observations 
of quasars and
radio galaxies have been 
made primarily to monitor the Earth's rotation
(``VLBI'' in Figure~\ref{gammavalues}). These measurements are
sensitive to the deflection of light over
almost the entire celestial sphere (at $90 ^\circ$ from the Sun, the
deflection is still 4 milli\-arcseconds).
A 2004 analysis of  almost 2 million VLBI observations
of 541 radio sources, made by 87 VLBI sites between 1979 and 1999
yielded 
$(1+\gamma)/2=0.99992 \pm 0.00023$, or equivalently,
$\gamma-1= (-1.7 \pm 4.5) \times 10^{-4}$~\cite{sshapiro04}.  Analyses that incorporated data through 2010 yielded $\gamma-1= (-0.8 \pm 1.2) \times 10^{-4}$~\cite{2011A&A...529A..70L}.

Analysis of observations made by the Hipparcos optical astrometry
satellite yielded a test at the level of 0.3
percent, while the GAIA mission, launched in 2013, is expected to reach the parts per million level.

Finally, a remarkable measurement of $\gamma$ on
galactic scales was reported in 2006~\cite{2006PhRvD..74f1501B}. It used data on gravitational lensing by 15 elliptical
galaxies, collected by the Sloan Digital Sky Survey. The Newtonian
potential $U$ of each lensing galaxy (including the contribution from
dark matter) was derived from
the observed velocity dispersion of stars in the galaxy. Comparing
the observed lensing with the lensing predicted by the models provided a
10 percent bound on $\gamma$, in agreement with general
relativity. Unlike the much tighter bounds described previously, which 
were obtained on the scale of the solar system, this bound was obtained
on a galactic scale.  

The results of light-deflection measurements are summarized in
Figure~\ref{gammavalues}.


\subsubsection{The time delay of light}
\label{timedelay}

A radar signal sent across the solar system past the Sun to a
planet or satellite and returned to the Earth suffers an
additional non-Newtonian delay in its round-trip travel time,
given by
\begin{equation}
  \delta t = 2 (1 + \gamma) \frac{GM_\odot}{c^3}
  \ln \left( \frac{(r_\oplus + {\bf x}_\oplus \cdot {\bf n})
  (r_\mathrm{e} - {\bf x}_\mathrm{e} \cdot {\bf n})}{d^2} \right),
  \label{E31}
\end{equation}
where $ {\bf x}_\mathrm{e} $ ($ {\bf x}_\oplus $) are the vectors, and
$ r_\mathrm{e} $ ($ r_\oplus $) are the distances from the Sun
to the source (Earth), respectively (TEGP~7.2). For a ray
which passes close to the Sun,
\begin{equation}
  \delta t \approx \frac{1}{2} (1 + \gamma)
  \left( 240 - 20 \ln \frac{d^2}{r} \right) \mathrm{\ \mu s},
  \label{E32}
\end{equation}
where $d$ is the distance of closest approach of the ray in solar
radii, and $r$ is the distance of the planet or satellite from the
Sun, in astronomical units.

In the two decades following Irwin Shapiro's 1964 discovery of
this effect as a theoretical consequence of general relativity,
several high-precision measurements were made
using radar ranging to targets passing through superior
conjunction. Since one does not have access to a ``Newtonian''
signal against which to compare the round-trip travel time of the
observed signal, it is necessary to do a differential measurement
of the variations in round-trip travel times as the target passes
through superior conjunction, and to look for the logarithmic
behavior of Equation~(\ref{E32}). In order to do this accurately however,
one must take into account the variations in round-trip travel
time due to the orbital motion of the target relative to the
Earth. This is done by using radar-ranging (and possibly other)
data on the target taken when it is far from superior conjunction
(i.e.\ when the time-delay term is negligible) to determine
an accurate ephemeris for the target, using the ephemeris to
predict the PPN coordinate trajectory ${\bf x}_\mathrm{e} (t)$ near
superior conjunction, then combining that trajectory with the
trajectory of the Earth ${\bf x}_\oplus (t)$ to determine the
Newtonian round-trip time and the logarithmic term in Equation~(\ref{E32}).
The resulting predicted round-trip travel times in terms of the
unknown coefficient $\frac{1}{2}(1+ \gamma)$
are then fit to the measured travel times using the method
of least-squares, and an estimate obtained for
$\frac{1}{2}(1+ \gamma)$.

The targets employed included
planets, such as Mercury or Venus, used as passive reflectors of
the radar signals, and
artificial satellites, such as Mariners~6 and 7, Voyager~2,
the Viking
Mars landers and orbiters, and the Cassini spacecraft to Saturn,
used as
active retransmitters of the radar signals.

The results for the coefficient $\frac{1}{2}(1+ \gamma)$
of all radar time-delay measurements
performed to date (including a measurement of the one-way time delay induced by the Sun on
signals from the millisecond pulsar PSR 1937+21)
are shown in Figure~\ref{gammavalues}. 

The best bound comes from Doppler tracking of the Cassini
spacecraft while it was on its way to Saturn~\cite{bertotti03}, with a
result $\gamma -1 = (2.1 \pm 2.3) \times 10^{-5}$. This was made possible
by the ability to do Doppler measurements using both X-band (7175~MHz) and
Ka-band (34316~MHz) radar, thereby significantly reducing the dispersive
effects of the
solar corona. In addition, the 2002 superior conjunction of Cassini was
particularly favorable: with the spacecraft at 8.43 astronomical units from
the Sun, the distance of closest approach of the radar signals to the Sun
was only $1.6 \, R_\odot$.

The Shapiro time delay plays a key role in determining the parameters of binary pulsar orbits, notably in the ``double pulsar'' system (see Sec.\ \ref{population}).


\begin{table}[t]
  \caption[Current limits on the PPN parameters.]{Current limits on
    the PPN parameters. }
  \label{ppnlimits}
  \renewcommand{\arraystretch}{1.2}
  \centering
  \begin{tabular}{l|lrl}
    \hline \hline
    Parameter &
    \multicolumn{1}{c}{Effect} &
    \multicolumn{1}{c}{Limit} &
    \multicolumn{1}{c}{Remarks} \\
    \hline \hline
    $\gamma-1$ &
    time delay &
    $2.3 \times 10^{-5 \phantom{0}}$ &
    Cassini tracking \\
    & light deflection &
    $2 \times 10^{-4 \phantom{0}}$ &
    VLBI \\
    $\beta-1$ &
    perihelion shift &
    $8 \times 10^{-5 \phantom{0}}$ &
    $J_{2\odot}=(2.2 \pm 0.1) \times10^{-7}$  \\
    & Nordtvedt effect &
    $2.3 \times 10^{-4 \phantom{0}}$ &
    $\eta_\mathrm{N}=4\beta-\gamma-3$ assumed \\
    $\xi$ &
    spin precession &
    $4 \times  10^{-9 \phantom{0}}$ &
    millisecond pulsars \\
    $\alpha_1$ &
    orbital polarization &
    $ 10^{-4 \phantom{0}}$ &
    Lunar laser ranging \\
    & & $4 \times 10^{-5 \phantom{0}}$ &
    PSR J1738+0333 \\
    $\alpha_2$ &
    spin precession &
    $2 \times 10^{-9 \phantom{0}}$ &
    millisecond pulsars \\
    $\alpha_3$ &
    pulsar acceleration &
    $4 \times 10^{-20}$ &
    pulsar $\dot P$ statistics \\
        $\zeta_1$ &
    \multicolumn{1}{c}{---} &
    $2 \times 10^{-2 \phantom{0}}$ &
    combined PPN bounds \\
    $\zeta_2$ &
    binary acceleration &
    $4 \times 10^{-5 \phantom{0}}$ &
    $\ddot P_\mathrm{p}$ for PSR 1913+16 \\
    $\zeta_3$ &
    Newton's 3rd law &
    $10^{-8 \phantom{0}}$ &
    lunar acceleration \\
    $\zeta_4$ &
    \multicolumn{1}{c}{---} &
    \multicolumn{1}{c}{---} &
    not independent \\
    \hline \hline
  \end{tabular}
  \renewcommand{\arraystretch}{1.0}
\end{table}


\subsection{The perihelion shift of Mercury}
\label{perihelion}

The explanation of the anomalous perihelion shift of Mercury's
orbit was another of the triumphs of general relativity. This had
been an unsolved problem in celestial mechanics for over half a
century, since the announcement by Le Verrier in 1859 that, after
the perturbing effects of the planets on Mercury's orbit had been
accounted for,  there remained an unexplained advance
in the perihelion of Mercury. The modern value for this
discrepancy is 42.98 arcseconds per century. A number of \emph{ad
hoc} proposals were made in an attempt to account for this
excess, including,  the existence of a new planet
Vulcan near the Sun, a ring of planetoids, a solar quadrupole
moment and a deviation from the inverse-square law of gravitation,
but none was successful. General relativity accounted
for the anomalous shift in a natural way without disturbing the
agreement with other planetary observations.

The predicted advance per orbit $\Delta \tilde \omega$, including
both relativistic PPN  contributions and the Newtonian
contribution resulting from a
possible solar quadru\-pole moment, is given by
\begin{equation}
  \Delta \tilde \omega = 6 \pi \frac{Gm}{pc^2}
 \left ( \frac{2 + 2 \gamma - \beta}{3}  \right )+
  3\pi J_2  \left (\frac{R}{p} \right )^2 \,,
  \label{E33}
\end{equation}
where $m$ is the total mass of the Sun-Mercury system; $p \equiv a(1-e^2 )$ is the semi-latus rectum of
the orbit, with the semi-major axis $a$ and the eccentricity $e$; $R$ is
the mean radius of the Sun; and $J_2$ is a
dimensionless measure of its quadrupole moment (for details of the derivation see TEGP~7.3).
We have here restricted attention to fully-conservative theories of gravity.

Through observations of the normal modes of solar oscillations
(helioseismology) it is now known that $J_2 =
(2.2 \pm 0.1) \times 10^{-7}$, comparable to what would be estimated from a uniformly rotating solar model.
Substituting standard orbital elements and physical
constants for Mercury and the Sun we obtain the rate of
perihelion shift $\dot {\tilde \omega}$, in seconds of arc per
century,
\begin{equation}
  \dot{\tilde{\omega}} = 42.''98
  \left( \frac{2 + 2 \gamma - \beta}{3} +
  3 \times 10^{-4} \frac{J_2}{10^{-7}} \right).
  \label{E34}
\end{equation}
The most recent fits to planetary data include data from the Messenger spacecraft that orbited Mercury, thereby significantly improving knowledge of its orbit.  Adopting the Cassini bound on $\gamma$ {\em a priori}, these analyses yield a bound on $\beta$  given by $|\beta -1| = (-4.1 \pm 7.8) \times 10^{-5}$~\cite{2011CeMDA.111..363F,2014A&A...561A.115V}.   Further analysis could push this bound even lower, although knowledge of $J_2$ would have to improve simultaneously.

Laser tracking of the Earth-orbiting satellite LAGEOS II led to a measurement of its relativistic perigee precession ($3.4$ arcseconds per year) in agreement with general relativity to two percent~\cite{2014PhRvD..89h2002L}.


\subsection{Tests of the strong equivalence principle}
\label{septests}

The next class of solar-system experiments that test relativistic
gravitational effects may be called tests of the strong
equivalence principle (SEP). In Sec.~\ref{metrictheories} we pointed out that many metric
theories of gravity (perhaps all except general relativity) can be
expected to violate one or more aspects of SEP. Among the
testable violations of SEP are a
violation of the weak equivalence principle for gravitating bodies
that leads to perturbations in the Earth-Moon
orbit, preferred-location and preferred-frame effects in orbital dynamics or in the structure of bodies such as the Earth, and possible variations in the
gravitational constant over cosmological timescales.


\subsubsection{The Nordtvedt effect}
\label{Nordtvedteffect}

In a pioneering calculation using his early form of the PPN
formalism, Kenneth Nord\-tvedt~\cite{nordtvedt1} showed that many metric theories
of gravity predict that massive bodies violate the weak
equivalence principle -- that is, fall with different
accelerations depending on their gravitational self-energy. 
Dicke~\cite{dicke1} argued that such an effect would occur in theories with
a spatially
varying gravitational constant, such as scalar-tensor
gravity.  In the PPN framework, the acceleration of a massive body in an
external gravitational potential $U$ has the form
\begin{equation}
  \begin{array}{rcl}
    {\bf a} & = & \displaystyle \left (1 - \eta_\mathrm{N} \frac{E_\mathrm{g}}{mc^2} \right ) \nabla U,
    \\ [1.0 em]
    \eta_\mathrm{N} & = & \displaystyle
    4 \beta - \gamma - 3 - \frac{10}{3} \xi -
    \alpha_1 + \frac{2}{3} \alpha_2 - \frac{2}{3} \zeta_1 -
    \frac{1}{3} \zeta_2,
  \end{array}
  \label{E35}
\end{equation}
where $E_\mathrm{g}$ is the negative of the gravitational self-energy
of the body ($E_\mathrm{g} >0$). This violation of the massive-body
equivalence principle is known as the ``Nordtvedt effect''. The
effect is absent in general relativity ($ \eta_\mathrm{N} = 0$) but present
in scalar-tensor theory ($ \eta_\mathrm{N} =  (1+2\lambda)/(2+ \omega )$). The existence
of the Nordtvedt effect does not violate the results of laboratory
E\"otv\"os experiments, since for laboratory-sized objects
$E_\mathrm{g} /mc^2 \le 10^{-27}$, far below the sensitivity of
current or future experiments. However, for astronomical bodies,
$E_\mathrm{g} /mc^2$ may be significant ($3.6 \times 10^{-6}$ for the Sun, $4.6 \times 10^{-10}$  for the Earth, $0.2 \times 10^{-10}$
for the Moon, $\sim 0.2$ for neutron stars). If the Nordtvedt effect is present
($\eta_\mathrm{N} \ne 0$) then the Earth should fall toward the Sun with a
slightly different acceleration than the Moon. This perturbation
in the Earth-Moon orbit leads to a polarization of the orbit that
is directed toward the Sun as it moves around the Earth-Moon
system, as seen from Earth. This polarization represents a
perturbation in the Earth-Moon distance of the form
$ \delta r = 13.1 \, \eta_\mathrm{N}
  \cos[( \omega_0 -\omega_\mathrm{s}) t] $ meters, where $\omega_0$ and $\omega_\mathrm{s}$ are the angular frequencies
of the orbits of the Moon and Sun around the Earth (see
TEGP~8.1 for detailed derivations).

Since August 1969, when the first successful acquisition was made
of a laser signal reflected from the Apollo~11 retroreflector on
the Moon, the LLR experiment has made
regular measurements of the round-trip travel times of laser
pulses between a network of observatories
and the lunar retroreflectors, with accuracies that are
approaching the 5~ps (1~mm) level. These measurements are fit
using the method of least-squares to a theoretical model for the
lunar motion that takes into account perturbations due to the Sun
and the other planets, tidal interactions, and post-Newtonian
gravitational effects. The predicted round-trip travel times
between retroreflector and telescope also take into account the
librations of the Moon, the orientation of the Earth, the
location of the observatories, and atmospheric effects on the
signal propagation. The ``Nordtvedt'' parameter $\eta_\mathrm{N}$ along with
several other important parameters of the model are then
estimated in the least-squares method.  For a review of lunar laser ranging, see \cite{2010LRR....13....7M}.

Numerous ongoing analyses of the data find no evidence, within
experimental uncertainty, for the Nordtvedt 
effect~\cite{williams04ijmp}.  
These results represent a limit on a possible violation of WEP for
massive bodies of about 
1.4 parts in $10^{13}$ (compare Figure~\ref{wepfig}). 

However, 
at this level of precision, one cannot regard the results of LLR
as a ``clean'' test of SEP until one eliminates the
possibility of a compensating violation of WEP for the two bodies,
because the chemical compositions of the Earth
and Moon differ in the relative fractions of iron and silicates. To
this end, the E{\"o}t-Wash group carried out an improved test of WEP
for laboratory bodies whose chemical compositions mimic that of the
Earth and Moon. The resulting bound of 1.4 parts 
in $10^{13}$~\cite{baessler99} from composition effects
reduces the ambiguity in the LLR bound, and establishes the firm SEP test
at the level of about 2 parts in $10^{13}$. These results
can be summarized by the Nordtvedt parameter bound
$| \eta_\mathrm{N} | = (4.4 \pm 4.5) \times 10^{-4}$.

APOLLO, the Apache Point Observatory for Lunar Laser ranging
Operation, a joint effort by researchers from the
Universities of Washington, Seattle, and California, San Diego, has achieved mm ranging precision using enhanced laser and telescope technology, together with a good,
high-altitude site in New Mexico.  However models of the lunar orbit must be improved in parallel in order to achieve an order-of-magnitude improvement in the test of the Nordtvedt effect~\cite{2012CQGra..29r4005M}.
This effort will be aided by the fortuitous 2010 discovery by the Lunar Reconnaissance Orbiter of the precise landing site of the Soviet Lunokhod I rover, which deployed a retroreflector in 1970.  Its uncertain location made it effectively ``lost'' to lunar laser ranging for almost 40 years.  Its location on the lunar surface will make it useful in improving models of the lunar 
libration.

Tests of the Nordtvedt effect for neutron stars
have also been carried out using
a class of systems known as wide-orbit binary millisecond pulsars (WBMSP),
which are 
pulsar--white-dwarf binary systems with very small orbital eccentricities.
In the gravitational field
of the galaxy, a non-zero Nordtvedt effect can induce an apparent anomalous
eccentricity pointed toward the galactic center,
which can be bounded using statistical methods, given enough WBMSPs. Using
data from 21 WBMSPs, including recently discovered highly circular systems,
Stairs et al.~\cite{stairs05} obtained the bound $\Delta < 5.6 \times
10^{-3}$, where $\Delta = \eta_\mathrm{N} (E_\mathrm{g}/M)_\mathrm{NS}$. Because 
$(E_\mathrm{g}/M)_\mathrm{NS} \sim 0.1$ for typical neutron stars, this bound does not
compete with the bound on $\eta_\mathrm{N}$ from LLR; on the
other hand, it does test SEP in the strong-field regime because of the
presence of the neutron star in each system.  The 2013 discovery of a millisecond pulsar in orbit with {\em two} white dwarfs in very circular, coplanar orbits~\cite{2014Natur.505..520R} may lead to a test of the Nordtvedt effect in the strong-field regime that surpasses the precision of lunar laser ranging by a substantial factor.


\subsubsection{Preferred-frame and preferred-location effects}
\label{preferred}

Some theories of gravity violate SEP by predicting that the
outcomes of local gravitational experiments may depend on the
velocity of the laboratory relative to the mean rest frame of the
universe (preferred-frame effects) or on the location of the
laboratory relative to a nearby gravitating body
(preferred-location effects). In the post-Newtonian limit,
preferred-frame effects are governed by the values of the PPN
parameters $\alpha_1$, $\alpha_2$, and $\alpha_3$, and some
preferred-location effects are governed by $\xi$ (see Table~\ref{ppnmeaning}).

The most important such effects are variations and anisotropies
in the locally-measured value of the gravitational constant which
lead to anomalous Earth tides and variations in the Earth's
rotation rate, anomalous contributions to the
orbital dynamics of planets, the Moon and binary pulsars; self-accelerations of
isolated pulsars; and anomalous torques on spinning bodies such as the Sun or pulsars  (see TEGP~8.2, 8.3, 9.3, and 14.3~(c)). 
A tight bound on $\alpha_3$ was obtained  from the period derivatives of 21 millisecond pulsars~\cite{stairs05}.  The best bound on
$\alpha_1$, comes from the orbit of the pulsar--white-dwarf system J1738+0333; while the best bounds on $\alpha_2$ and $\xi$ come from bounding torques on the solitary millisecond pulsars B1937+21 and J1744--1134~\cite{2012CQGra..29u5018S,2013CQGra..30p5019S,2013CQGra..30p5020S}.  Because these bounds involved systems with strong internal gravity of the neutron stars, they should strictly speaking be regarded as bounds on ``strong field'' analogues of the PPN parameters.  Here we will treat them as bounds on the standard
PPN  parameters, as shown in Table~\ref{ppnlimits}.

\subsubsection{Constancy of the Newtonian gravitational constant}
\label{bigG}

Most theories of gravity that violate SEP predict that the locally
measured Newtonian gravitational constant may vary with time as
the universe evolves. For the scalar-tensor theories listed in
Table~\ref{ppnvalues}, for example,
the predictions for $\dot{G}/G$ can be written in terms of time
derivatives of the asymptotic scalar field.
Where $G$
does change with cosmic evolution, its rate of variation should
be related to the expansion rate of the universe,
i.e.\ $\dot{G}/G \sim H_0$, where $H_0$ is the Hubble expansion parameter
and is given by
$H_0 =  73 \pm 3 \mathrm{\ km\ s}^{-1}\mathrm{\ Mpc}^{-1} \approx 0.75
\times 10^{-10}  \mathrm{\ yr}^{-1}$.

Several observational constraints can be placed on $\dot{G}/G$, one kind
coming from bounding the present rate of variation, another from bounding a
difference between the present value and a past value.
The first type of bound typically comes from LLR measurements,
planetary radar-ranging measurements, and pulsar timing data.  The best limits come from improvements in the ephemeris of Mars using range and Doppler data from the Mars Global
Surveyor (1998\,--\,2006), Mars Odyssey (2002\,--\,2008), and Mars Reconnaissance Orbiter 
(2006\,--\,2008), together with improved data and modeling of the effects of the asteroid belt~\cite{2011Icar..211..401K}.  Since the bound is actually on variations of $GM_\odot$, any future improvements in $\dot{G}/G$ beyond a part in $10^{13}$ will have to take into account models of the actual mass loss from the Sun, due to radiation of light and neutrinos ($\sim 0.7 \times 10^{-13} \, {\rm yr}^{-1}$) and due to the solar wind ($\sim 0.2 \times 10^{-13} \, {\rm yr}^{-1}$).
The
second type of bound comes from studies of the evolution of the Sun, stars and
the Earth, big-bang nucleosynthesis, and analyses of ancient eclipse data.
Selected results are shown in Table~\ref{Gdottable}; a thorough review is given in~\cite{2011LRR....14....2U}.

\begin{table}[t]
  \caption{Constancy of the gravitational constant.}
  \label{Gdottable}
  \renewcommand{\arraystretch}{1.2}
  \centering
  \begin{tabular}{l|cl}
    \hline \hline
    Method &
    $\rule{0 cm}{1.2 em}\dot{G}/G$ &
    Reference \\
    & ($10^{-13} \mathrm{\ yr}^{-1}$) \\
    \hline \hline
    Mars ephemeris&$0.1 \pm 1.6$&\cite{2011Icar..211..401K}\\
    Lunar laser ranging & $ 4 \pm 9 $ & \cite{williams04} \\
    Binary pulsars& $ -7 \pm 33 $ & \cite{2008ApJ...685L..67D,2009MNRAS.400..805L} \\
    Helioseismology & $ \phantom{0}0 \pm 16 $ & \cite{guenther98} \\
    Big Bang nucleosynthesis & $ 0 \pm 4 $ & \cite{copi04,bambi04} \\
    \hline \hline
  \end{tabular}
  \renewcommand{\arraystretch}{1.0}
\end{table}



\subsection{The search for gravitomagnetism}
\label{gravitomagnetism}

According to general relativity, moving or rotating matter should
produce a contribution to the gravitational field that is the analogue
of the magnetic field of a moving charge or a magnetic dipole. 
From the geometrical point of view, rotating matter produces a ``dragging of inertial
frames'' around the body, 
also called the Lense--Thirring effect. One consequence of this phenomenon
is a precession of a gyroscope's spin $\bf S$ according to 
\begin{equation}
  \frac{d{\bf S}}{dt} =
  {\bf \Omega}_\mathrm{LT} \times {\bf S},
  \qquad
  {\bf \Omega}_\mathrm{LT} =
  - \frac{1}{2} \left( 1 + \gamma + \frac{1}{4} \alpha_1 \right) 
  \frac{G\left ({\bf J} - 3 {\bf n} ({\bf n} \cdot {\bf J})\right )}{c^2r^3},
  \label{E42}
\end{equation}
where $\bf n$ is a
unit radial vector, and $r$ is the distance from the center of the
body with angular momentum $\bf J$ (TEGP~9.1).  

Another effect on the spin of a gyroscope arises from curved spacetime around the central body, called the ``geodetic precession'', given by
\begin{equation}
  \frac{d{\bf S}}{d\tau} =
  {\bf \Omega}_\mathrm{G} \times {\bf S},
  \qquad
  {\bf \Omega}_\mathrm{G} =
  \left( \gamma + \frac{1}{2} \right) \frac{{\bf v} \times \nabla U}{c^2},
  \label{E41}
\end{equation}
where $\bf {v}$ is the velocity of the gyroscope, and $U$ is the
Newtonian gravitational potential of the source (TEGP~9.1).

The Relativity
Gyroscope Experiment (Gravity Probe~B or GPB)
carried out by Stanford University, 
NASA  and Lockheed--Martin Corporation, was a space mission designed to detect both precessional effects.
Gravity Probe B will very likely go down in the history of science as
one of the most ambitious, difficult, expensive, and controversial
relativity experiments ever performed.\footnote{Full disclosure: The author served as Chair of an external NASA Science Advisory Committee
  for Gravity Probe B from 1998 to 2010.}  It was almost 50 years from inception to completion, although only about half of that time was spent as a full-fledged, approved space program.

The GPB spacecraft was launched on April 20, 2004 into an almost perfectly
circular polar orbit at an altitude of $642$ km, with the orbital plane
parallel to the direction of a guide star known as {\em IM Pegasi} (HR 8703). 
The spacecraft contained four spheres made of fuzed quartz, all
spinning about the same axis (two were spun in the opposite
direction), which was oriented to be in the orbital plane, pointing
toward the guide star.  An onboard telescope pointed continuously at
the guide star, and the direction of each spin was compared with the
direction to the star, which was at a declination of $16^{\rm o}$
relative to the Earth's equatorial plane.  With these conditions, the predicted precessions were 
$6630$ milliarcsecond per year for the geodetic effect, and 
 $38$ milliarcsecond per year for frame dragging, the 
former in the orbital plane (in the north-south direction) and
the latter perpendicular to it (in the east-west direction).    

In order to reduce the non-relativistic torques on the
rotors to an acceptable level, the rotors were fabricated to be both
spherical and homogenous to better than a few parts in 10 million.
Each rotor was coated with a thin film of niobium, and the experiment
was conducted at cryogenic temperatures inside a dewar containing 2200 
litres of superfluid liquid helium. As the niobium film becomes a
superconductor, each rotor develops a magnetic moment parallel to its
spin axis.  Variations in the direction of the magnetic moment
relative to the spacecraft were then measured using superconducting current loops
surrounding each rotor.  As the spacecraft orbits the Earth, the
aberration of light 
from the guide star causes an artificial but
predictable change in direction between the rotors and the on-board
telescope; this was an essential tool for calibrating the conversion
between the voltages read by the current loops and the actual angle
between the rotors and the guide star.  

The mission
ended in September 2005, as scheduled, when the last of the
liquid helium boiled off.  Although all subsystems of the spacecraft 
and the apparatus performed extremely well, they were not perfect.
Calibration measurements carried out during the mission, both before
and after the science phase, revealed unexpectedly large torques on
the rotors.  Numerous diagnostic tests worthy of a detective novel showed that these were caused by electrostatic interactions
between surface imperfections (``patch effect'') on the niobium films and the spherical
housings surrounding each rotor.  These effects and other anomalies
greatly contaminated the data and complicated its analysis, but
finally, in October 2010, the Gravity Probe B team announced that the
experiment had successfully measured both the geodetic and
frame-dragging precessions. The outcome was in agreement with general
relativity, with a precision of $0.3$ percent for the geodetic precession, 
and $20$ percent for the frame-dragging effect~\cite{2011PhRvL.106v1101E}. 

Another way to look for frame-dragging is to
measure the precession of orbital planes of bodies circling a rotating
body.  One implementation of this idea is to
measure the relative precession, at about 31 milliarcseconds per year,
of the line of nodes of a pair
of laser-ranged geodynamics satellites (LAGEOS), ideally with supplementary
inclination angles; the inclinations must be supplementary in order
to cancel the huge (126 degrees per year)
nodal precession caused by the Earth's
Newtonian gravitational multipole moments. Unfortunately, the two
existing LAGEOS satellites are not in appropriately inclined orbits.
 Nevertheless, Ciufolini et al.~\cite{ciufolini04,2006NewA...11..527C} 
combined 
nodal precession data
from LAGEOS I and II with improved models for the Earth's multipole moments
provided by 
two recent orbiting geodesy satellites, Europe's
CHAMP (Challenging Minisatellite Payload)
and NASA's GRACE (Gravity Recovery and Climate Experiment),
and reported a 5\,--\,10 percent confirmation of general relativity. 

On February 13, 2012, a third laser-ranged satellite, known as LARES
(Laser Relativity Satellite) was launched by the Italian Space Agency.
Its inclination was very close to the required
supplementary angle relative to LAGEOS I, and its eccentricity was
very nearly zero.  However, because its semimajor axis is only $2/3$ that of either LAGEOS I or II, 
and because the Newtonian precession rate is proportional to $a^{-3/2}$, LARES does not provide a cancellation of the Newtonian precession.  Combining data from
all three satellites with continually improving Earth data from GRACE,
the LARES team hopes to achieve a test of
frame-dragging at the one percent level. 


\section{Binary-pulsar tests of gravitational theory}
\label{stellar}

The 1974 discovery of the binary pulsar B1913+16 by Joseph Taylor
and Russell Hulse during a routine search for new pulsars was a milestone in the history of general relativity.  It led to the first confirmation of the existence of gravitational radiation and to a Nobel Prize for Taylor and Hulse, and it opened a new field for testing gravitational theories in the strong-field and gravitational-wave regimes.

Following that discovery, only two new binary pulsars were discovered with interesting relativistic properties during the next 15 years (out of a total of 14 systems), but thanks to improved radio-telescope sensitivity, better techniques for removing the effects of interstellar dispersion, better timing capabilities (exploiting GPS time transfer, for example) and improved detection algorithms, the pace of discovery of binary pulsars picked up steadily.  Today, over 220 binary pulsars are known, with 23 discovered in 2012 alone.  To be sure, the vast majority of these are of little relativistic interest, either because they are so widely separated that relativistic effects are negligible, because they involve significant mass-transfer from the companion (X-ray binary pulsars), or because the timing characteristics of the pulsar do not meet the stability demanded to measure relativistic effects.  

Nevertheless, close to a dozen binary pulsars are relativistically interesting for one reason or another, and this zoo of systems contains some fascinating beasts, including the famous ``double pulsar'', systems with white-dwarf companions, a system with the most massive neutron star known, and systems with extraordinarily circular orbits.

\subsection{The binary pulsar and general relativity}
\label{binarypulsars}

We will begin by reviewing the iconic Hulse-Taylor binary pulsar, by far the best studied of all the systems.   Pulse arrival-time measurements have been made regularly (apart from a four-year break during upgrading of the Arecibo radio telescope) for over 30 years.  Table \ref{bpdata} lists the current values of the key orbital and relativistic parameters, from analysis of data through 2006~\cite{2010ApJ...722.1030W}.

\begin{table}[t]
  \caption[Parameters of the binary pulsar B1913+16.]{Parameters of
    the binary pulsar B1913+16. The numbers in parentheses denote
    errors in the last digit.}
  \label{bpdata}
  \renewcommand{\arraystretch}{1.2}
  \centering
  \begin{tabular}{l|ll}
    \hline \hline
    Parameter &
    \multicolumn{1}{c}{Symbol} &
    \multicolumn{1}{c}{Value} \\
    & \multicolumn{1}{c}{(units)} \\
    \hline \hline
    \phantom{ii}(i) Astrometric and spin parameters: \\ [0.3 em]
    \phantom{(iii)~} Right Ascension &
    $\alpha$ &
    $19^\mathrm{h} 15^\mathrm{m}27.^\mathrm{s} 99928(9)$ \\
    \phantom{(iii)~} Declination &
    $\delta$ &
    $16^\circ 06'27.''3871(13)$ \\
    \phantom{(iii)~} Pulsar period &
    $P_\mathrm{p}$ (ms) &
    $59.0299983444181(5)$ \\
    \phantom{(iii)~} Derivative of period &
    $\dot P_\mathrm{p}$ &
    $8.62713(8) \times 10^{-18}$ \\ [1 em]
    \phantom{i}(ii) ``Keplerian'' parameters: \\ [0.3 em]
    \phantom{(iii)~} Projected semimajor axis &
    $a_\mathrm{p} \sin i$ (s) &
    $2.341782(3)$ \\
    \phantom{(iii)~} Eccentricity &
    $e$ &
    $0.6171334(5)$ \\
    \phantom{(iii)~} Orbital period &
    $P_\mathrm{b}$ (day) &
    $0.322997448911(4)$ \\
    \phantom{(iii)~} Longitude of periastron &
    $\omega_0$ (${}^\circ$) &
    $292.54472(6)$ \\
    \phantom{(iii)~} Julian date of periastron &
    $T_0$ (MJD) &
    $52144.90097841(4)$ \\ [1 em]
    (iii) ``Post-Keplerian'' parameters: \\ [0.3 em]
    \phantom{(iii)~} Mean rate of periastron advance &
    $\langle \dot\omega \rangle$ ($ {}^\circ \mathrm{\ yr}^{-1} $) &
    $4.226598(5)$ \\
    \phantom{(iii)~} Redshift/time dilation &
    $\gamma'$ (ms) &
    $4.2992(8)$ \\
    \phantom{(iii)~} Orbital period derivative &
    $\dot P_\mathrm{b}$ ($10^{-12}$) &
    $-2.423(1)$ \\
    \hline \hline
  \end{tabular}
  \renewcommand{\arraystretch}{1.2}
\end{table}

The observational parameters that are obtained from a least-squares
solution of the arrival-time data fall into three groups:
\begin{enumerate}
\item non-orbital parameters, such as the pulsar period and its rate
  of change (defined at a given epoch), and the position of the pulsar
  on the sky;
\item five ``Keplerian'' parameters, most closely related to those
  appropriate for standard Newtonian binary systems, such as the
  eccentricity $e$, the orbital period $P_\mathrm{b}$, and the
  semi-major axis of the pulsar projected along the line of sight,
  $a_\mathrm{p} \sin i$; and
\item five ``post-Keplerian'' parameters.
\end{enumerate}
The five post-Keplerian parameters are: $\langle \dot \omega \rangle$,
the average rate of periastron advance; $\gamma'$, the amplitude of
delays in arrival of pulses caused by the varying effects of the
gravitational redshift and time dilation as the pulsar moves in its
elliptical orbit at varying distances from the companion and with
varying speeds (not to be confused with the PPN parameter $\gamma$); $\dot P_\mathrm{b}$, the rate of change of orbital
period, caused predominantly by gravitational radiation damping; and
$r$ and $s = \sin i$, respectively the ``range'' and ``shape'' of the
Shapiro time delay of the pulsar signal as it propagates through the
curved spacetime region near the companion, where $i$ is the angle of
inclination of the orbit relative to the plane of the sky. An
additional 14 relativistic parameters are measurable in
principle~\cite{DamourTaylor92}.

In general relativity, the five post-Keplerian parameters can be related
to the masses of the two bodies and to measured Keplerian parameters
by the equations (TEGP~12.1, 14.6~(a))
\begin{eqnarray}
    \langle \dot \omega \rangle & = &
    \frac{6 \pi}{P_\mathrm{b}} \left ( \frac{2\pi Gm \omega_\mathrm{b}}{c^3 } \right)^{2/3} \frac{1}{1-e^2},
    \\ 
    \gamma' & = & \displaystyle e \frac{P_\mathrm{b}}{2\pi}
     \left ( \frac{2\pi Gm \omega_\mathrm{b}}{c^3} \right)^{2/3} \frac{m_2}{m}
    \left( 1 + \frac{m_2}{m} \right),
    \\ [0.5 em]
    \dot P_\mathrm{b} & = & \displaystyle
    - \frac{192 \pi}{5}  \left ( \frac{2\pi G{\cal M} \omega_\mathrm{b}}{c^3} \right)^{5/3} F(e),
    \\ [0.5 em]
    s & = & \sin i,
    \\ [0.3 em]
    r & = & m_2,
  \label{pkparameters}
\end{eqnarray}
where $\omega_\mathrm{b} \equiv 2\pi/P_\mathrm{b}$ is the orbital angular frequency; 
$m_1$ and $m_2$ are the pulsar and companion masses, respectively; $m$ is the total mass; $\cal M$ is the so-called ``chirp'' mass, given by ${\cal M} \equiv (m_1m_2/m^2)^{3/5} m$; and 
$F (e)  =  (1 - e^2)^{-7/2}
  ( 1 + {73e^2}/{24}  + {37e^4}/{96} )$.

Notice that, by virtue of Kepler's 
third
law,  $(2\pi Gm \omega_\mathrm{b}/c^3)^{2/3} = Gm/ac^2 \sim
\epsilon$, thus the first two post-Keplerian parameters can be seen
as ${\cal O} (\epsilon)$, or 1PN corrections to the underlying variable, while the
third is an ${\cal O} (\epsilon^{5/2})$, or 2.5PN correction.
The parameters $r$ and $s$ are not separately
measurable with interesting accuracy for B1913+16 because the
orbit's $47 ^\circ$ inclination does not lead to a substantial Shapiro
delay.

Because $P_\mathrm{b}$ and $e$ are separately measured parameters, the
measurement of the three post-Keplerian parameters provide three
constraints on the two unknown masses. The periastron shift measures
the total mass of the system, $\dot P_\mathrm{b}$ measures the chirp mass, and
$\gamma'$ measures a complicated function of the masses.
General relativity passes the test if it provides a consistent solution to these
constraints, within the measurement errors.

\begin{figure}[t]
  \centerline{\includegraphics[scale=0.45]{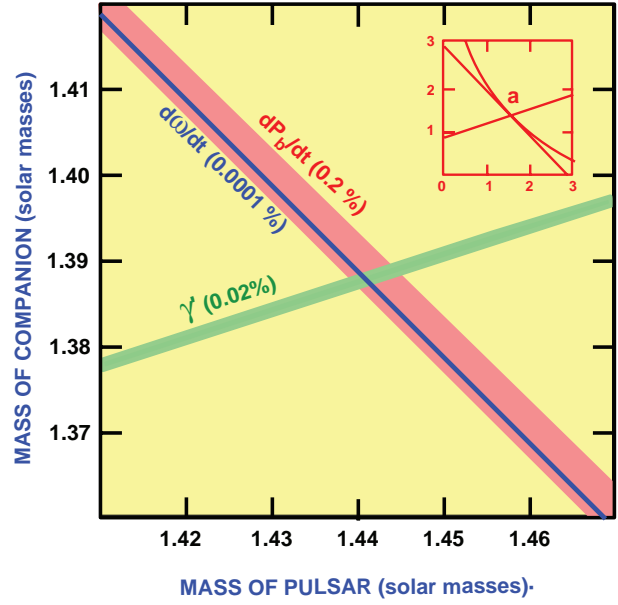}}
  \caption{Constraints on masses of the pulsar and its companion
    from data on B1913+16, assuming general relativity to be valid. The width of
    each strip in the plane reflects observational accuracy, shown as
    a percentage. An inset shows the three constraints on the full
    mass plane;  the intersection region (a) has been magnified 400
    times for the full figure.}
  \label{bpfigure1}
\end{figure}

From the intersection of the  $\langle \dot \omega \rangle$
and $\gamma' $ constraints we obtain the values
$m_1 = 1.4398 \pm 0.0002 M_\odot$ and
$m_2 = 1.3886 \pm 0.0002 M_\odot$. The third of Eqs.~(\ref{pkparameters})
then predicts the value
$\dot P_\mathrm{b} = -2.402531 \pm 0.000014 \times 10^{-12}$.
In order to compare the predicted
value for $\dot P_\mathrm{b}$ with the observed value of Table~\ref{bpdata}, it
is necessary to
take into account the small kinematic effect of a relative acceleration between the
binary pulsar system and the solar system caused by the differential
rotation of the galaxy.  Using data
on the location and proper motion of the pulsar, combined with the best
information available on galactic rotation; the current value of this effect
is
 $\dot P_\mathrm{b}^\mathrm{gal} \simeq -(0.027 \pm 0.005) \times 10^{-12}$.
Subtracting this from the observed $\dot P_\mathrm{b}$ (see Table~\ref{bpdata})
gives the corrected 
$\dot P_\mathrm{b}^\mathrm{corr} = -(2.396 \pm 0.005)
\times 10^{-12}$,
which agrees with the prediction within the errors. In other words,
\begin{equation}
  \frac{\dot P_\mathrm{b}^\mathrm{corr}}{\dot P_\mathrm{b}^\mathrm{GR}} =
  0.997 \pm 0.002.
  \label{Pdotcompare}
\end{equation}
The consistency among the measurements is displayed in Figure~\ref{bpfigure1},
in which the regions allowed by the three most precise constraints
have a single common overlap.
Uncertainties in the parameters that go into the galactic correction are now
the limiting factor in the accuracy of the test of gravitational damping.

\begin{figure}[t]
  \centerline{\includegraphics[scale=0.45]{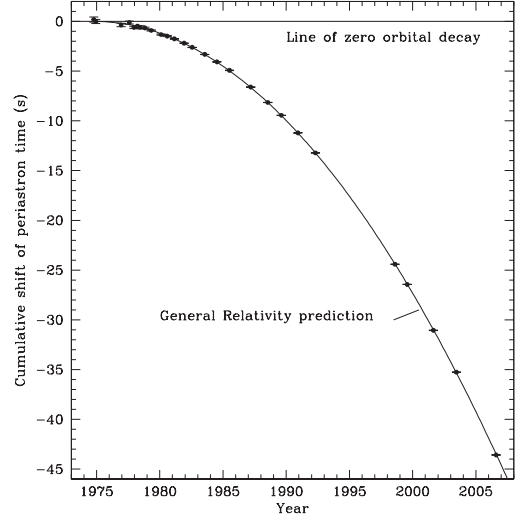}}
  \caption{Plot of the cumulative shift of the periastron time
    from 1975\,--\,2005. The points are data, the curve is the general relativity
    prediction. The gap during the middle 1990s was caused by a
    closure of Arecibo for upgrading~\cite{2010ApJ...722.1030W}.}
  \label{bpfigure2}
\end{figure}

A third way to display the agreement with general relativity is by
plotting the cumulative shift of the time of periastron passage caused by the changing orbital period.  Figure~\ref{bpfigure2} shows the results: the
dots are the data points, while the curve is the predicted difference
using the measured masses and the quadrupole formula for gravitational radiation damping~\cite{2010ApJ...722.1030W}.

The consistency among the constraints
provides a test of the assumption that the two bodies
behave as ``point'' masses, without complicated tidal effects, obeying
the general relativistic equations of motion including
gravitational radiation. It is also a test of strong gravity,
in that the highly relativistic internal structure of the
neutron stars does not influence their orbital motion, as predicted by
the SEP of general relativity.

Observations of  
variations in the pulse profile suggest that the pulsar is
undergoing geodetic precession on a 300-year timescale
as it moves through the curved spacetime
generated by its companion (see Section~\ref{gravitomagnetism}).
The amount is consistent with general relativity, assuming that the pulsar's
spin is suitably misaligned with the orbital angular momentum~\cite{kramer, WeisbergTaylor02}.
Unfortunately, the evidence suggests that the pulsar beam may precess
out of our line of sight by 2025.


\subsection{A zoo of binary pulsars}
\label{population}

Over 70 binary neutron star systems with orbital periods less than
a day are now known.  While some are less interesting for
testing relativity, some have yielded interesting tests, and others, notably
the recently discovered ``double pulsar'' are likely to continue to produce significant
results well into the future. Here we describe some of the more interesting
or best studied cases.

\begin{figure}[t]
  \centerline{\includegraphics[scale=0.45]{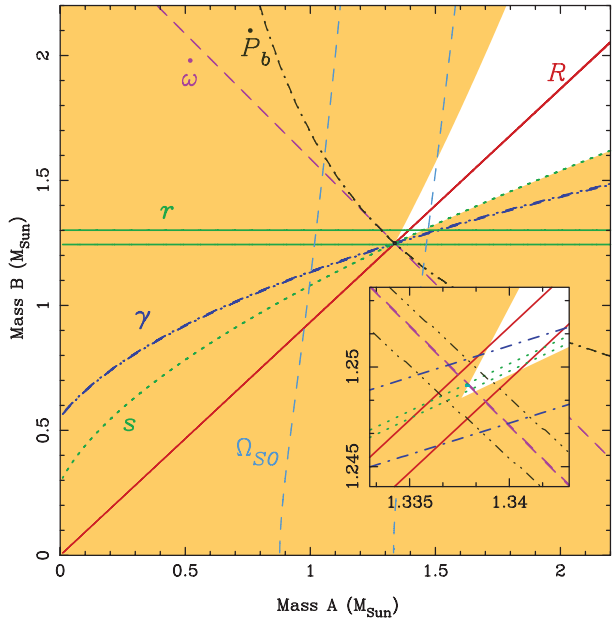}}
  \caption{Constraints on masses of the pulsar and its companion
    from data on J0737-3039A,B, assuming general relativity to be valid.  The inset shows the intersection region magnified by a factor of 80.  (Figure courtesy of M. Kramer)
  }
  \label{bpfigure3}
\end{figure}

\medskip
\noindent
{\bf The ``double'' pulsar: J0737-3039A, B.}  This binary pulsar system, discovered in 2003~\cite{burgay03}, was
  already remarkable for its extraordinarily short orbital period (0.1
  days) and large periastron advance
  ($16.8995^\circ \mathrm{\ yr}^{-1}$), but then the companion was also
  discovered to be a pulsar~\cite{lyne04}.  Because two projected
  semi-major axes could be measured,  the mass ratio was obtained
  directly from the ratio of the two values of $a_\mathrm{p} \sin i$,
  and thereby the two masses could be obtained by combining that ratio with the
  periastron advance, assuming general relativity. The results are
  $m_A = 1.3381 \pm 0.0007 \, M_\odot$ and
  $m_B = 1.2489 \pm 0.0007 \, M_\odot$, where $A$ denotes the primary
  (first) pulsar. From these values, one finds that the orbit is
  nearly edge-on, with $\sin i = 0.9997$, a value which is
  completely consistent with that inferred from the Shapiro delay
  parameter. In fact, the five measured
  post-Keplerian parameters plus the ratio of the projected semi-major
  axes give six constraints on the masses (assuming general relativity): as seen in Fig.\ \ref{bpfigure3}, all six
  overlap within their measurement errors~\cite{2006Sci...314...97K}.  Because of the location of the system, galactic proper motion
  effects play a significantly smaller role in the interpretation
  of $\dot P_\mathrm{b}$ measurements than they did in B1913+16; this and the reduced effect of interstellar dispersion means that the accuracy of measuring the gravitational-wave damping may soon beat that from the Hulse-Taylor system.   The geodetic precession of pulsar B's spin axis has also been measured by monitoring changes in the patterns of eclipses of the signal from pulsar A, with a result in agreement with general relativity to about 13 percent~\cite{2008Sci...321..104B}; the constraint on the masses from that effect (assuming general relativity to be correct) is also shown in Fig.\  \ref{bpfigure3}.

\medskip
\noindent
{\bf J1738+0333: A white-dwarf companion.}
This is a low-eccentricity, $8.5$-hour period system in which the white-dwarf companion is bright enough to permit detailed spectroscopy, allowing the companion mass to be determined directly to be $0.181 \, M_\odot$.  The mass ratio is determined from Doppler shifts of the spectral lines of the companion and of the pulsar period, giving the pulsar mass $1.46 \, M_\odot$.  Ten years of observation of the system yielded both a measurement of the apparent orbital period decay, and enough information about its parallax and proper motion to account for the substantial galactic kinematic effect to give a value of the intrinsic period decay of $\dot{P}_\mathrm{b} =  (-25.9 \pm 3.2) \times 10^{-15} \mathrm{ s \, s^{-1}}$ in agreement with the predicted effect~ \cite{2012MNRAS.423.3328F}.   But because of the asymmetry of the system, the result also places a significant bound on the existence of dipole radiation, predicted by many alternative theories of gravity (see Sec. \ref{binarypulsarsalt} below for discussion).  Data from this system were also used to place the tight bound on the PPN parameter $\alpha_1$ shown in Table \ref{ppnlimits}.

\medskip
\noindent
{\bf J1141-6545: A white-dwarf companion.}
This system is similar in some ways to the Hulse-Taylor binary: short orbital period ($0.20$ days), significant orbital eccentricity ($0.172$), rapid periastron advance ($5.3$ degrees per year) and massive components ($M_p = 1.4 M_\odot$, $M_c = 1.0 M_\odot$).  The key difference is that the companion is again a white dwarf.  The intrinsic orbit period decay has been measured in agreement with general relativity to about six percent, again placing limits on dipole gravitational radiation~\cite{2008PhRvD..77l4017B}.

\medskip
\noindent
{\bf J0348+0432: The most massive neutron star.}
Discovered in 2011, this is another neutron-star white-dwarf system, in a very short period ($0.1$ day), low eccentricity ($2 \times 10^{-6}$) orbit.  Timing of the neutron star and spectroscopy of the white dwarf have led to mass values of $0.172 \, M_\odot$ for the white dwarf and $2.01 \pm 0.04 \, M_\odot$ for the pulsar, making it the most massive accurately measured neutron star yet.  Along with an earlier measurement of a $2 \, M_\odot$ pulsar, this ruled out a number of heretofore viable soft equations of state for nuclear matter.   The orbit period decay agrees with the general relativity prediction within 20 percent and is expected to improve steadily with time. 

\medskip
\noindent
{\bf J0337+1715: Two white-dwarf companions.}
This remarkable system was reported in 2014~\cite{2014Natur.505..520R}.  It consists of a 2.73 millisecond pulsar ($M=1.44 \, M_\odot$) with extremely good timing precision, accompanied by {\em two} white dwarfs in coplanar circular orbits.
The inner white dwarf ($M = 0.1975 \, M_\odot$) has an orbital period of $1.629$ days, with $e = 6.918 \times 10^{-4}$, and the outer white dwarf ($M = 0.41 \, M_\odot$) has a period of $327.26$ days, with $e = 3.536 \times 10^{-2}$.   This is an ideal system for testing the Nordtvedt effect in the strong-field regime.  Here the inner system is the analogue of the Earth-Moon system, and the outer white dwarf plays the role of the Sun.  Because the outer semi-major axis is about 1/3 of an astronomical unit, the basic driving perturbation is comparable to that provided by the Sun.  However, the self-gravitational binding energy per unit mass of the neutron star is almost a billion times larger than that of the Earth, greatly amplifying the size of the Nordtvedt effect.  Depending on the details, this system could exceed lunar laser ranging in testing the Nordtvedt effect by several orders of magnitude.

\medskip
\noindent
{\bf Other binary pulsars.}
Two of the earliest binary pulsars, B1534+12 and B2127+11C, discovered in 1990, failed to live up to their early promise despite being similar to the Hulse-Taylor system in most respects (both were believe to be double neutron-star systems).  The main reason was the significant uncertainty in the kinematic effect on $\dot{P}_\mathrm{b}$  of local accelerations, galactic in the case of B1534+12, and those arising from the globular cluster that was home to B2127+11C.


\subsection{Binary pulsars and alternative theories}
\label{binarypulsarsalt}

Soon after the discovery of the binary pulsar it was widely hailed as
a new testing ground for relativistic gravitational effects.
As we have seen in the case of general relativity, in most respects,
the system has lived up to, indeed exceeded, the early expectations.

In another respect, however, the system has only partially lived up to its
promise, namely as a direct testing ground for alternative theories of
gravity. The origin of this promise was the discovery
that alternative theories of gravity generically predict the emission
of dipole gravitational radiation from binary star systems.
In general relativity, there is no dipole radiation because the
``dipole moment'' (center of mass) of isolated systems is
uniform in time (conservation
of momentum), and because the ``inertial mass'' that determines the
dipole moment is the same as the mass that generates gravitational
waves (SEP). In other theories, while the
inertial dipole moment may remain uniform, the ``gravity wave'' dipole
moment need not, because the mass that generates gravitational waves
depends differently on the internal
gravitational binding energy of each body than does the inertial mass
(violation of SEP).

In theories that violate SEP,
the difference between gravitational wave mass and inertial mass is a
function of the internal gravitational binding energy of the bodies.
This additional form of gravitational radiation damping could,
at least in principle, be significantly stronger than the usual quadrupole
damping, because it depends on fewer powers of the orbital velocity $v$,
and it depends on the gravitational binding energy per unit mass of
the bodies, which, for neutron stars, could be as large as 20~percent
(see TEGP~10 for further details).
As one fulfillment of this promise, Will and Eardley worked out in
detail the effects of dipole gravitational radiation in the bimetric theory
of Rosen, and, when the first observation of the decrease of
the orbital period was announced in 1979, the Rosen theory
suffered a terminal blow.  A wide
class of alternative theories also failed the binary pulsar test because
of dipole gravitational radiation (TEGP~12.3).

On the other hand, the early observations of PSR 1913+16
already indicated that, in
general relativity, the masses of the two bodies were nearly equal, so
that, in theories of gravity that are in some sense ``close'' to
general relativity, dipole gravitational radiation would not be a
strong effect, because of the apparent symmetry of the system.
Thus, despite the presence of dipole gravitational radiation,
the Hulse-Taylor binary pulsar provides at present only a weak test of Brans--Dicke
theory, not competitive with solar-system tests.

However, in some 
generalized scalar-tensor theories, the strong internal gravity of the neutron star can induce a phenomenon called ``spontaneous scalarization'' whereby the behavior can be significantly different from either general relativity or from a scalar-tensor theory with a nominally large value of $\omega_0$.  This is true of theories with negative values of the parameter
$\beta_0$ in Eq.\ (\ref{alphaexpand})~\cite{DamourEspo98}.  Furthermore, the recently discovered mixed binary pulsar systems J1141-6545 and J1738+0333 have been exploited using precise timing of the pulsar, spectroscopic observations of the white-dwarf companion, and the strong dipole gravitational radiation effect to yield  stringent bounds.  Indeed, the latter system surpasses the Cassini bound for $\beta_0 > 1$ and $\beta_0 < -2$, and is close to that bound for the pure Brans-Dicke case $\beta_0 =0$~\cite{2012MNRAS.423.3328F}.

\section{Testing general relativity in its second century}
\label{S4}

\subsection{Gravitational-wave tests}
\label{gwaves}

Shortly after the centenary year of general relativity, a new opportunity for testing
relativistic gravity will be realized, 
when a worldwide network of advanced kilometer-scale, laser interferometric
gravitational wave observatories in the U.S.\ (LIGO pro\-ject), and Europe
(VIRGO and GEO600 projects),  begins regular
detection and analysis of gravitational wave signals from astrophysical
sources. These will be followed by advanced detectors in Japan and India.
They will have the capability of detecting and
measuring the gravitational waveforms from astronomical sources in a
frequency band between about 10~Hz  and
5000~Hz, with a maximum
sensitivity to strain at around a hectahertz (100~Hz). 
The most promising source for detection and study of
the gravitational wave signal is the ``inspiralling compact binary''
-- a binary system of neutron stars or black holes (or one of each) in
the final minutes of a death spiral leading to a violent merger.
Such is the fate, for example, of the Hulse--Taylor binary pulsar 
B1913+16 in about 300~Myr, or the ``double pulsar'' J0737-3039
in about 85~Myr.  Given the expected sensitivity of 
``advanced LIGO'' (around 2016), which could see such sources out to
many hundreds of megaparsecs, it has been estimated that from 10 to
several hundred annual inspiral events could be detectable.
Other sources, such as binary black-hole mergers, supernova core collapse events, instabilities
in rapidly rotating newborn neutron stars, signals from
non-axisymmetric pulsars, and a stochastic background of waves, may be
detectable.  For a review of gravitational-wave theory and detection, see~\cite{CreightonAnderson}.

In addition, plans are being developed for an orbiting laser
interferometer space antenna. Such a system,
consisting of three spacecraft orbiting the sun in a triangular
formation separated from each other by a million kilometers,
would be sensitive primarily in the very low frequency band between
$10^{-4}$ and $10^{-1} \mathrm{\ Hz}$~\cite{2012CQGra..29l4016A}.  Such a mission, dubbed eLISA, is a leading candidate to address the ``gravitational universe'' science theme that has been selected by the European Space Agency for a mission to be launched around 2034.

A third approach that focuses on the ultra low-frequency band (nanohertz) is that of Pulsar Timing Arrays (PTA), whereby a network of highly stable millisecond pulsars is monitored in a coherent way using radio telescopes, in hopes of detecting the fluctuations in arrival times induced by passing gravitational waves.

In addition to opening a new astronomical window, the
detailed observation of gravitational waves by such observatories may
provide the means to test general relativistic predictions for the
polarization and speed of the waves, for gravitational radiation
damping and for strong-field gravity.


\subsubsection{Polarization of gravitational waves}

\begin{figure}[t]
  \centerline{\includegraphics[scale=0.45]{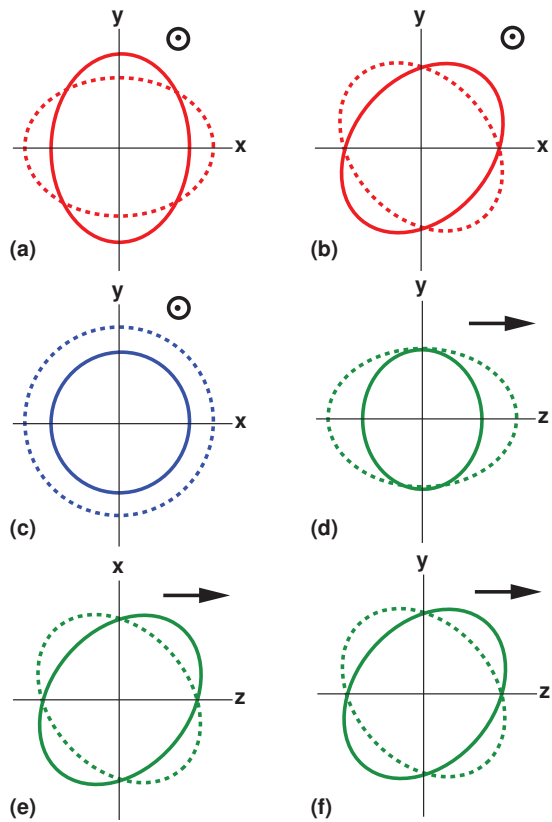}}
  \caption{The six polarization modes for gravitational waves
    permitted in any metric theory of gravity. Shown is the
    displacement that each mode induces on a ring of test particles.
    The wave propagates in the $+z$ direction. There is no
    displacement out of the plane of the picture. In (a), (b), and
    (c), the wave propagates out of the plane; in (d), (e), and (f),
    the wave propagates in the plane.}
  \label{wavemodes}
\end{figure}

Like electromagnetic waves, gravitational waves can be characterized by  polarizations.
In general metric theories of gravity there are six independent polarization modes. Three are transverse to the direction of
propagation, with two representing quadrupolar deformations and one
representing a monopolar ``breathing'' deformation. Three modes are
longitudinal, with one an axially symmetric
stretching mode in the propagation direction,
and one quadrupolar mode in each of the two orthogonal planes containing the
propagation direction. Figure~\ref{wavemodes} shows the displacements
induced on a ring of freely falling test particles by each of these
modes.
General relativity predicts only the first two
transverse quadrupolar modes (a) and (b) independently of the source.
Massless scalar-tensor gravitational waves
can in addition contain the transverse breathing mode (c).  More general
metric theories predict additional longitudinal modes,
up to the full complement of
six (TEGP~10.2).

A suitable array of gravitational antennas could delineate or limit
the number of modes present in a given wave; a space antenna could do likewise by virtue of its changing orientation during the passage of a sufficiently long wave train.
If
distinct evidence were found of any mode other than the two
transverse quadrupolar modes of general relativity, the result would be disastrous for
general relativity.  

\subsubsection{Gravitational radiation back-reaction}
\label{backreaction}

In the binary pulsar, a test of general relativity was made possible by measuring at
least three relativistic effects that depended upon only two unknown
masses. The evolution of the orbital phase under the damping effect
of gravitational radiation played a crucial role. Another situation
in which measurement of orbital phase can lead to tests of general relativity is that
of the inspiralling compact binary system. The key differences are
that here gravitational radiation itself is the detected signal,
rather than radio pulses, and the phase evolution alone carries all
the information. In the binary pulsar, the first derivative of the
binary frequency $\dot \omega_\mathrm{b}$ was measured; here the full nonlinear
variation of $\omega_\mathrm{b}$ as a function of time is measured.

Broad-band laser interferometers
are especially sensitive to the phase evolution of the gravitational
waves, which carry the information about the orbital phase evolution.
The analysis of gravitational wave data from such sources will involve
some form of matched filtering of the noisy detector output against an
ensemble of theoretical ``template'' waveforms which depend on the
intrinsic parameters of the inspiralling binary, such as the component
masses, spins, and so on, and on its inspiral evolution.

But for laser interferometric observations of gravitational waves,
the bottom line is that, in order to measure the astrophysical
parameters of the source and to test the properties of the
gravitational waves, it is necessary to derive the
gravitational waveform and the resulting radiation back-reaction on the orbit
phasing to many post-Newtonian orders
beyond the quadrupole
approximation.  Two decades of intensive work by many groups have led to the development of waveforms in general relativity that are accurate to 3.5PN order, and for some specific effects, such as those related to spin, to 4.5PN order
(see~\cite{BlanchetLRR} for a thorough review).

On the other hand, alternative theories of gravity are likely to predict rather different gravitational waveforms, notably via the addition of
dipole gravitational radiation.  Thus tests of alternative theories might be possible using the gravitational-wave signals.  Several attempts have been made to parametrize the gravitational-wave signal from binary inspiral, in order to encompass alternative theories of gravity, in the spirit of the PPN framework~\cite{2009PhRvD..80l2003Y,2010PhRvD..82f4010M}.


\subsubsection{Speed of gravitational waves}

According to general relativity, in the limit in which the wavelength of gravitational
waves is small compared to the radius of curvature of the background
spacetime, the waves propagate along null geodesics of the background
spacetime, i.e.\ they have the same speed $c$ as light (in this
section, we do not set $c=1$). In other
theories, the speed could differ from $c$ because of coupling of
gravitation to other gravitational fields, 
or if gravitation were propagated by a massive field (a
massive graviton).  In the latter case $v_\mathrm{g}$ would
be given by, in a local inertial frame,
\begin{equation}
  \frac{v_\mathrm{g}^2}{c^2} = 1 - \frac{m_\mathrm{g}^2 c^4}{E^2},
  \label{eq1}
\end{equation}
where $m_\mathrm{g}$ and $E$ are the graviton rest mass and energy,
respectively.

The most obvious way to test this is to
compare the arrival times of a gravitational wave and an
electromagnetic
wave from the same event, e.g., a supernova, or a binary inspiral with an associated electromagnetic signal.
For a source at a distance $D$, the
resulting value of the difference $1-v_\mathrm{g}/c$ is
\begin{equation}
  1 - \frac{v_\mathrm{g}}{c} = 5 \times 10^{-17}
  \left( \frac{200 \mathrm{\ Mpc}}{D} \right)
  \left( \frac{\Delta t}{1 \mathrm{\ s}} \right),
  \label{eq2}
\end{equation}
where $\Delta t \equiv \Delta t_\mathrm{a} - (1+Z) \Delta t_\mathrm{e} $
is the ``time difference'',
where $\Delta t_\mathrm{a}$ and $\Delta t_\mathrm{e}$ are the differences
in arrival time and emission time of the
two signals, respectively, and $Z$ is the redshift of the source.
In many cases, $\Delta t_\mathrm{e}$ is unknown,
so that the best one can do is employ an upper bound on
$\Delta t_\mathrm{e}$ based on observation or modelling.
The result will then be a bound on $1-v_\mathrm{g}/c$.

The foregoing discussion assumes that the source emits \emph{both}
gravitational and electromagnetic radiation in detectable amounts, and
that the relative time of emission can be established
to sufficient accuracy, or can be shown to be sufficiently
small.

However, there is a situation in which a bound on the graviton mass
can be set using gravitational radiation alone~\cite{graviton}.
That is the case of
the inspiralling compact binary. Because the frequency of the
gravitational radiation sweeps from low frequency at the initial
moment of observation to higher frequency at the final moment, the
speed of the gravitons emitted will vary, from lower speeds initially
to higher speeds (closer to $c$) at the end. This will cause a
distortion of the observed phasing of the waves that can be detected or bounded
through matched filtering.  

These and other tests of gravitational theory using gravitational waves are thoroughly reviewed by~\cite{2013LRR....16....7G} and by~\cite{2013LRR....16....9Y}.


\subsection{Strong-field tests}
\label{other}

One of the central difficulties of testing general relativity in the
strong-field regime is the possibility of contamination by uncertain
or complex physics. In the solar system, weak-field gravitational effects
can
in most cases be measured cleanly and separately from
non-gravitational effects. The remarkable cleanliness of  binary
pulsars permits precise measurements of gravitational phenomena
in a partially strong-field context.

Unfortunately, nature is rarely so kind.
Still, under suitable conditions, qualitative and even quantitative
strong-field tests of general relativity could be carried out.

One example is the exploration of the spacetime near black holes
and neutron stars. 
Studies of certain kinds of accretion known as advection-dominated
accretion flow (ADAF) in low-luminosity binary X-ray sources
may yield the signature of the black hole event
horizon~\cite{2008NewAR..51..733N}.
The spectrum of frequencies of quasi-periodic oscillations (QPO) from galactic
black hole binaries may permit measurement of the spins of the black
holes~\cite{Psaltis04}.
Aspects of strong-field gravity and frame-dragging may be revealed in
spectral shapes of iron fluorescence lines from the inner regions of
accretion disks~\cite{2013SSRv..tmp...81R,2013CQGra..30x4004R}.  Using submm VLBI, a collaboration dubbed the Event Horizon Telescope could image our galactic center black hole SgrA* and the black hole in M87 with horizon-scale angular resolution; observation of accretion phenomena at these angular resolutions could provide tests of the spacetime geometry very close to the black hole~\cite{2009astro2010S..68D}.
Tracking of hypothetical stars whose orbits are within a fraction of a milliparsec of SgrA* could test the black hole ``ho-hair'' theorem, via a direct measurement of both the angular momentum $J$ and quadrupole moment $Q$ of the black hole, and a test of the requirement that $Q=-J^2/M$~\cite{2008ApJ...674L..25W}.

Because of uncertainties in the detailed models, the results to date of 
studies like these are
suggestive at best, but the combination of future higher-resolution
observations and better modelling could lead to striking tests of
strong-field predictions of general relativity.

For a detailed review of strong-field tests of general relativity using electromagnetic observations, see~\cite{2008LRR....11....9P}.


\section{Conclusions}
\label{S5}

General relativity has held up under extensive experimental scrutiny.
The question then arises, why bother to continue to test it?  One
reason is that gravity is a fundamental interaction of nature, and as
such requires the most solid empirical underpinning we can provide.
Another is that all attempts to quantize gravity and to unify it with
the other forces suggest that the standard general relativity of Einstein may not be the last word.
Furthermore, the predictions of general relativity are fixed;
the pure theory contains
no adjustable constants so nothing can be changed. Thus every test
of the theory is either a potentially deadly test or a possible probe for
new physics. Although it is remarkable
that this theory, born 100 years ago out of almost pure thought,
has managed to survive every test, the possibility of finding
a discrepancy will continue to drive experiments for years to come.
These experiments will search for new physics beyond Einstein at many different scales: the large distance scales of the astrophysical, galactic, and cosmological realms; scales of very short distances or high energy; and scales related to strong or dynamical gravity.

Having spent almost half of the century of general relativity's existence being astonished by its continuing agreement with observation, I might be permitted
a personal reflection at this point on the future of the subject: {\em It would not at all surprise me if general relativity turned out to be perfectly valid at all scales, from the cosmological to the astrophysical to the microscopic, failing only at the Planck scale where one naturally expects quantum gravity to take over.}

Of course, having made that statement, I am reminded of the story about Yogi Berra, the famous New York Yankees baseball player and manager (one of my childhood heroes), and one blessed with the most sublimely illogical mind.   At some point after his retirement from baseball, his wife Carmen asked him:  ``Yogi, you were born in St. Louis, you played ball in New York and we now live in New Jersey.  If you should die first, where would you like me to bury you?''  Yogi's answer:  ``Surprise me!''


\section*{Acknowledgments}

This work was supported 
in part by the National Science Foundation,
Grant Numbers PHY 12-60995 and 13-06069.  
I am grateful for the hospitality of the Institut d'Astrophysique de Paris,
where this article was prepared.

\bibliography{refs}

\begin{thebibliography}{115}
\expandafter\ifx\csname natexlab\endcsname\relax\def\natexlab#1{#1}\fi
\expandafter\ifx\csname bibnamefont\endcsname\relax
  \def\bibnamefont#1{#1}\fi
\expandafter\ifx\csname bibfnamefont\endcsname\relax
  \def\bibfnamefont#1{#1}\fi
\expandafter\ifx\csname citenamefont\endcsname\relax
  \def\citenamefont#1{#1}\fi
\expandafter\ifx\csname url\endcsname\relax
  \def\url#1{\texttt{#1}}\fi
\expandafter\ifx\csname urlprefix\endcsname\relax\def\urlprefix{URL }\fi
\providecommand{\bibinfo}[2]{#2}
\providecommand{\eprint}[2][]{\url{#2}}

\bibitem[{\citenamefont{Will}(1993{\natexlab{a}})}]{WER}
\bibinfo{author}{\bibfnamefont{C.}~\bibnamefont{Will}},
  \emph{\bibinfo{title}{Was Einstein Right?: Putting General Relativity to the
  Test}} (\bibinfo{publisher}{Basic Books}, \bibinfo{address}{New York},
  \bibinfo{year}{1993}{\natexlab{a}}), \bibinfo{edition}{2nd} ed.

\bibitem[{\citenamefont{Will}(1993{\natexlab{b}})}]{tegp}
\bibinfo{author}{\bibfnamefont{C.}~\bibnamefont{Will}},
  \emph{\bibinfo{title}{Theory and Experiment in Gravitational Physics}}
  (\bibinfo{publisher}{Cambridge University Press},
  \bibinfo{address}{Cambridge; New York}, \bibinfo{year}{1993}{\natexlab{b}}),
  \bibinfo{edition}{2nd} ed.

\bibitem[{\citenamefont{Will}(2006)}]{WillLivrev}
\bibinfo{author}{\bibfnamefont{C.}~\bibnamefont{Will}},
  \bibinfo{journal}{Living Rev. Relativity} \textbf{\bibinfo{volume}{9}}
  (\bibinfo{year}{2006}),
  \bibinfo{note}{http://www.living\-reviews.\-org/lrr-2006-3}.

\bibitem[{\citenamefont{Will}(2014)}]{2014arXiv1403.7377W}
\bibinfo{author}{\bibfnamefont{C.}~\bibnamefont{Will}},
  \bibinfo{journal}{Living Rev. Relativity} \textbf{\bibinfo{volume}{17}}
  (\bibinfo{year}{2014}),
  \bibinfo{note}{http://www.living\-reviews.\-org/lrr-2014-4}.

\bibitem[{\citenamefont{{Will}}(2010)}]{2010AmJPh..78.1240W}
\bibinfo{author}{\bibfnamefont{C.}~\bibnamefont{{Will}}}, \bibinfo{journal}{Am.
  J. Phys.} \textbf{\bibinfo{volume}{78}}, \bibinfo{pages}{1240}
  (\bibinfo{year}{2010}), \eprint{1008.0296}.

\bibitem[{\citenamefont{Dicke}(1964)}]{dicke64}
\bibinfo{author}{\bibfnamefont{R.}~\bibnamefont{Dicke}}, in
  \emph{\bibinfo{booktitle}{Relativity, Groups and Topology}}, edited by
  \bibinfo{editor}{\bibfnamefont{C.}~\bibnamefont{DeWitt}} \bibnamefont{and}
  \bibinfo{editor}{\bibfnamefont{B.}~\bibnamefont{DeWitt}}
  (\bibinfo{publisher}{Gordon and Breach}, \bibinfo{address}{New York; London},
  \bibinfo{year}{1964}), pp. \bibinfo{pages}{165--313}.

\bibitem[{\citenamefont{{Schlamminger}
  et~al.}(2008)\citenamefont{{Schlamminger}, {Choi}, {Wagner}, {Gundlach}, and
  {Adelberger}}}]{2008PhRvL.100d1101S}
\bibinfo{author}{\bibfnamefont{S.}~\bibnamefont{{Schlamminger}}},
  \bibinfo{author}{\bibfnamefont{K.-Y.} \bibnamefont{{Choi}}},
  \bibinfo{author}{\bibfnamefont{T.~A.} \bibnamefont{{Wagner}}},
  \bibinfo{author}{\bibfnamefont{J.~H.} \bibnamefont{{Gundlach}}},
  \bibnamefont{and} \bibinfo{author}{\bibfnamefont{E.~G.}
  \bibnamefont{{Adelberger}}}, \bibinfo{journal}{Phys. Rev. Lett.}
  \textbf{\bibinfo{volume}{100}}, \bibinfo{eid}{041101} (\bibinfo{year}{2008}),
  \eprint{0712.0607}.

\bibitem[{\citenamefont{{Merlet} et~al.}(2010)\citenamefont{{Merlet}, {Bodart},
  {Malossi}, {Landragin}, {Pereira Dos Santos}, {Gitlein}, and
  {Timmen}}}]{2010Metro..47L...9M}
\bibinfo{author}{\bibfnamefont{S.}~\bibnamefont{{Merlet}}},
  \bibinfo{author}{\bibfnamefont{Q.}~\bibnamefont{{Bodart}}},
  \bibinfo{author}{\bibfnamefont{N.}~\bibnamefont{{Malossi}}},
  \bibinfo{author}{\bibfnamefont{A.}~\bibnamefont{{Landragin}}},
  \bibinfo{author}{\bibfnamefont{F.}~\bibnamefont{{Pereira Dos Santos}}},
  \bibinfo{author}{\bibfnamefont{O.}~\bibnamefont{{Gitlein}}},
  \bibnamefont{and} \bibinfo{author}{\bibfnamefont{L.}~\bibnamefont{{Timmen}}},
  \bibinfo{journal}{Metrologia} \textbf{\bibinfo{volume}{47}},
  \bibinfo{pages}{L9} (\bibinfo{year}{2010}), \eprint{1005.0357}.

\bibitem[{\citenamefont{{M{\"u}ller} et~al.}(2010)\citenamefont{{M{\"u}ller},
  {Peters}, and {Chu}}}]{2010Natur.463..926M}
\bibinfo{author}{\bibfnamefont{H.}~\bibnamefont{{M{\"u}ller}}},
  \bibinfo{author}{\bibfnamefont{A.}~\bibnamefont{{Peters}}}, \bibnamefont{and}
  \bibinfo{author}{\bibfnamefont{S.}~\bibnamefont{{Chu}}},
  \bibinfo{journal}{Nature} \textbf{\bibinfo{volume}{463}},
  \bibinfo{pages}{926} (\bibinfo{year}{2010}).

\bibitem[{\citenamefont{{Wolf} et~al.}(2011)\citenamefont{{Wolf}, {Blanchet},
  {Bord{\'e}}, {Reynaud}, {Salomon}, and
  {Cohen-Tannoudji}}}]{2011CQGra..28n5017W}
\bibinfo{author}{\bibfnamefont{P.}~\bibnamefont{{Wolf}}},
  \bibinfo{author}{\bibfnamefont{L.}~\bibnamefont{{Blanchet}}},
  \bibinfo{author}{\bibfnamefont{C.~J.} \bibnamefont{{Bord{\'e}}}},
  \bibinfo{author}{\bibfnamefont{S.}~\bibnamefont{{Reynaud}}},
  \bibinfo{author}{\bibfnamefont{C.}~\bibnamefont{{Salomon}}},
  \bibnamefont{and}
  \bibinfo{author}{\bibfnamefont{C.}~\bibnamefont{{Cohen-Tannoudji}}},
  \bibinfo{journal}{Class. Quantum Grav.} \textbf{\bibinfo{volume}{28}},
  \bibinfo{eid}{145017} (\bibinfo{year}{2011}), \eprint{1012.1194}.

\bibitem[{\citenamefont{{Hogan} et~al.}(2009)\citenamefont{{Hogan}, {Johnson},
  and {Kasevich}}}]{2009aosp.conf..411H}
\bibinfo{author}{\bibfnamefont{J.~M.} \bibnamefont{{Hogan}}},
  \bibinfo{author}{\bibfnamefont{D.~M.~S.} \bibnamefont{{Johnson}}},
  \bibnamefont{and} \bibinfo{author}{\bibfnamefont{M.~A.}
  \bibnamefont{{Kasevich}}}, in \emph{\bibinfo{booktitle}{Atom Optics and Space
  Physics: Proceedings of the International School of Physics "Enrico Fermi",
  Course 168}}, edited by
  \bibinfo{editor}{\bibfnamefont{E.}~\bibnamefont{{Arimondo}}},
  \bibinfo{editor}{\bibfnamefont{W.}~\bibnamefont{{Ertmer}}},
  \bibinfo{editor}{\bibfnamefont{W.~P.} \bibnamefont{{Schleich}}},
  \bibnamefont{and} \bibinfo{editor}{\bibfnamefont{E.~M.}
  \bibnamefont{{Rasel}}} (\bibinfo{publisher}{IOS Press Amsterdam},
  \bibinfo{year}{2009}), p. \bibinfo{pages}{411}, \eprint{0806.3261}.

\bibitem[{\citenamefont{Fischbach et~al.}(1992)\citenamefont{Fischbach,
  Gillies, Krause, Schwan, and Talmadge}}]{fischbach92}
\bibinfo{author}{\bibfnamefont{E.}~\bibnamefont{Fischbach}},
  \bibinfo{author}{\bibfnamefont{G.}~\bibnamefont{Gillies}},
  \bibinfo{author}{\bibfnamefont{D.}~\bibnamefont{Krause}},
  \bibinfo{author}{\bibfnamefont{J.}~\bibnamefont{Schwan}}, \bibnamefont{and}
  \bibinfo{author}{\bibfnamefont{C.}~\bibnamefont{Talmadge}},
  \bibinfo{journal}{Metrologia} \textbf{\bibinfo{volume}{29}},
  \bibinfo{pages}{213} (\bibinfo{year}{1992}).

\bibitem[{\citenamefont{Speake and Will}(2012)}]{0264-9381-29-18-180301}
\bibinfo{author}{\bibfnamefont{C.}~\bibnamefont{Speake}} \bibnamefont{and}
  \bibinfo{author}{\bibfnamefont{C.}~\bibnamefont{Will}},
  \bibinfo{journal}{Class. Quantum Grav.} \textbf{\bibinfo{volume}{29}},
  \bibinfo{pages}{180301} (\bibinfo{year}{2012}),
  \urlprefix\url{http://stacks.iop.org/0264-9381/29/i=18/a=180301}.

\bibitem[{\citenamefont{{Adelberger} et~al.}(2009)\citenamefont{{Adelberger},
  {Gundlach}, {Heckel}, {Hoedl}, and {Schlamminger}}}]{2009PrPNP..62..102A}
\bibinfo{author}{\bibfnamefont{E.~G.} \bibnamefont{{Adelberger}}},
  \bibinfo{author}{\bibfnamefont{J.~H.} \bibnamefont{{Gundlach}}},
  \bibinfo{author}{\bibfnamefont{B.~R.} \bibnamefont{{Heckel}}},
  \bibinfo{author}{\bibfnamefont{S.}~\bibnamefont{{Hoedl}}}, \bibnamefont{and}
  \bibinfo{author}{\bibfnamefont{S.}~\bibnamefont{{Schlamminger}}},
  \bibinfo{journal}{Prog. Part. Nucl. Phys.} \textbf{\bibinfo{volume}{62}},
  \bibinfo{pages}{102} (\bibinfo{year}{2009}).

\bibitem[{\citenamefont{{Will}}(2006)}]{2006eins.book...33W}
\bibinfo{author}{\bibfnamefont{C.}~\bibnamefont{{Will}}}, in
  \emph{\bibinfo{booktitle}{Einstein, 1905-2005: Poincar{\'e} Seminar 2005}},
  edited by \bibinfo{editor}{\bibfnamefont{T.}~\bibnamefont{{Damour}}},
  \bibinfo{editor}{\bibfnamefont{O.}~\bibnamefont{{Darrigol}}},
  \bibinfo{editor}{\bibfnamefont{B.}~\bibnamefont{{Duplantier}}},
  \bibnamefont{and}
  \bibinfo{editor}{\bibfnamefont{V.}~\bibnamefont{{Rivasseau}}}
  (\bibinfo{publisher}{Birk\"auser Verlag}, \bibinfo{address}{Basel},
  \bibinfo{year}{2006}), p.~\bibinfo{pages}{33}.

\bibitem[{\citenamefont{Mattingly}(2005)}]{mattingly}
\bibinfo{author}{\bibfnamefont{D.}~\bibnamefont{Mattingly}},
  \bibinfo{journal}{Living Rev. Relativity} \textbf{\bibinfo{volume}{8}},
  \bibinfo{eid}{lrr-2005-5} (\bibinfo{year}{2005}),
  \bibinfo{note}{http://www.living\-reviews.\-org/lrr-2005-5},
  \eprint{gr-qc/0502097}.

\bibitem[{\citenamefont{Liberati}(2013)}]{Liberati2013}
\bibinfo{author}{\bibfnamefont{S.}~\bibnamefont{Liberati}},
  \bibinfo{journal}{Class. Quantum Grav.} \textbf{\bibinfo{volume}{30}},
  \bibinfo{pages}{133001} (\bibinfo{year}{2013}),
  \urlprefix\url{http://stacks.iop.org/0264-9381/30/i=13/a=133001}.

\bibitem[{\citenamefont{Kosteleck\'y and Russell}(2011)}]{RevModPhys.83.11}
\bibinfo{author}{\bibfnamefont{V.}~\bibnamefont{Kosteleck\'y}}
  \bibnamefont{and} \bibinfo{author}{\bibfnamefont{N.}~\bibnamefont{Russell}},
  \bibinfo{journal}{Rev. Mod. Phys.} \textbf{\bibinfo{volume}{83}},
  \bibinfo{pages}{11} (\bibinfo{year}{2011}),
  \urlprefix\url{http://link.aps.org/doi/10.1103/RevModPhys.83.11}.

\bibitem[{\citenamefont{Kosteleck\'y and Tasson}(2011)}]{PhysRevD.83.016013}
\bibinfo{author}{\bibfnamefont{V.}~\bibnamefont{Kosteleck\'y}}
  \bibnamefont{and} \bibinfo{author}{\bibfnamefont{J.}~\bibnamefont{Tasson}},
  \bibinfo{journal}{Phys. Rev. D} \textbf{\bibinfo{volume}{83}},
  \bibinfo{pages}{016013} (\bibinfo{year}{2011}),
  \urlprefix\url{http://link.aps.org/doi/10.1103/PhysRevD.83.016013}.

\bibitem[{\citenamefont{LoPresto et~al.}(1991)\citenamefont{LoPresto, Schrader,
  and Pierce}}]{lopresto91}
\bibinfo{author}{\bibfnamefont{J.}~\bibnamefont{LoPresto}},
  \bibinfo{author}{\bibfnamefont{C.}~\bibnamefont{Schrader}}, \bibnamefont{and}
  \bibinfo{author}{\bibfnamefont{A.}~\bibnamefont{Pierce}},
  \bibinfo{journal}{Astrophys. J.} \textbf{\bibinfo{volume}{376}},
  \bibinfo{pages}{757} (\bibinfo{year}{1991}).

\bibitem[{\citenamefont{Gu{\'e}na et~al.}(2012)\citenamefont{Gu{\'e}na,
  Abgrall, Rovera, Rosenbusch, Tobar, Laurent, Clairon, and
  Bize}}]{2012PhRvL.109h0801G}
\bibinfo{author}{\bibfnamefont{J.}~\bibnamefont{Gu{\'e}na}},
  \bibinfo{author}{\bibfnamefont{M.}~\bibnamefont{Abgrall}},
  \bibinfo{author}{\bibfnamefont{D.}~\bibnamefont{Rovera}},
  \bibinfo{author}{\bibfnamefont{P.}~\bibnamefont{Rosenbusch}},
  \bibinfo{author}{\bibfnamefont{M.~E.} \bibnamefont{Tobar}},
  \bibinfo{author}{\bibfnamefont{P.}~\bibnamefont{Laurent}},
  \bibinfo{author}{\bibfnamefont{A.}~\bibnamefont{Clairon}}, \bibnamefont{and}
  \bibinfo{author}{\bibfnamefont{S.}~\bibnamefont{Bize}},
  \bibinfo{journal}{Phys. Rev. Lett.} \textbf{\bibinfo{volume}{109}},
  \bibinfo{eid}{080801} (\bibinfo{year}{2012}), \eprint{1205.4235}.

\bibitem[{\citenamefont{{Peil} et~al.}(2013)\citenamefont{{Peil}, {Crane},
  {Hanssen}, {Swanson}, and {Ekstrom}}}]{2013PhRvA..87a0102P}
\bibinfo{author}{\bibfnamefont{S.}~\bibnamefont{{Peil}}},
  \bibinfo{author}{\bibfnamefont{S.}~\bibnamefont{{Crane}}},
  \bibinfo{author}{\bibfnamefont{J.~L.} \bibnamefont{{Hanssen}}},
  \bibinfo{author}{\bibfnamefont{T.~B.} \bibnamefont{{Swanson}}},
  \bibnamefont{and} \bibinfo{author}{\bibfnamefont{C.~R.}
  \bibnamefont{{Ekstrom}}}, \bibinfo{journal}{\pra}
  \textbf{\bibinfo{volume}{87}}, \bibinfo{eid}{010102} (\bibinfo{year}{2013}),
  \eprint{1301.6145}.

\bibitem[{\citenamefont{Leefer et~al.}(2013)\citenamefont{Leefer, Weber,
  Cing{\"o}z, Torgerson, and Budker}}]{2013PhRvL.111f0801L}
\bibinfo{author}{\bibfnamefont{N.}~\bibnamefont{Leefer}},
  \bibinfo{author}{\bibfnamefont{C.~T.~M.} \bibnamefont{Weber}},
  \bibinfo{author}{\bibfnamefont{A.}~\bibnamefont{Cing{\"o}z}},
  \bibinfo{author}{\bibfnamefont{J.~R.} \bibnamefont{Torgerson}},
  \bibnamefont{and} \bibinfo{author}{\bibfnamefont{D.}~\bibnamefont{Budker}},
  \bibinfo{journal}{Phys. Rev. Lett.} \textbf{\bibinfo{volume}{111}},
  \bibinfo{eid}{060801} (\bibinfo{year}{2013}), \eprint{1304.6940}.

\bibitem[{\citenamefont{Ashby}(2003)}]{ashby2}
\bibinfo{author}{\bibfnamefont{N.}~\bibnamefont{Ashby}},
  \bibinfo{journal}{Living Rev. Relativity} \textbf{\bibinfo{volume}{6}},
  \bibinfo{eid}{lrr-2003-1} (\bibinfo{year}{2003}),
  \bibinfo{note}{http://www.living\-reviews.\-org/lrr-2003-1}.

\bibitem[{\citenamefont{Will}(2000)}]{physicscentral}
\bibinfo{author}{\bibfnamefont{C.}~\bibnamefont{Will}},
  \emph{\bibinfo{title}{Einstein's relativity and everyday life}}
  (\bibinfo{year}{2000}),
  \bibinfo{note}{http://www.\-physics\-central.\-com/writers/writers-00-2.html}.

\bibitem[{\citenamefont{Dyson}(1972)}]{dyson72}
\bibinfo{author}{\bibfnamefont{F.}~\bibnamefont{Dyson}}, in
  \emph{\bibinfo{booktitle}{Aspects of Quantum Theory}}, edited by
  \bibinfo{editor}{\bibfnamefont{A.}~\bibnamefont{Salam}} \bibnamefont{and}
  \bibinfo{editor}{\bibfnamefont{E.}~\bibnamefont{Wigner}}
  (\bibinfo{publisher}{Cambridge University Press},
  \bibinfo{address}{Cambridge; New York}, \bibinfo{year}{1972}), pp.
  \bibinfo{pages}{213--236}.

\bibitem[{\citenamefont{{Uzan}}(2011)}]{2011LRR....14....2U}
\bibinfo{author}{\bibfnamefont{J.-P.} \bibnamefont{{Uzan}}},
  \bibinfo{journal}{Living Rev. Relativity} \textbf{\bibinfo{volume}{14}}
  (\bibinfo{year}{2011}),
  \bibinfo{note}{http://www.living\-reviews.\-org/lrr-2011-2},
  \eprint{1009.5514}.

\bibitem[{\citenamefont{{King} et~al.}(2012)\citenamefont{{King}, {Webb},
  {Murphy}, {Flambaum}, {Carswell}, {Bainbridge}, {Wilczynska}, and
  {Koch}}}]{2012MNRAS.422.3370K}
\bibinfo{author}{\bibfnamefont{J.~A.} \bibnamefont{{King}}},
  \bibinfo{author}{\bibfnamefont{J.~K.} \bibnamefont{{Webb}}},
  \bibinfo{author}{\bibfnamefont{M.~T.} \bibnamefont{{Murphy}}},
  \bibinfo{author}{\bibfnamefont{V.~V.} \bibnamefont{{Flambaum}}},
  \bibinfo{author}{\bibfnamefont{R.~F.} \bibnamefont{{Carswell}}},
  \bibinfo{author}{\bibfnamefont{M.~B.} \bibnamefont{{Bainbridge}}},
  \bibinfo{author}{\bibfnamefont{M.~R.} \bibnamefont{{Wilczynska}}},
  \bibnamefont{and} \bibinfo{author}{\bibfnamefont{F.~E.}
  \bibnamefont{{Koch}}}, \bibinfo{journal}{\mnras}
  \textbf{\bibinfo{volume}{422}}, \bibinfo{pages}{3370} (\bibinfo{year}{2012}),
  \eprint{1202.4758}.

\bibitem[{\citenamefont{{Kanekar} et~al.}(2012)\citenamefont{{Kanekar},
  {Langston}, {Stocke}, {Carilli}, and {Menten}}}]{2012ApJ...746L..16K}
\bibinfo{author}{\bibfnamefont{N.}~\bibnamefont{{Kanekar}}},
  \bibinfo{author}{\bibfnamefont{G.~I.} \bibnamefont{{Langston}}},
  \bibinfo{author}{\bibfnamefont{J.~T.} \bibnamefont{{Stocke}}},
  \bibinfo{author}{\bibfnamefont{C.~L.} \bibnamefont{{Carilli}}},
  \bibnamefont{and} \bibinfo{author}{\bibfnamefont{K.~M.}
  \bibnamefont{{Menten}}}, \bibinfo{journal}{\apjl}
  \textbf{\bibinfo{volume}{746}}, \bibinfo{eid}{L16} (\bibinfo{year}{2012}),
  \eprint{1201.3372}.

\bibitem[{\citenamefont{{Lentati} et~al.}(2013)\citenamefont{{Lentati},
  {Carilli}, {Alexander}, {Maiolino}, {Wang}, {Cox}, {Downes}, {McMahon},
  {Menten}, {Neri} et~al.}}]{2013MNRAS.430.2454L}
\bibinfo{author}{\bibfnamefont{L.}~\bibnamefont{{Lentati}}},
  \bibinfo{author}{\bibfnamefont{C.}~\bibnamefont{{Carilli}}},
  \bibinfo{author}{\bibfnamefont{P.}~\bibnamefont{{Alexander}}},
  \bibinfo{author}{\bibfnamefont{R.}~\bibnamefont{{Maiolino}}},
  \bibinfo{author}{\bibfnamefont{R.}~\bibnamefont{{Wang}}},
  \bibinfo{author}{\bibfnamefont{P.}~\bibnamefont{{Cox}}},
  \bibinfo{author}{\bibfnamefont{D.}~\bibnamefont{{Downes}}},
  \bibinfo{author}{\bibfnamefont{R.}~\bibnamefont{{McMahon}}},
  \bibinfo{author}{\bibfnamefont{K.~M.} \bibnamefont{{Menten}}},
  \bibinfo{author}{\bibfnamefont{R.}~\bibnamefont{{Neri}}},
  \bibnamefont{et~al.}, \bibinfo{journal}{\mnras}
  \textbf{\bibinfo{volume}{430}}, \bibinfo{pages}{2454} (\bibinfo{year}{2013}),
  \eprint{1211.3316}.

\bibitem[{\citenamefont{{Turyshev} and {Toth}}(2010)}]{2010LRR....13....4T}
\bibinfo{author}{\bibfnamefont{S.~G.} \bibnamefont{{Turyshev}}}
  \bibnamefont{and} \bibinfo{author}{\bibfnamefont{V.~T.}
  \bibnamefont{{Toth}}}, \bibinfo{journal}{Living Rev. Relativity}
  \textbf{\bibinfo{volume}{13}}, \bibinfo{eid}{lrr-2010-4}
  (\bibinfo{year}{2010}),
  \bibinfo{note}{http://www.living\-reviews.\-org/lrr-2010-4},
  \eprint{1001.3686}.

\bibitem[{\citenamefont{{Turyshev} et~al.}(2012)\citenamefont{{Turyshev},
  {Toth}, {Kinsella}, {Lee}, {Lok}, and {Ellis}}}]{2012PhRvL.108x1101T}
\bibinfo{author}{\bibfnamefont{S.~G.} \bibnamefont{{Turyshev}}},
  \bibinfo{author}{\bibfnamefont{V.~T.} \bibnamefont{{Toth}}},
  \bibinfo{author}{\bibfnamefont{G.}~\bibnamefont{{Kinsella}}},
  \bibinfo{author}{\bibfnamefont{S.-C.} \bibnamefont{{Lee}}},
  \bibinfo{author}{\bibfnamefont{S.~M.} \bibnamefont{{Lok}}}, \bibnamefont{and}
  \bibinfo{author}{\bibfnamefont{J.}~\bibnamefont{{Ellis}}},
  \bibinfo{journal}{Phys. Rev. Lett.} \textbf{\bibinfo{volume}{108}},
  \bibinfo{eid}{241101} (\bibinfo{year}{2012}), \eprint{1204.2507}.

\bibitem[{\citenamefont{{Schutz}}(2009)}]{2009fcgr.book.....S}
\bibinfo{author}{\bibfnamefont{B.}~\bibnamefont{{Schutz}}},
  \emph{\bibinfo{title}{A First Course in General Relativity}}
  (\bibinfo{publisher}{Cambridge University Press},
  \bibinfo{address}{Cambridge}, \bibinfo{year}{2009}).

\bibitem[{\citenamefont{Poisson and Will}(2014)}]{PW2014}
\bibinfo{author}{\bibfnamefont{E.}~\bibnamefont{Poisson}} \bibnamefont{and}
  \bibinfo{author}{\bibfnamefont{C.}~\bibnamefont{Will}},
  \emph{\bibinfo{title}{Gravity: Newtonian, Post-Newtonian, Relativistic}}
  (\bibinfo{publisher}{Cambridge University Press},
  \bibinfo{address}{Cambridge}, \bibinfo{year}{2014}).

\bibitem[{\citenamefont{Nordtvedt~Jr.}(1968{\natexlab{a}})}]{nordtvedt2}
\bibinfo{author}{\bibfnamefont{K.}~\bibnamefont{Nordtvedt~Jr.}},
  \bibinfo{journal}{Phys. Rev.} \textbf{\bibinfo{volume}{169}},
  \bibinfo{pages}{1017} (\bibinfo{year}{1968}{\natexlab{a}}).

\bibitem[{\citenamefont{Will}(1971)}]{Will71a}
\bibinfo{author}{\bibfnamefont{C.}~\bibnamefont{Will}},
  \bibinfo{journal}{Astrophys. J.} \textbf{\bibinfo{volume}{163}},
  \bibinfo{pages}{611} (\bibinfo{year}{1971}).

\bibitem[{\citenamefont{Will and Nordtvedt~Jr.}(1972)}]{willnordtvedt72}
\bibinfo{author}{\bibfnamefont{C.}~\bibnamefont{Will}} \bibnamefont{and}
  \bibinfo{author}{\bibfnamefont{K.}~\bibnamefont{Nordtvedt~Jr.}},
  \bibinfo{journal}{Astrophys. J.} \textbf{\bibinfo{volume}{177}},
  \bibinfo{pages}{757} (\bibinfo{year}{1972}).

\bibitem[{\citenamefont{Misner et~al.}(1973)\citenamefont{Misner, Thorne, and
  Wheeler}}]{MTW}
\bibinfo{author}{\bibfnamefont{C.}~\bibnamefont{Misner}},
  \bibinfo{author}{\bibfnamefont{K.}~\bibnamefont{Thorne}}, \bibnamefont{and}
  \bibinfo{author}{\bibfnamefont{J.}~\bibnamefont{Wheeler}},
  \emph{\bibinfo{title}{Gravitation}} (\bibinfo{publisher}{W.H. Freeman},
  \bibinfo{address}{San Francisco}, \bibinfo{year}{1973}).

\bibitem[{\citenamefont{Damour and Esposito-Far{\`{e}}se}(1992)}]{DamourEspo92}
\bibinfo{author}{\bibfnamefont{T.}~\bibnamefont{Damour}} \bibnamefont{and}
  \bibinfo{author}{\bibfnamefont{G.}~\bibnamefont{Esposito-Far{\`{e}}se}},
  \bibinfo{journal}{Class. Quantum Grav.} \textbf{\bibinfo{volume}{9}},
  \bibinfo{pages}{2093} (\bibinfo{year}{1992}).

\bibitem[{\citenamefont{Damour and
  Nordtvedt~Jr.}(1993{\natexlab{a}})}]{DamourNord93a}
\bibinfo{author}{\bibfnamefont{T.}~\bibnamefont{Damour}} \bibnamefont{and}
  \bibinfo{author}{\bibfnamefont{K.}~\bibnamefont{Nordtvedt~Jr.}},
  \bibinfo{journal}{Phys. Rev. Lett.} \textbf{\bibinfo{volume}{70}},
  \bibinfo{pages}{2217} (\bibinfo{year}{1993}{\natexlab{a}}).

\bibitem[{\citenamefont{Damour and
  Nordtvedt~Jr.}(1993{\natexlab{b}})}]{DamourNord93b}
\bibinfo{author}{\bibfnamefont{T.}~\bibnamefont{Damour}} \bibnamefont{and}
  \bibinfo{author}{\bibfnamefont{K.}~\bibnamefont{Nordtvedt~Jr.}},
  \bibinfo{journal}{Phys. Rev. D} \textbf{\bibinfo{volume}{48}},
  \bibinfo{pages}{3436} (\bibinfo{year}{1993}{\natexlab{b}}).

\bibitem[{\citenamefont{{Fujii} and {Maeda}}(2007)}]{2007sttg.book.....F}
\bibinfo{author}{\bibfnamefont{Y.}~\bibnamefont{{Fujii}}} \bibnamefont{and}
  \bibinfo{author}{\bibfnamefont{K.-I.} \bibnamefont{{Maeda}}},
  \emph{\bibinfo{title}{{The Scalar-Tensor Theory of Gravitation}}}
  (\bibinfo{publisher}{Cambridge University Press},
  \bibinfo{address}{Cambridge}, \bibinfo{year}{2007}).

\bibitem[{\citenamefont{{Sotiriou} and {Faraoni}}(2010)}]{2010RvMP...82..451S}
\bibinfo{author}{\bibfnamefont{T.~P.} \bibnamefont{{Sotiriou}}}
  \bibnamefont{and}
  \bibinfo{author}{\bibfnamefont{V.}~\bibnamefont{{Faraoni}}},
  \bibinfo{journal}{Rev. Mod. Phys.} \textbf{\bibinfo{volume}{82}},
  \bibinfo{pages}{451} (\bibinfo{year}{2010}), \eprint{0805.1726}.

\bibitem[{\citenamefont{De~Felice and Tsujikawa}(2010)}]{lrr-2010-3}
\bibinfo{author}{\bibfnamefont{A.}~\bibnamefont{De~Felice}} \bibnamefont{and}
  \bibinfo{author}{\bibfnamefont{S.}~\bibnamefont{Tsujikawa}},
  \bibinfo{journal}{Living Rev. Relativity} \textbf{\bibinfo{volume}{13}}
  (\bibinfo{year}{2010}),
  \bibinfo{note}{http://www.living\-reviews.\-org/lrr-2010-3},
  \urlprefix\url{http://www.living\-reviews.\-org/lrr-2010-3}.

\bibitem[{\citenamefont{Hellings and Nordtvedt~Jr.}(1973)}]{hellings73}
\bibinfo{author}{\bibfnamefont{R.}~\bibnamefont{Hellings}} \bibnamefont{and}
  \bibinfo{author}{\bibfnamefont{K.}~\bibnamefont{Nordtvedt~Jr.}},
  \bibinfo{journal}{Phys. Rev. D} \textbf{\bibinfo{volume}{7}},
  \bibinfo{pages}{3593} (\bibinfo{year}{1973}).

\bibitem[{\citenamefont{Jacobson and Mattingly}(2001)}]{jacobson01}
\bibinfo{author}{\bibfnamefont{T.}~\bibnamefont{Jacobson}} \bibnamefont{and}
  \bibinfo{author}{\bibfnamefont{D.}~\bibnamefont{Mattingly}},
  \bibinfo{journal}{Phys. Rev. D} \textbf{\bibinfo{volume}{64}},
  \bibinfo{eid}{024028} (\bibinfo{year}{2001}), \eprint{gr-qc/0007031}.

\bibitem[{\citenamefont{Mattingly and Jacobson}(2002)}]{mattingly02}
\bibinfo{author}{\bibfnamefont{D.}~\bibnamefont{Mattingly}} \bibnamefont{and}
  \bibinfo{author}{\bibfnamefont{T.}~\bibnamefont{Jacobson}}, in
  \emph{\bibinfo{booktitle}{CPT and Lorentz Symmetry II}}, edited by
  \bibinfo{editor}{\bibfnamefont{V.}~\bibnamefont{Kosteleck{\'{y}}}}
  (\bibinfo{publisher}{World Scientific}, \bibinfo{address}{Singapore; River
  Edge}, \bibinfo{year}{2002}), pp. \bibinfo{pages}{331--335},
  \eprint{gr-qc/0112012}.

\bibitem[{\citenamefont{Jacobson and Mattingly}(2004)}]{jacobson04}
\bibinfo{author}{\bibfnamefont{T.}~\bibnamefont{Jacobson}} \bibnamefont{and}
  \bibinfo{author}{\bibfnamefont{D.}~\bibnamefont{Mattingly}},
  \bibinfo{journal}{Phys. Rev. D} \textbf{\bibinfo{volume}{70}},
  \bibinfo{eid}{024003} (\bibinfo{year}{2004}), \eprint{gr-qc/0402005}.

\bibitem[{\citenamefont{Eling and Jacobson}(2004)}]{eling04}
\bibinfo{author}{\bibfnamefont{C.}~\bibnamefont{Eling}} \bibnamefont{and}
  \bibinfo{author}{\bibfnamefont{T.}~\bibnamefont{Jacobson}},
  \bibinfo{journal}{Phys. Rev. D} \textbf{\bibinfo{volume}{69}},
  \bibinfo{eid}{064005} (\bibinfo{year}{2004}), \eprint{gr-qc/0310044}.

\bibitem[{\citenamefont{Foster and Jacobson}(2006)}]{foster05}
\bibinfo{author}{\bibfnamefont{B.}~\bibnamefont{Foster}} \bibnamefont{and}
  \bibinfo{author}{\bibfnamefont{T.}~\bibnamefont{Jacobson}},
  \bibinfo{journal}{Phys. Rev. D} \textbf{\bibinfo{volume}{73}},
  \bibinfo{eid}{064015} (\bibinfo{year}{2006}), \eprint{gr-qc/0509083}.

\bibitem[{\citenamefont{{Ho{\v r}ava}}(2009)}]{2009PhRvD..79h4008H}
\bibinfo{author}{\bibfnamefont{P.}~\bibnamefont{{Ho{\v r}ava}}},
  \bibinfo{journal}{\prd} \textbf{\bibinfo{volume}{79}}, \bibinfo{eid}{084008}
  (\bibinfo{year}{2009}), \eprint{0901.3775}.

\bibitem[{\citenamefont{{Blas} et~al.}(2010)\citenamefont{{Blas},
  {Pujol{\`a}s}, and {Sibiryakov}}}]{2010PhRvL.104r1302B}
\bibinfo{author}{\bibfnamefont{D.}~\bibnamefont{{Blas}}},
  \bibinfo{author}{\bibfnamefont{O.}~\bibnamefont{{Pujol{\`a}s}}},
  \bibnamefont{and}
  \bibinfo{author}{\bibfnamefont{S.}~\bibnamefont{{Sibiryakov}}},
  \bibinfo{journal}{Phys. Rev. Lett.} \textbf{\bibinfo{volume}{104}},
  \bibinfo{eid}{181302} (\bibinfo{year}{2010}), \eprint{0909.3525}.

\bibitem[{\citenamefont{{Blas} et~al.}(2011)\citenamefont{{Blas},
  {Pujol{\`a}s}, and {Sibiryakov}}}]{2011JHEP...04..018B}
\bibinfo{author}{\bibfnamefont{D.}~\bibnamefont{{Blas}}},
  \bibinfo{author}{\bibfnamefont{O.}~\bibnamefont{{Pujol{\`a}s}}},
  \bibnamefont{and}
  \bibinfo{author}{\bibfnamefont{S.}~\bibnamefont{{Sibiryakov}}},
  \bibinfo{journal}{J. High Energy Phys.} \textbf{\bibinfo{volume}{4}},
  \bibinfo{eid}{18} (\bibinfo{year}{2011}), \eprint{1007.3503}.

\bibitem[{\citenamefont{{Jacobson}}(2014)}]{2014PhRvD..89h1501J}
\bibinfo{author}{\bibfnamefont{T.}~\bibnamefont{{Jacobson}}},
  \bibinfo{journal}{\prd} \textbf{\bibinfo{volume}{89}}, \bibinfo{eid}{081501}
  (\bibinfo{year}{2014}), \eprint{1310.5115}.

\bibitem[{\citenamefont{Bekenstein}(2004)}]{PhysRevD.70.083509}
\bibinfo{author}{\bibfnamefont{J.~D.} \bibnamefont{Bekenstein}},
  \bibinfo{journal}{Phys. Rev. D} \textbf{\bibinfo{volume}{70}},
  \bibinfo{pages}{083509} (\bibinfo{year}{2004}),
  \urlprefix\url{http://link.aps.org/doi/10.1103/PhysRevD.70.083509}.

\bibitem[{\citenamefont{Skordis}(2008)}]{PhysRevD.77.123502}
\bibinfo{author}{\bibfnamefont{C.}~\bibnamefont{Skordis}},
  \bibinfo{journal}{Phys. Rev. D} \textbf{\bibinfo{volume}{77}},
  \bibinfo{pages}{123502} (\bibinfo{year}{2008}),
  \urlprefix\url{http://link.aps.org/doi/10.1103/PhysRevD.77.123502}.

\bibitem[{\citenamefont{Sagi}(2009)}]{PhysRevD.80.044032}
\bibinfo{author}{\bibfnamefont{E.}~\bibnamefont{Sagi}}, \bibinfo{journal}{Phys.
  Rev. D} \textbf{\bibinfo{volume}{80}}, \bibinfo{pages}{044032}
  (\bibinfo{year}{2009}),
  \urlprefix\url{http://link.aps.org/doi/10.1103/PhysRevD.80.044032}.

\bibitem[{\citenamefont{{Skordis}}(2009)}]{2009CQGra..26n3001S}
\bibinfo{author}{\bibfnamefont{C.}~\bibnamefont{{Skordis}}},
  \bibinfo{journal}{Class. Quantum Grav.} \textbf{\bibinfo{volume}{26}},
  \bibinfo{eid}{143001} (\bibinfo{year}{2009}), \eprint{0903.3602}.

\bibitem[{\citenamefont{{Famaey} and {McGaugh}}(2012)}]{2012LRR....15...10F}
\bibinfo{author}{\bibfnamefont{B.}~\bibnamefont{{Famaey}}} \bibnamefont{and}
  \bibinfo{author}{\bibfnamefont{S.~S.} \bibnamefont{{McGaugh}}},
  \bibinfo{journal}{Living Rev. Relativity} \textbf{\bibinfo{volume}{15}}
  (\bibinfo{year}{2012}),
  \bibinfo{note}{http://www.living\-reviews.\-org/lrr-2012-10},
  \eprint{1112.3960}.

\bibitem[{\citenamefont{{Hinterbichler}}(2012)}]{2012RvMP...84..671H}
\bibinfo{author}{\bibfnamefont{K.}~\bibnamefont{{Hinterbichler}}},
  \bibinfo{journal}{Rev. Mod. Phys.} \textbf{\bibinfo{volume}{84}},
  \bibinfo{pages}{671} (\bibinfo{year}{2012}), \eprint{1105.3735}.

\bibitem[{\citenamefont{Will}(1988)}]{willcavendish}
\bibinfo{author}{\bibfnamefont{C.}~\bibnamefont{Will}}, \bibinfo{journal}{Am.
  J. Phys.} \textbf{\bibinfo{volume}{56}}, \bibinfo{pages}{413}
  (\bibinfo{year}{1988}).

\bibitem[{\citenamefont{Shapiro et~al.}(2004)\citenamefont{Shapiro, Davis,
  Lebach, and Gregory}}]{sshapiro04}
\bibinfo{author}{\bibfnamefont{S.}~\bibnamefont{Shapiro}},
  \bibinfo{author}{\bibfnamefont{J.}~\bibnamefont{Davis}},
  \bibinfo{author}{\bibfnamefont{D.}~\bibnamefont{Lebach}}, \bibnamefont{and}
  \bibinfo{author}{\bibfnamefont{J.}~\bibnamefont{Gregory}},
  \bibinfo{journal}{Phys. Rev. Lett.} \textbf{\bibinfo{volume}{92}},
  \bibinfo{eid}{121101} (\bibinfo{year}{2004}).

\bibitem[{\citenamefont{{Lambert} and {Le
  Poncin-Lafitte}}(2011)}]{2011A&A...529A..70L}
\bibinfo{author}{\bibfnamefont{S.~B.} \bibnamefont{{Lambert}}}
  \bibnamefont{and} \bibinfo{author}{\bibfnamefont{C.}~\bibnamefont{{Le
  Poncin-Lafitte}}}, \bibinfo{journal}{\aap} \textbf{\bibinfo{volume}{529}},
  \bibinfo{eid}{A70} (\bibinfo{year}{2011}).

\bibitem[{\citenamefont{{Bolton} et~al.}(2006)\citenamefont{{Bolton},
  {Rappaport}, and {Burles}}}]{2006PhRvD..74f1501B}
\bibinfo{author}{\bibfnamefont{A.~S.} \bibnamefont{{Bolton}}},
  \bibinfo{author}{\bibfnamefont{S.}~\bibnamefont{{Rappaport}}},
  \bibnamefont{and} \bibinfo{author}{\bibfnamefont{S.}~\bibnamefont{{Burles}}},
  \bibinfo{journal}{\prd} \textbf{\bibinfo{volume}{74}}, \bibinfo{eid}{061501}
  (\bibinfo{year}{2006}), \eprint{arXiv:astro-ph/0607657}.

\bibitem[{\citenamefont{Bertotti et~al.}(2003)\citenamefont{Bertotti, Iess, and
  Tortora}}]{bertotti03}
\bibinfo{author}{\bibfnamefont{B.}~\bibnamefont{Bertotti}},
  \bibinfo{author}{\bibfnamefont{L.}~\bibnamefont{Iess}}, \bibnamefont{and}
  \bibinfo{author}{\bibfnamefont{P.}~\bibnamefont{Tortora}},
  \bibinfo{journal}{Nature} \textbf{\bibinfo{volume}{425}},
  \bibinfo{pages}{374} (\bibinfo{year}{2003}).

\bibitem[{\citenamefont{{Fienga} et~al.}(2011)\citenamefont{{Fienga}, {Laskar},
  {Kuchynka}, {Manche}, {Desvignes}, {Gastineau}, {Cognard}, and
  {Theureau}}}]{2011CeMDA.111..363F}
\bibinfo{author}{\bibfnamefont{A.}~\bibnamefont{{Fienga}}},
  \bibinfo{author}{\bibfnamefont{J.}~\bibnamefont{{Laskar}}},
  \bibinfo{author}{\bibfnamefont{P.}~\bibnamefont{{Kuchynka}}},
  \bibinfo{author}{\bibfnamefont{H.}~\bibnamefont{{Manche}}},
  \bibinfo{author}{\bibfnamefont{G.}~\bibnamefont{{Desvignes}}},
  \bibinfo{author}{\bibfnamefont{M.}~\bibnamefont{{Gastineau}}},
  \bibinfo{author}{\bibfnamefont{I.}~\bibnamefont{{Cognard}}},
  \bibnamefont{and}
  \bibinfo{author}{\bibfnamefont{G.}~\bibnamefont{{Theureau}}},
  \bibinfo{journal}{Cel. Mech. Dyn. Astron.} \textbf{\bibinfo{volume}{111}},
  \bibinfo{pages}{363} (\bibinfo{year}{2011}), \eprint{1108.5546}.

\bibitem[{\citenamefont{{Verma} et~al.}(2014)\citenamefont{{Verma}, {Fienga},
  {Laskar}, {Manche}, and {Gastineau}}}]{2014A&A...561A.115V}
\bibinfo{author}{\bibfnamefont{A.~K.} \bibnamefont{{Verma}}},
  \bibinfo{author}{\bibfnamefont{A.}~\bibnamefont{{Fienga}}},
  \bibinfo{author}{\bibfnamefont{J.}~\bibnamefont{{Laskar}}},
  \bibinfo{author}{\bibfnamefont{H.}~\bibnamefont{{Manche}}}, \bibnamefont{and}
  \bibinfo{author}{\bibfnamefont{M.}~\bibnamefont{{Gastineau}}},
  \bibinfo{journal}{\aap} \textbf{\bibinfo{volume}{561}}, \bibinfo{eid}{A115}
  (\bibinfo{year}{2014}), \eprint{1306.5569}.

\bibitem[{\citenamefont{{Lucchesi} and {Peron}}(2014)}]{2014PhRvD..89h2002L}
\bibinfo{author}{\bibfnamefont{D.~M.} \bibnamefont{{Lucchesi}}}
  \bibnamefont{and} \bibinfo{author}{\bibfnamefont{R.}~\bibnamefont{{Peron}}},
  \bibinfo{journal}{\prd} \textbf{\bibinfo{volume}{89}}, \bibinfo{eid}{082002}
  (\bibinfo{year}{2014}).

\bibitem[{\citenamefont{Nordtvedt~Jr.}(1968{\natexlab{b}})}]{nordtvedt1}
\bibinfo{author}{\bibfnamefont{K.}~\bibnamefont{Nordtvedt~Jr.}},
  \bibinfo{journal}{Phys. Rev.} \textbf{\bibinfo{volume}{169}},
  \bibinfo{pages}{1014} (\bibinfo{year}{1968}{\natexlab{b}}).

\bibitem[{\citenamefont{Dicke}(1970)}]{dicke1}
\bibinfo{author}{\bibfnamefont{R.}~\bibnamefont{Dicke}},
  \emph{\bibinfo{title}{Gravitation and the universe}},
  vol.~\bibinfo{volume}{78} of \emph{\bibinfo{series}{Memoirs of the American
  Philosophical Society. Jayne Lecture for 1969}} (\bibinfo{publisher}{American
  Philosophical Society}, \bibinfo{address}{Philadelphia},
  \bibinfo{year}{1970}).

\bibitem[{\citenamefont{{Merkowitz}}(2010)}]{2010LRR....13....7M}
\bibinfo{author}{\bibfnamefont{S.~M.} \bibnamefont{{Merkowitz}}},
  \bibinfo{journal}{Living Rev. Relativity} \textbf{\bibinfo{volume}{13}}
  (\bibinfo{year}{2010}),
  \bibinfo{note}{http://www.living\-reviews.\-org/lrr-2010-7}.

\bibitem[{\citenamefont{Williams
  et~al.}(2004{\natexlab{a}})\citenamefont{Williams, Turyshev, and
  Murphy~Jr}}]{williams04ijmp}
\bibinfo{author}{\bibfnamefont{J.}~\bibnamefont{Williams}},
  \bibinfo{author}{\bibfnamefont{S.}~\bibnamefont{Turyshev}}, \bibnamefont{and}
  \bibinfo{author}{\bibfnamefont{T.}~\bibnamefont{Murphy~Jr}},
  \bibinfo{journal}{Int. J. Mod. Phys. D} \textbf{\bibinfo{volume}{13}},
  \bibinfo{pages}{567} (\bibinfo{year}{2004}{\natexlab{a}}),
  \eprint{gr-qc/0311021}.

\bibitem[{\citenamefont{Bae{\ss}ler et~al.}(1999)\citenamefont{Bae{\ss}ler,
  Heckel, Adelberger, Gundlach, Schmidt, and Swanson}}]{baessler99}
\bibinfo{author}{\bibfnamefont{S.}~\bibnamefont{Bae{\ss}ler}},
  \bibinfo{author}{\bibfnamefont{B.}~\bibnamefont{Heckel}},
  \bibinfo{author}{\bibfnamefont{E.}~\bibnamefont{Adelberger}},
  \bibinfo{author}{\bibfnamefont{J.}~\bibnamefont{Gundlach}},
  \bibinfo{author}{\bibfnamefont{U.}~\bibnamefont{Schmidt}}, \bibnamefont{and}
  \bibinfo{author}{\bibfnamefont{H.}~\bibnamefont{Swanson}},
  \bibinfo{journal}{Phys. Rev. Lett.} \textbf{\bibinfo{volume}{83}},
  \bibinfo{pages}{3585} (\bibinfo{year}{1999}).

\bibitem[{\citenamefont{{Murphy} et~al.}(2012)\citenamefont{{Murphy},
  {Adelberger}, {Battat}, {Hoyle}, {Johnson}, {McMillan}, {Stubbs}, and
  {Swanson}}}]{2012CQGra..29r4005M}
\bibinfo{author}{\bibfnamefont{T.~W.} \bibnamefont{{Murphy}},
  \bibfnamefont{Jr.}}, \bibinfo{author}{\bibfnamefont{E.~G.}
  \bibnamefont{{Adelberger}}}, \bibinfo{author}{\bibfnamefont{J.~B.~R.}
  \bibnamefont{{Battat}}}, \bibinfo{author}{\bibfnamefont{C.~D.}
  \bibnamefont{{Hoyle}}}, \bibinfo{author}{\bibfnamefont{N.~H.}
  \bibnamefont{{Johnson}}}, \bibinfo{author}{\bibfnamefont{R.~J.}
  \bibnamefont{{McMillan}}}, \bibinfo{author}{\bibfnamefont{C.~W.}
  \bibnamefont{{Stubbs}}}, \bibnamefont{and}
  \bibinfo{author}{\bibfnamefont{H.~E.} \bibnamefont{{Swanson}}},
  \bibinfo{journal}{Class. Quantum Grav.} \textbf{\bibinfo{volume}{29}},
  \bibinfo{eid}{184005} (\bibinfo{year}{2012}).

\bibitem[{\citenamefont{Stairs et~al.}(2005)\citenamefont{Stairs, Faulkner,
  Lyne, Kramer, Lorimer, McLaughlin, Manchester, Hobbs, Camilo, Possenti
  et~al.}}]{stairs05}
\bibinfo{author}{\bibfnamefont{I.}~\bibnamefont{Stairs}},
  \bibinfo{author}{\bibfnamefont{A.}~\bibnamefont{Faulkner}},
  \bibinfo{author}{\bibfnamefont{A.}~\bibnamefont{Lyne}},
  \bibinfo{author}{\bibfnamefont{M.}~\bibnamefont{Kramer}},
  \bibinfo{author}{\bibfnamefont{D.}~\bibnamefont{Lorimer}},
  \bibinfo{author}{\bibfnamefont{M.}~\bibnamefont{McLaughlin}},
  \bibinfo{author}{\bibfnamefont{R.}~\bibnamefont{Manchester}},
  \bibinfo{author}{\bibfnamefont{G.}~\bibnamefont{Hobbs}},
  \bibinfo{author}{\bibfnamefont{F.}~\bibnamefont{Camilo}},
  \bibinfo{author}{\bibfnamefont{A.}~\bibnamefont{Possenti}},
  \bibnamefont{et~al.}, \bibinfo{journal}{Astrophys. J.}
  \textbf{\bibinfo{volume}{632}}, \bibinfo{pages}{1060} (\bibinfo{year}{2005}),
  \eprint{astro-ph/0506188}.

\bibitem[{\citenamefont{{Ransom} et~al.}(2014)\citenamefont{{Ransom}, {Stairs},
  {Archibald}, {Hessels}, {Kaplan}, {van Kerkwijk}, {Boyles}, {Deller},
  {Chatterjee}, {Schechtman-Rook} et~al.}}]{2014Natur.505..520R}
\bibinfo{author}{\bibfnamefont{S.~M.} \bibnamefont{{Ransom}}},
  \bibinfo{author}{\bibfnamefont{I.~H.} \bibnamefont{{Stairs}}},
  \bibinfo{author}{\bibfnamefont{A.~M.} \bibnamefont{{Archibald}}},
  \bibinfo{author}{\bibfnamefont{J.~W.~T.} \bibnamefont{{Hessels}}},
  \bibinfo{author}{\bibfnamefont{D.~L.} \bibnamefont{{Kaplan}}},
  \bibinfo{author}{\bibfnamefont{M.~H.} \bibnamefont{{van Kerkwijk}}},
  \bibinfo{author}{\bibfnamefont{J.}~\bibnamefont{{Boyles}}},
  \bibinfo{author}{\bibfnamefont{A.~T.} \bibnamefont{{Deller}}},
  \bibinfo{author}{\bibfnamefont{S.}~\bibnamefont{{Chatterjee}}},
  \bibinfo{author}{\bibfnamefont{A.}~\bibnamefont{{Schechtman-Rook}}},
  \bibnamefont{et~al.}, \bibinfo{journal}{\nat} \textbf{\bibinfo{volume}{505}},
  \bibinfo{pages}{520} (\bibinfo{year}{2014}), \eprint{1401.0535}.

\bibitem[{\citenamefont{{Shao} and {Wex}}(2012)}]{2012CQGra..29u5018S}
\bibinfo{author}{\bibfnamefont{L.}~\bibnamefont{{Shao}}} \bibnamefont{and}
  \bibinfo{author}{\bibfnamefont{N.}~\bibnamefont{{Wex}}},
  \bibinfo{journal}{Class. Quantum Grav.} \textbf{\bibinfo{volume}{29}},
  \bibinfo{pages}{215018} (\bibinfo{year}{2012}), \eprint{1209.4503}.

\bibitem[{\citenamefont{{Shao} et~al.}(2013)\citenamefont{{Shao}, {Caballero},
  {Kramer}, {Wex}, {Champion}, and {Jessner}}}]{2013CQGra..30p5019S}
\bibinfo{author}{\bibfnamefont{L.}~\bibnamefont{{Shao}}},
  \bibinfo{author}{\bibfnamefont{R.~N.} \bibnamefont{{Caballero}}},
  \bibinfo{author}{\bibfnamefont{M.}~\bibnamefont{{Kramer}}},
  \bibinfo{author}{\bibfnamefont{N.}~\bibnamefont{{Wex}}},
  \bibinfo{author}{\bibfnamefont{D.~J.} \bibnamefont{{Champion}}},
  \bibnamefont{and}
  \bibinfo{author}{\bibfnamefont{A.}~\bibnamefont{{Jessner}}},
  \bibinfo{journal}{Class. Quantum Grav.} \textbf{\bibinfo{volume}{30}},
  \bibinfo{eid}{165019} (\bibinfo{year}{2013}), \eprint{1307.2552}.

\bibitem[{\citenamefont{{Shao} and {Wex}}(2013)}]{2013CQGra..30p5020S}
\bibinfo{author}{\bibfnamefont{L.}~\bibnamefont{{Shao}}} \bibnamefont{and}
  \bibinfo{author}{\bibfnamefont{N.}~\bibnamefont{{Wex}}},
  \bibinfo{journal}{Class. Quantum Grav.} \textbf{\bibinfo{volume}{30}},
  \bibinfo{eid}{165020} (\bibinfo{year}{2013}), \eprint{1307.2637}.

\bibitem[{\citenamefont{{Konopliv} et~al.}(2011)\citenamefont{{Konopliv},
  {Asmar}, {Folkner}, {Karatekin}, {Nunes}, {Smrekar}, {Yoder}, and
  {Zuber}}}]{2011Icar..211..401K}
\bibinfo{author}{\bibfnamefont{A.~S.} \bibnamefont{{Konopliv}}},
  \bibinfo{author}{\bibfnamefont{S.~W.} \bibnamefont{{Asmar}}},
  \bibinfo{author}{\bibfnamefont{W.~M.} \bibnamefont{{Folkner}}},
  \bibinfo{author}{\bibfnamefont{{\"O}.}~\bibnamefont{{Karatekin}}},
  \bibinfo{author}{\bibfnamefont{D.~C.} \bibnamefont{{Nunes}}},
  \bibinfo{author}{\bibfnamefont{S.~E.} \bibnamefont{{Smrekar}}},
  \bibinfo{author}{\bibfnamefont{C.~F.} \bibnamefont{{Yoder}}},
  \bibnamefont{and} \bibinfo{author}{\bibfnamefont{M.~T.}
  \bibnamefont{{Zuber}}}, \bibinfo{journal}{Icarus}
  \textbf{\bibinfo{volume}{211}}, \bibinfo{pages}{401} (\bibinfo{year}{2011}).

\bibitem[{\citenamefont{Williams
  et~al.}(2004{\natexlab{b}})\citenamefont{Williams, Turyshev, and
  Boggs}}]{williams04}
\bibinfo{author}{\bibfnamefont{J.}~\bibnamefont{Williams}},
  \bibinfo{author}{\bibfnamefont{S.}~\bibnamefont{Turyshev}}, \bibnamefont{and}
  \bibinfo{author}{\bibfnamefont{D.}~\bibnamefont{Boggs}},
  \bibinfo{journal}{Phys. Rev. Lett.} \textbf{\bibinfo{volume}{93}},
  \bibinfo{eid}{261101} (\bibinfo{year}{2004}{\natexlab{b}}),
  \eprint{gr-qc/0411113}.

\bibitem[{\citenamefont{{Deller} et~al.}(2008)\citenamefont{{Deller},
  {Verbiest}, {Tingay}, and {Bailes}}}]{2008ApJ...685L..67D}
\bibinfo{author}{\bibfnamefont{A.~T.} \bibnamefont{{Deller}}},
  \bibinfo{author}{\bibfnamefont{J.~P.~W.} \bibnamefont{{Verbiest}}},
  \bibinfo{author}{\bibfnamefont{S.~J.} \bibnamefont{{Tingay}}},
  \bibnamefont{and} \bibinfo{author}{\bibfnamefont{M.}~\bibnamefont{{Bailes}}},
  \bibinfo{journal}{\apjl} \textbf{\bibinfo{volume}{685}}, \bibinfo{pages}{L67}
  (\bibinfo{year}{2008}), \eprint{0808.1594}.

\bibitem[{\citenamefont{{Lazaridis} et~al.}(2009)\citenamefont{{Lazaridis},
  {Wex}, {Jessner}, {Kramer}, {Stappers}, {Janssen}, {Desvignes}, {Purver},
  {Cognard}, {Theureau} et~al.}}]{2009MNRAS.400..805L}
\bibinfo{author}{\bibfnamefont{K.}~\bibnamefont{{Lazaridis}}},
  \bibinfo{author}{\bibfnamefont{N.}~\bibnamefont{{Wex}}},
  \bibinfo{author}{\bibfnamefont{A.}~\bibnamefont{{Jessner}}},
  \bibinfo{author}{\bibfnamefont{M.}~\bibnamefont{{Kramer}}},
  \bibinfo{author}{\bibfnamefont{B.~W.} \bibnamefont{{Stappers}}},
  \bibinfo{author}{\bibfnamefont{G.~H.} \bibnamefont{{Janssen}}},
  \bibinfo{author}{\bibfnamefont{G.}~\bibnamefont{{Desvignes}}},
  \bibinfo{author}{\bibfnamefont{M.~B.} \bibnamefont{{Purver}}},
  \bibinfo{author}{\bibfnamefont{I.}~\bibnamefont{{Cognard}}},
  \bibinfo{author}{\bibfnamefont{G.}~\bibnamefont{{Theureau}}},
  \bibnamefont{et~al.}, \bibinfo{journal}{\mnras}
  \textbf{\bibinfo{volume}{400}}, \bibinfo{pages}{805} (\bibinfo{year}{2009}),
  \eprint{0908.0285}.

\bibitem[{\citenamefont{Guenther et~al.}(1998)\citenamefont{Guenther, Krauss,
  and Demarque}}]{guenther98}
\bibinfo{author}{\bibfnamefont{D.}~\bibnamefont{Guenther}},
  \bibinfo{author}{\bibfnamefont{L.}~\bibnamefont{Krauss}}, \bibnamefont{and}
  \bibinfo{author}{\bibfnamefont{P.}~\bibnamefont{Demarque}},
  \bibinfo{journal}{Astrophys. J.} \textbf{\bibinfo{volume}{498}},
  \bibinfo{pages}{871} (\bibinfo{year}{1998}).

\bibitem[{\citenamefont{Copi et~al.}(2004)\citenamefont{Copi, Davis, and
  Krauss}}]{copi04}
\bibinfo{author}{\bibfnamefont{C.}~\bibnamefont{Copi}},
  \bibinfo{author}{\bibfnamefont{A.}~\bibnamefont{Davis}}, \bibnamefont{and}
  \bibinfo{author}{\bibfnamefont{L.}~\bibnamefont{Krauss}},
  \bibinfo{journal}{Phys. Rev. Lett.} \textbf{\bibinfo{volume}{92}},
  \bibinfo{eid}{171301} (\bibinfo{year}{2004}), \eprint{astro-ph/0311334}.

\bibitem[{\citenamefont{Bambi et~al.}(2005)\citenamefont{Bambi, Giannotti, and
  Villante}}]{bambi04}
\bibinfo{author}{\bibfnamefont{C.}~\bibnamefont{Bambi}},
  \bibinfo{author}{\bibfnamefont{M.}~\bibnamefont{Giannotti}},
  \bibnamefont{and} \bibinfo{author}{\bibfnamefont{F.}~\bibnamefont{Villante}},
  \bibinfo{journal}{Phys. Rev. D} \textbf{\bibinfo{volume}{71}},
  \bibinfo{pages}{123524} (\bibinfo{year}{2005}), \eprint{astro-ph/0503502}.

\bibitem[{\citenamefont{{Everitt} et~al.}(2011)\citenamefont{{Everitt},
  {Debra}, {Parkinson}, {Turneaure}, {Conklin}, {Heifetz}, {Keiser},
  {Silbergleit}, {Holmes}, {Kolodziejczak} et~al.}}]{2011PhRvL.106v1101E}
\bibinfo{author}{\bibfnamefont{C.~W.~F.} \bibnamefont{{Everitt}}},
  \bibinfo{author}{\bibfnamefont{D.~B.} \bibnamefont{{Debra}}},
  \bibinfo{author}{\bibfnamefont{B.~W.} \bibnamefont{{Parkinson}}},
  \bibinfo{author}{\bibfnamefont{J.~P.} \bibnamefont{{Turneaure}}},
  \bibinfo{author}{\bibfnamefont{J.~W.} \bibnamefont{{Conklin}}},
  \bibinfo{author}{\bibfnamefont{M.~I.} \bibnamefont{{Heifetz}}},
  \bibinfo{author}{\bibfnamefont{G.~M.} \bibnamefont{{Keiser}}},
  \bibinfo{author}{\bibfnamefont{A.~S.} \bibnamefont{{Silbergleit}}},
  \bibinfo{author}{\bibfnamefont{T.}~\bibnamefont{{Holmes}}},
  \bibinfo{author}{\bibfnamefont{J.}~\bibnamefont{{Kolodziejczak}}},
  \bibnamefont{et~al.}, \bibinfo{journal}{Phys. Rev. Lett.}
  \textbf{\bibinfo{volume}{106}}, \bibinfo{eid}{221101} (\bibinfo{year}{2011}),
  \eprint{1105.3456}.

\bibitem[{\citenamefont{Ciufolini and Pavlis}(2004)}]{ciufolini04}
\bibinfo{author}{\bibfnamefont{I.}~\bibnamefont{Ciufolini}} \bibnamefont{and}
  \bibinfo{author}{\bibfnamefont{E.}~\bibnamefont{Pavlis}},
  \bibinfo{journal}{Nature} \textbf{\bibinfo{volume}{431}},
  \bibinfo{pages}{958} (\bibinfo{year}{2004}).

\bibitem[{\citenamefont{{Ciufolini} et~al.}(2006)\citenamefont{{Ciufolini},
  {Pavlis}, and {Peron}}}]{2006NewA...11..527C}
\bibinfo{author}{\bibfnamefont{I.}~\bibnamefont{{Ciufolini}}},
  \bibinfo{author}{\bibfnamefont{E.~C.} \bibnamefont{{Pavlis}}},
  \bibnamefont{and} \bibinfo{author}{\bibfnamefont{R.}~\bibnamefont{{Peron}}},
  \bibinfo{journal}{\na} \textbf{\bibinfo{volume}{11}}, \bibinfo{pages}{527}
  (\bibinfo{year}{2006}).

\bibitem[{\citenamefont{{Weisberg} et~al.}(2010)\citenamefont{{Weisberg},
  {Nice}, and {Taylor}}}]{2010ApJ...722.1030W}
\bibinfo{author}{\bibfnamefont{J.~M.} \bibnamefont{{Weisberg}}},
  \bibinfo{author}{\bibfnamefont{D.~J.} \bibnamefont{{Nice}}},
  \bibnamefont{and} \bibinfo{author}{\bibfnamefont{J.~H.}
  \bibnamefont{{Taylor}}}, \bibinfo{journal}{\apj}
  \textbf{\bibinfo{volume}{722}}, \bibinfo{pages}{1030} (\bibinfo{year}{2010}),
  \eprint{1011.0718}.

\bibitem[{\citenamefont{Damour and Taylor}(1992)}]{DamourTaylor92}
\bibinfo{author}{\bibfnamefont{T.}~\bibnamefont{Damour}} \bibnamefont{and}
  \bibinfo{author}{\bibfnamefont{J.}~\bibnamefont{Taylor}},
  \bibinfo{journal}{Phys. Rev. D} \textbf{\bibinfo{volume}{45}},
  \bibinfo{pages}{1840} (\bibinfo{year}{1992}).

\bibitem[{\citenamefont{Kramer}(1998)}]{kramer}
\bibinfo{author}{\bibfnamefont{M.}~\bibnamefont{Kramer}},
  \bibinfo{journal}{Astrophys. J.} \textbf{\bibinfo{volume}{509}},
  \bibinfo{pages}{856} (\bibinfo{year}{1998}), \eprint{astro-ph/9808127}.

\bibitem[{\citenamefont{Weisberg and Taylor}(2002)}]{WeisbergTaylor02}
\bibinfo{author}{\bibfnamefont{J.}~\bibnamefont{Weisberg}} \bibnamefont{and}
  \bibinfo{author}{\bibfnamefont{J.}~\bibnamefont{Taylor}},
  \bibinfo{journal}{Astrophys. J.} \textbf{\bibinfo{volume}{576}},
  \bibinfo{pages}{942} (\bibinfo{year}{2002}), \eprint{astro-ph/0205280}.

\bibitem[{\citenamefont{Burgay et~al.}(2003)\citenamefont{Burgay, D'Amico,
  Possenti, Manchester, Lyne, Joshi, McLaughlin, Kramer, Sarkissian, Camilo
  et~al.}}]{burgay03}
\bibinfo{author}{\bibfnamefont{M.}~\bibnamefont{Burgay}},
  \bibinfo{author}{\bibfnamefont{N.}~\bibnamefont{D'Amico}},
  \bibinfo{author}{\bibfnamefont{A.}~\bibnamefont{Possenti}},
  \bibinfo{author}{\bibfnamefont{R.}~\bibnamefont{Manchester}},
  \bibinfo{author}{\bibfnamefont{A.}~\bibnamefont{Lyne}},
  \bibinfo{author}{\bibfnamefont{B.}~\bibnamefont{Joshi}},
  \bibinfo{author}{\bibfnamefont{M.}~\bibnamefont{McLaughlin}},
  \bibinfo{author}{\bibfnamefont{M.}~\bibnamefont{Kramer}},
  \bibinfo{author}{\bibfnamefont{J.}~\bibnamefont{Sarkissian}},
  \bibinfo{author}{\bibfnamefont{F.}~\bibnamefont{Camilo}},
  \bibnamefont{et~al.}, \bibinfo{journal}{Nature}
  \textbf{\bibinfo{volume}{426}}, \bibinfo{pages}{531} (\bibinfo{year}{2003}),
  \eprint{astro-ph/0312071}.

\bibitem[{\citenamefont{Lyne et~al.}(2004)\citenamefont{Lyne, Burgay, Kramer,
  Possenti, Manchester, Camilo, McLaughlin, Lorimer, D'Amico, Joshi
  et~al.}}]{lyne04}
\bibinfo{author}{\bibfnamefont{A.}~\bibnamefont{Lyne}},
  \bibinfo{author}{\bibfnamefont{M.}~\bibnamefont{Burgay}},
  \bibinfo{author}{\bibfnamefont{M.}~\bibnamefont{Kramer}},
  \bibinfo{author}{\bibfnamefont{A.}~\bibnamefont{Possenti}},
  \bibinfo{author}{\bibfnamefont{R.}~\bibnamefont{Manchester}},
  \bibinfo{author}{\bibfnamefont{F.}~\bibnamefont{Camilo}},
  \bibinfo{author}{\bibfnamefont{M.}~\bibnamefont{McLaughlin}},
  \bibinfo{author}{\bibfnamefont{D.}~\bibnamefont{Lorimer}},
  \bibinfo{author}{\bibfnamefont{N.}~\bibnamefont{D'Amico}},
  \bibinfo{author}{\bibfnamefont{B.}~\bibnamefont{Joshi}},
  \bibnamefont{et~al.}, \bibinfo{journal}{Science}
  \textbf{\bibinfo{volume}{303}}, \bibinfo{pages}{1153} (\bibinfo{year}{2004}),
  \eprint{astro-ph/0401086}.

\bibitem[{\citenamefont{{Kramer} et~al.}(2006)\citenamefont{{Kramer}, {Stairs},
  {Manchester}, {McLaughlin}, {Lyne}, {Ferdman}, {Burgay}, {Lorimer},
  {Possenti}, {D'Amico} et~al.}}]{2006Sci...314...97K}
\bibinfo{author}{\bibfnamefont{M.}~\bibnamefont{{Kramer}}},
  \bibinfo{author}{\bibfnamefont{I.~H.} \bibnamefont{{Stairs}}},
  \bibinfo{author}{\bibfnamefont{R.~N.} \bibnamefont{{Manchester}}},
  \bibinfo{author}{\bibfnamefont{M.~A.} \bibnamefont{{McLaughlin}}},
  \bibinfo{author}{\bibfnamefont{A.~G.} \bibnamefont{{Lyne}}},
  \bibinfo{author}{\bibfnamefont{R.~D.} \bibnamefont{{Ferdman}}},
  \bibinfo{author}{\bibfnamefont{M.}~\bibnamefont{{Burgay}}},
  \bibinfo{author}{\bibfnamefont{D.~R.} \bibnamefont{{Lorimer}}},
  \bibinfo{author}{\bibfnamefont{A.}~\bibnamefont{{Possenti}}},
  \bibinfo{author}{\bibfnamefont{N.}~\bibnamefont{{D'Amico}}},
  \bibnamefont{et~al.}, \bibinfo{journal}{Science}
  \textbf{\bibinfo{volume}{314}}, \bibinfo{pages}{97} (\bibinfo{year}{2006}),
  \eprint{arXiv:astro-ph/0609417}.

\bibitem[{\citenamefont{{Breton} et~al.}(2008)\citenamefont{{Breton}, {Kaspi},
  {Kramer}, {McLaughlin}, {Lyutikov}, {Ransom}, {Stairs}, {Ferdman}, {Camilo},
  and {Possenti}}}]{2008Sci...321..104B}
\bibinfo{author}{\bibfnamefont{R.~P.} \bibnamefont{{Breton}}},
  \bibinfo{author}{\bibfnamefont{V.~M.} \bibnamefont{{Kaspi}}},
  \bibinfo{author}{\bibfnamefont{M.}~\bibnamefont{{Kramer}}},
  \bibinfo{author}{\bibfnamefont{M.~A.} \bibnamefont{{McLaughlin}}},
  \bibinfo{author}{\bibfnamefont{M.}~\bibnamefont{{Lyutikov}}},
  \bibinfo{author}{\bibfnamefont{S.~M.} \bibnamefont{{Ransom}}},
  \bibinfo{author}{\bibfnamefont{I.~H.} \bibnamefont{{Stairs}}},
  \bibinfo{author}{\bibfnamefont{R.~D.} \bibnamefont{{Ferdman}}},
  \bibinfo{author}{\bibfnamefont{F.}~\bibnamefont{{Camilo}}}, \bibnamefont{and}
  \bibinfo{author}{\bibfnamefont{A.}~\bibnamefont{{Possenti}}},
  \bibinfo{journal}{Science} \textbf{\bibinfo{volume}{321}},
  \bibinfo{pages}{104} (\bibinfo{year}{2008}), \eprint{0807.2644}.

\bibitem[{\citenamefont{{Freire} et~al.}(2012)\citenamefont{{Freire}, {Wex},
  {Esposito-Far{\`e}se}, {Verbiest}, {Bailes}, {Jacoby}, {Kramer}, {Stairs},
  {Antoniadis}, and {Janssen}}}]{2012MNRAS.423.3328F}
\bibinfo{author}{\bibfnamefont{P.~C.~C.} \bibnamefont{{Freire}}},
  \bibinfo{author}{\bibfnamefont{N.}~\bibnamefont{{Wex}}},
  \bibinfo{author}{\bibfnamefont{G.}~\bibnamefont{{Esposito-Far{\`e}se}}},
  \bibinfo{author}{\bibfnamefont{J.~P.~W.} \bibnamefont{{Verbiest}}},
  \bibinfo{author}{\bibfnamefont{M.}~\bibnamefont{{Bailes}}},
  \bibinfo{author}{\bibfnamefont{B.~A.} \bibnamefont{{Jacoby}}},
  \bibinfo{author}{\bibfnamefont{M.}~\bibnamefont{{Kramer}}},
  \bibinfo{author}{\bibfnamefont{I.~H.} \bibnamefont{{Stairs}}},
  \bibinfo{author}{\bibfnamefont{J.}~\bibnamefont{{Antoniadis}}},
  \bibnamefont{and} \bibinfo{author}{\bibfnamefont{G.~H.}
  \bibnamefont{{Janssen}}}, \bibinfo{journal}{\mnras}
  \textbf{\bibinfo{volume}{423}}, \bibinfo{pages}{3328} (\bibinfo{year}{2012}),
  \eprint{1205.1450}.

\bibitem[{\citenamefont{{Bhat} et~al.}(2008)\citenamefont{{Bhat}, {Bailes}, and
  {Verbiest}}}]{2008PhRvD..77l4017B}
\bibinfo{author}{\bibfnamefont{N.~D.~R.} \bibnamefont{{Bhat}}},
  \bibinfo{author}{\bibfnamefont{M.}~\bibnamefont{{Bailes}}}, \bibnamefont{and}
  \bibinfo{author}{\bibfnamefont{J.~P.~W.} \bibnamefont{{Verbiest}}},
  \bibinfo{journal}{\prd} \textbf{\bibinfo{volume}{77}}, \bibinfo{eid}{124017}
  (\bibinfo{year}{2008}), \eprint{0804.0956}.

\bibitem[{\citenamefont{Damour and Esposito-Far{\`{e}}se}(1998)}]{DamourEspo98}
\bibinfo{author}{\bibfnamefont{T.}~\bibnamefont{Damour}} \bibnamefont{and}
  \bibinfo{author}{\bibfnamefont{G.}~\bibnamefont{Esposito-Far{\`{e}}se}},
  \bibinfo{journal}{Phys. Rev. D} \textbf{\bibinfo{volume}{58}},
  \bibinfo{eid}{042001} (\bibinfo{year}{1998}), \eprint{gr-qc/9803031}.

\bibitem[{\citenamefont{Creighton and Anderson}(2011)}]{CreightonAnderson}
\bibinfo{author}{\bibfnamefont{J.}~\bibnamefont{Creighton}} \bibnamefont{and}
  \bibinfo{author}{\bibfnamefont{W.}~\bibnamefont{Anderson}},
  \emph{\bibinfo{title}{Gravitational-Wave Physics and Astronomy: An
  Introduction to Theory, Experiment and Data Analysis}}
  (\bibinfo{publisher}{Wiley}, \bibinfo{address}{Cambridge; New York},
  \bibinfo{year}{2011}).

\bibitem[{\citenamefont{{Amaro-Seoane}
  et~al.}(2012)\citenamefont{{Amaro-Seoane}, {Aoudia}, {Babak}, {Bin{\'e}truy},
  {Berti}, {Boh{\'e}}, {Caprini}, {Colpi}, {Cornish}, {Danzmann}
  et~al.}}]{2012CQGra..29l4016A}
\bibinfo{author}{\bibfnamefont{P.}~\bibnamefont{{Amaro-Seoane}}},
  \bibinfo{author}{\bibfnamefont{S.}~\bibnamefont{{Aoudia}}},
  \bibinfo{author}{\bibfnamefont{S.}~\bibnamefont{{Babak}}},
  \bibinfo{author}{\bibfnamefont{P.}~\bibnamefont{{Bin{\'e}truy}}},
  \bibinfo{author}{\bibfnamefont{E.}~\bibnamefont{{Berti}}},
  \bibinfo{author}{\bibfnamefont{A.}~\bibnamefont{{Boh{\'e}}}},
  \bibinfo{author}{\bibfnamefont{C.}~\bibnamefont{{Caprini}}},
  \bibinfo{author}{\bibfnamefont{M.}~\bibnamefont{{Colpi}}},
  \bibinfo{author}{\bibfnamefont{N.~J.} \bibnamefont{{Cornish}}},
  \bibinfo{author}{\bibfnamefont{K.}~\bibnamefont{{Danzmann}}},
  \bibnamefont{et~al.}, \bibinfo{journal}{Class. Quantum Grav.}
  \textbf{\bibinfo{volume}{29}}, \bibinfo{eid}{124016} (\bibinfo{year}{2012}),
  \eprint{1202.0839}.

\bibitem[{\citenamefont{{Blanchet}}(2006)}]{BlanchetLRR}
\bibinfo{author}{\bibfnamefont{L.}~\bibnamefont{{Blanchet}}},
  \bibinfo{journal}{Living Rev. Relativity} \textbf{\bibinfo{volume}{9}}
  (\bibinfo{year}{2006}),
  \bibinfo{note}{http://www.\-living\-reviews.\-org/lrr-2006-4}.

\bibitem[{\citenamefont{{Yunes} and {Pretorius}}(2009)}]{2009PhRvD..80l2003Y}
\bibinfo{author}{\bibfnamefont{N.}~\bibnamefont{{Yunes}}} \bibnamefont{and}
  \bibinfo{author}{\bibfnamefont{F.}~\bibnamefont{{Pretorius}}},
  \bibinfo{journal}{\prd} \textbf{\bibinfo{volume}{80}}, \bibinfo{eid}{122003}
  (\bibinfo{year}{2009}), \eprint{0909.3328}.

\bibitem[{\citenamefont{{Mishra} et~al.}(2010)\citenamefont{{Mishra}, {Arun},
  {Iyer}, and {Sathyaprakash}}}]{2010PhRvD..82f4010M}
\bibinfo{author}{\bibfnamefont{C.~K.} \bibnamefont{{Mishra}}},
  \bibinfo{author}{\bibfnamefont{K.~G.} \bibnamefont{{Arun}}},
  \bibinfo{author}{\bibfnamefont{B.~R.} \bibnamefont{{Iyer}}},
  \bibnamefont{and} \bibinfo{author}{\bibfnamefont{B.~S.}
  \bibnamefont{{Sathyaprakash}}}, \bibinfo{journal}{\prd}
  \textbf{\bibinfo{volume}{82}}, \bibinfo{eid}{064010} (\bibinfo{year}{2010}),
  \eprint{1005.0304}.

\bibitem[{\citenamefont{Will}(1998)}]{graviton}
\bibinfo{author}{\bibfnamefont{C.}~\bibnamefont{Will}}, \bibinfo{journal}{Phys.
  Rev. D} \textbf{\bibinfo{volume}{57}}, \bibinfo{pages}{2061}
  (\bibinfo{year}{1998}), \eprint{gr-qc/9709011}.

\bibitem[{\citenamefont{{Gair} et~al.}(2013)\citenamefont{{Gair}, {Vallisneri},
  {Larson}, and {Baker}}}]{2013LRR....16....7G}
\bibinfo{author}{\bibfnamefont{J.~R.} \bibnamefont{{Gair}}},
  \bibinfo{author}{\bibfnamefont{M.}~\bibnamefont{{Vallisneri}}},
  \bibinfo{author}{\bibfnamefont{S.~L.} \bibnamefont{{Larson}}},
  \bibnamefont{and} \bibinfo{author}{\bibfnamefont{J.~G.}
  \bibnamefont{{Baker}}}, \bibinfo{journal}{Living Rev. Relativity}
  \textbf{\bibinfo{volume}{16}} (\bibinfo{year}{2013}),
  \bibinfo{note}{http://www.living\-reviews.\-org/lrr-2013-7},
  \eprint{1212.5575}.

\bibitem[{\citenamefont{{Yunes} and {Siemens}}(2013)}]{2013LRR....16....9Y}
\bibinfo{author}{\bibfnamefont{N.}~\bibnamefont{{Yunes}}} \bibnamefont{and}
  \bibinfo{author}{\bibfnamefont{X.}~\bibnamefont{{Siemens}}},
  \bibinfo{journal}{Living Rev. Relativity} \textbf{\bibinfo{volume}{16}},
  \bibinfo{eid}{lrr-2013-9} (\bibinfo{year}{2013}),
  \bibinfo{note}{http://www.living\-reviews.\-org/lrr-2013-9},
  \eprint{1304.3473}.

\bibitem[{\citenamefont{{Narayan} and
  {McClintock}}(2008)}]{2008NewAR..51..733N}
\bibinfo{author}{\bibfnamefont{R.}~\bibnamefont{{Narayan}}} \bibnamefont{and}
  \bibinfo{author}{\bibfnamefont{J.~E.} \bibnamefont{{McClintock}}},
  \bibinfo{journal}{New Astron. Rev.} \textbf{\bibinfo{volume}{51}},
  \bibinfo{pages}{733} (\bibinfo{year}{2008}), \eprint{0803.0322}.

\bibitem[{\citenamefont{Psaltis}(2004)}]{Psaltis04}
\bibinfo{author}{\bibfnamefont{D.}~\bibnamefont{Psaltis}}, in
  \emph{\bibinfo{booktitle}{X-Ray Timing 2003: Rossi and Beyond}}, edited by
  \bibinfo{editor}{\bibfnamefont{P.}~\bibnamefont{Kaaret}},
  \bibinfo{editor}{\bibfnamefont{F.}~\bibnamefont{Lamb}}, \bibnamefont{and}
  \bibinfo{editor}{\bibfnamefont{J.}~\bibnamefont{Swank}}
  (\bibinfo{publisher}{American Institute of Physics},
  \bibinfo{address}{Melville}, \bibinfo{year}{2004}), vol.
  \bibinfo{volume}{714} of \emph{\bibinfo{series}{AIP Conference Proceedings}},
  pp. \bibinfo{pages}{29--35}, \eprint{astro-ph/0402213}.

\bibitem[{\citenamefont{{Reynolds}}(2013{\natexlab{a}})}]{2013SSRv..tmp...81R}
\bibinfo{author}{\bibfnamefont{C.~S.} \bibnamefont{{Reynolds}}},
  \bibinfo{journal}{Space Sci. Rev. On Line} pp. \bibinfo{pages}{1--18}
  (\bibinfo{year}{2013}{\natexlab{a}}), \eprint{1302.3260},
  \urlprefix\url{http://adsabs.harvard.edu/abs/2013SSRv..tmp...81R}.

\bibitem[{\citenamefont{{Reynolds}}(2013{\natexlab{b}})}]{2013CQGra..30x4004R}
\bibinfo{author}{\bibfnamefont{C.~S.} \bibnamefont{{Reynolds}}},
  \bibinfo{journal}{Class. Quantum Grav.} \textbf{\bibinfo{volume}{30}},
  \bibinfo{eid}{244004} (\bibinfo{year}{2013}{\natexlab{b}}),
  \eprint{1307.3246}.

\bibitem[{\citenamefont{{Doeleman} et~al.}(2009)\citenamefont{{Doeleman},
  {Agol}, {Backer}, {Baganoff}, {Bower}, {Broderick}, {Fabian}, {Fish},
  {Gammie}, {Ho} et~al.}}]{2009astro2010S..68D}
\bibinfo{author}{\bibfnamefont{S.}~\bibnamefont{{Doeleman}}},
  \bibinfo{author}{\bibfnamefont{E.}~\bibnamefont{{Agol}}},
  \bibinfo{author}{\bibfnamefont{D.}~\bibnamefont{{Backer}}},
  \bibinfo{author}{\bibfnamefont{F.}~\bibnamefont{{Baganoff}}},
  \bibinfo{author}{\bibfnamefont{G.~C.} \bibnamefont{{Bower}}},
  \bibinfo{author}{\bibfnamefont{A.}~\bibnamefont{{Broderick}}},
  \bibinfo{author}{\bibfnamefont{A.}~\bibnamefont{{Fabian}}},
  \bibinfo{author}{\bibfnamefont{V.}~\bibnamefont{{Fish}}},
  \bibinfo{author}{\bibfnamefont{C.}~\bibnamefont{{Gammie}}},
  \bibinfo{author}{\bibfnamefont{P.}~\bibnamefont{{Ho}}}, \bibnamefont{et~al.},
  in \emph{\bibinfo{booktitle}{Astro2010: The Astronomy and Astrophysics
  Decadal Survey}} (\bibinfo{year}{2009}), p.~\bibinfo{pages}{68},
  \eprint{0906.3899}.

\bibitem[{\citenamefont{{Will}}(2008)}]{2008ApJ...674L..25W}
\bibinfo{author}{\bibfnamefont{C.}~\bibnamefont{{Will}}},
  \bibinfo{journal}{\apjl} \textbf{\bibinfo{volume}{674}}, \bibinfo{pages}{L25}
  (\bibinfo{year}{2008}), \eprint{0711.1677}.

\bibitem[{\citenamefont{{Psaltis}}(2008)}]{2008LRR....11....9P}
\bibinfo{author}{\bibfnamefont{D.}~\bibnamefont{{Psaltis}}},
  \bibinfo{journal}{Living Rev. Relativity} \textbf{\bibinfo{volume}{11}}
  (\bibinfo{year}{2008}),
  \bibinfo{note}{http://www.\-living\-reviews.org/lrr-2008-9},
  \eprint{0806.1531}.

\end{thebibliography}
\end{document}